\documentclass[12pt]{iopart}
\usepackage{iopams}
\usepackage{graphicx}
\usepackage{tensor}
\usepackage{color}
\usepackage{url}

\definecolor{red}{rgb}{1,0,0}

\eqnobysec

\begin{document}
\title{Chaotic exits from a weakly magnetized Schwarzschild black hole}

\author{Joshua Bautista and Ian Vega}
\address{National Institute of Physics, University of the Philippines, Diliman, Quezon City 1101, Philippines}
\eads{\mailto{jbautista@nip.upd.edu.ph}, \mailto{ivega@nip.upd.edu.ph}}

\begin{abstract}
A charged particle kicked from an initial circular orbit around a weakly magnetized Schwarzschild black hole undergoes transient chaotic motion before either getting captured by the black hole or escaping upstream or downstream with respect to the direction of the magnetic field. These final states form basins of attraction in the space of initial states. 
We provide a detailed numerical study of the basin structure of this initial state space. We find it to possess the peculiar Wada property: each of its basin boundaries is shared by all three basins. Using basin entropy as a measure, we show that uncertainty in predicting the final exit state increases with stronger magnetic interaction. We also present an approximate analytic expression of the critical escape energy for a vertically-kicked charged particle, and discuss how this depends on the strength of the magnetic interaction.
\end{abstract}

\maketitle

\section{Introduction}
Deterministic dynamical systems in nature can be extremely sensitive to initial conditions. The challenge this poses to predictability remains a staple concern in all fields of modern science \cite{Zeraoulia2012}. Within this vast subject the phenomenon of chaotic scattering concerns a particle that goes through a transient period of chaotic behavior as it interacts with some localized scattering potential before settling into some final state (see \cite{Seoane2013} for a recent review). These final states (or groups of final states) are so-called attractors, and in chaotic systems it can be difficult to predict which attractor results from a particular initial condition. Well-known examples of systems that exhibit chaotic scattering are the H\`{e}non-Heiles system \cite{Barrio2010}, the four-hill potential system \cite{Bleher1990}, and the three hard-disk scattering system \cite{Gaspard1998}.

One useful approach to studying chaotic scattering focuses on the structural properties of basins of attraction in phase space. A basin of attraction is a subset of phase space consisting of initial conditions that will asymptotically evolve into a common attractor. In practice, basins of attraction are commonly visualized and analyzed with so-called basin plots, which are constructed by taking a two dimensional slice of phase space and assigning a color to each basin. The resulting visualization provides the distribution of asymptotic states on a slice of phase space. A dynamical system can have multiple basins, separated by either regular or fractal boundaries.

Relativistic particle dynamics is a rich source of chaotic dynamical systems. In this paper, we revisit the nonlinear dynamics of a charged particle in a weakly magnetized Schwarzschild black hole with the goal of characterizing its basins of attraction. In particular, we consider a non-rotating black hole of mass $M$ immersed in a static axisymmetric test magnetic field that is homogeneous at infinity, where it has strength $B$. The magnetic field is considered weak because it does not affect the geometry around the black hole, even though it significantly affects the motion of the particle.

Al Zahrani et al \cite{AlZahrani2013} were the first to study this system with basin plots, showing that its boundaries are fractal. With these basin plots, they also determined the critical escape velocity for a charged particle that is vertically kicked from its innermost stable circular orbit (ISCO). A similar analysis was performed in \cite{AlZahrani2014} for the spinning black hole case. The present work extends \cite{AlZahrani2013} by adopting a general stable circular orbit as the starting orbit of our charge. In generalizing, we gain access to richer dynamics and are able to study how the critical escape energy depends on the strength of the interaction between the magnetic field and the charged particle.
In addition, using newly proposed concepts and methods from nonlinear dynamics \cite{Daza2016,Daza2018}, we study some topological and structural aspects of the basins that were left uninvestigated in \cite{AlZahrani2013}. In particular, with the merging method of \cite{Daza2018} we show that the basin boundaries are not just fractal, they also possess the peculiar Wada property, which means that they are boundaries shared by more than two basins.
We also quantify the ``uncertainty" of our dynamical system, using the concept of \emph{basin entropy} that was introduced in \cite{Daza2016}, and investigate how this uncertainty is affected by the strength of the interaction between the magnetic field and the charged particle.

Charged particle dynamics in magnetized black holes has inspired much work since Al Zahrani et al \cite{AlZahrani2013}. Stuchlík and Kolo\v{s} \cite{Stuchlik2016} as well as Kop\'{a}\v{c}ek and Karas \cite{Kopacek2018} studied the dynamics of initially neutral particles undergoing ionization in the vicinity of weakly magnetized rotating black holes. Both studies demonstrated that charged particles can be accelerated to very high speeds along the direction of the magnetic field, with stronger accelerations being achieved for larger black hole spins and stronger magnetic interactions. Tursunov et al \cite{Tursunov2018} provides an interesting discussion of acceleration mechanisms such as this, in the context of ultra-high-energy cosmic rays from supermassive black holes. Chaotic properties were also discussed in \cite{Stuchlik2016,Kopacek2018}. In particular, fractal features were shown to be present in the scattering functions of \cite{Stuchlik2016} as well as in the system basins of \cite{Kopacek2018}.

The study of magnetized black holes finds physical relevance in the study of astrophysical jets associated with black holes in active galactic nuclei \cite{Horowitz1999}, which involves numerical simulations based on complicated magnetohydrodynamic equations \cite{Punsly2009}.
The system we consider here is hugely simplified in comparison, but this loss of astrophysical realism allows us to exhibit interesting aspects that may be entirely buried in full magnetohydrodynamic simulations. Our computational experiments aim to contribute to the long and fruitful synergy between relativistic astrophysics and nonlinear dynamics.

The paper is organized as follows: In \sref{sec:particledynamics} we derive the equations of motion for a charged particle in the vicinity of a weakly magnetized Schwarzschild black hole.
In \sref{sec:effectivepotential}, we derive dimensionless forms of the equations of motion and present the effective potential of the system.
We show that the system contains three asymptotic states: escape to $z\rightarrow \pm \infty$ and black hole capture.
Then in \sref{sec:basinplotscase1}, we present the basin plots of a particular slice of phase space corresponding to a charged particle initially in a stable circular orbit experiencing a purely vertical kick.
We present an approximate analytic expression for the critical escape energy as a function of $b$.
\Sref{sec:wada} demonstrates the Wada property in our basins using the merging method, and \sref{sec:basinentropy} summarizes our calculations of the basin entropy.
In \sref{sec:basinplotscase2}, we explore a different slice of phase space corresponding to a different set of initial conditions. We verify that our statements about the Wada property and basin entropy for charged particle motion in this weakly magnetized environment are independent of phase space slicing.
In \sref{sec:summary}, we summarize and give a brief discussion of the overall results.

We follow the conventions used in \cite{Misner1973}.
In particular, we use the signature $(-,+,+,+)$ for the spacetime metric $g_{\mu\nu}$, and units in which $G=c=1$.
The Einstein summation for repeated indices is assumed throughout.

\section{Charged particle dynamics}\label{sec:particledynamics}

For completeness, we follow \cite{AlZahrani2013} (and many others) in their treatment of charged particle dynamics in a weakly magnetized Schwarzschild black hole.
We consider a particle with charge $q$ in the vicinity of a weakly magnetized Schwarzschild black hole.
The weak, static, axis-symmetric magnetic field exists in the exterior of the Schwarzschild black hole and is homogeneous at infinity. 
The magnetic field is weak in the sense that it does not affect the exterior geometry.
The Schwarzschild metric in $(t,r,\theta,\phi)$ coordinates is given by
\begin{equation}\label{eq:schdmetric}
\mathrm{d} s^{2}=-f(r)\, \mathrm{d} t^{2}+\frac{1}{f(r)}\, \mathrm{d} r^{2}+r^{2}\, \mathrm{d} \Omega^{2}
\end{equation}
where
\begin{equation}
f(r)=1-\frac{r_{g}}{r} , \quad r_{g}=2 G M
\end{equation}
and
\begin{equation}
\mathrm{d} \Omega^{2}=\mathrm{d} \theta^{2}+\sin ^{2} \theta\, \mathrm{d} \phi^{2}.
\end{equation}
The Maxwell tensor $F^{\mu\nu}$ that corresponds to a static, axis-symmetric magnetic field with strength $B$ can be obtained using the procedure done in \cite{Aliev1989,Wald1974}.
It relies on a theorem that states that a Killing vector $\xi^\mu$ in a vacuum spacetime ($R\indices{_\mu_\nu}=0$) generates a solution of the source-free Maxwell equations in that spacetime i.e., the tensor
\begin{equation}
F_{\mu \nu}=\xi_{\nu ; \mu}-\xi_{\mu ; \nu}=-2 \xi_{\mu ; \nu}
\end{equation}
satisfies
\begin{equation}
F\indices{^\mu^\nu_;_\nu}=0.
\end{equation}
It also then follows that any linear combination of such Killing vectors also generates a solution of the source-free Maxwell equations.
The Schwarzschild spacetime in \eref{eq:schdmetric} is an example of a vacuum spacetime. This spacetime has two Killing vectors: $\xi^\mu_{(t)}$ and $\xi^\mu_{(\phi)}$ corresponding to time translation symmetry and azimuthal symmetry respectively. 
It was shown in \cite{Aliev1989, Wald1974} that the 4-potential $A^\mu$ that corresponds to a weak axis-symmetric magnetic field with strength $B$ and is homogeneous at spatial infinity is
\begin{equation}\label{eq:4potential}
A^\mu = \frac{B}{2}\xi^\mu_{(\phi)}.
\end{equation}
We can see this by calculating the Maxwell tensor
\begin{equation}
F_{\mu\nu} = A_{\nu;\mu}-A_{\mu;\nu}
\end{equation}
corresponding to the 4-potential given in \eref{eq:4potential} and then calculating the magnetic field as seen by an observer with 4-velocity $u^\mu$
\begin{equation}\label{eq:magneticfield}
B^\mu = -\frac{1}{2}\frac{\epsilon^{\mu\nu\lambda\sigma}}{\sqrt{-g}} F_{\lambda\sigma} u_\nu
\end{equation}
where $\epsilon^{\mu\nu\lambda\sigma}$ is the familiar Levi-Civita symbol, and $g=\det{(g_{\mu\nu})}$.
For a local observer at rest in the Schwarzschild spacetime, the 4-velocity is given by $u_o^\mu = f^{-1/2}\xi^\mu_{(t)}$. With this, the magnetic field \eref{eq:magneticfield} is given by
\begin{equation}
B^{\mu}=B f^{1 / 2}\left(\cos \theta \delta_{r}^{\mu}-\frac{1}{r} \sin \theta \delta_{\theta}^{\mu}\right).
\end{equation}
We see that indeed at spatial infinity the magnetic field is directed along the vertical $z$ axis. 
Without loss of generality, we assume that the magnetic field points upwards by taking $B>0$.

In curved spacetime, a charged particle moving in an external electromagnetic field $F\indices{_\mu_\nu}$ obeys the equation
\begin{equation}\label{eq:lorentzforcelaw}
m \frac{\mathrm{D} u^{\mu}}{\mathrm{d} \tau}=q F\indices{^\mu_\nu} u^{\nu}
\end{equation}
where $\mathrm{D}/\mathrm{d} \tau$ is the covariant derivative with respect to the proper time $\tau$ corresponding to the metric \eref{eq:schdmetric}:
\begin{equation}
\frac{\mathrm{D} u^{\mu}}{\mathrm{d} \tau}=\frac{\mathrm{d}^{2} u^{\mu}}{\mathrm{d} \tau^{2}}+\Gamma\indices{^\mu_\alpha_\beta} u^{\alpha} u^{\beta}.
\end{equation}
For a charged particle moving in a magnetic field, it is well-known that the generalized canonical 4-momentum becomes $P_{\mu} \equiv m u_{\mu}+q A_{\mu}$. When $A_{\mu}$ respects the symmetries of the spacetime i.e., if it is Lie-dragged by Killing vectors $\xi^a$, $\mathcal{L}_\xi A_a = 0$, then $\xi^aP_a$ is constant along the trajectory of the charged particle. Corresponding to the temporal and azimuthal symmetries of the Schwarzschild metric, we thus define the conserved specific energy and azimuthal angular momentum as
\begin{equation}\label{eq:conservedE}
\mathcal{E} \equiv-\xi_{(t)}^{\mu} P_{\mu} / m = \dot{t} f
\end{equation}
and
\begin{equation}\label{eq:conservedLz}
\mathcal{L}_{z} \equiv \xi_{(\phi)}^{\mu} P_{\mu} / m = \left(\dot{\phi}+\mathcal{B}\right) r^{2} \sin ^{2} \theta
\end{equation}
respectively, where
\begin{equation}
\mathcal{B}=\frac{q B}{2 m}.
\end{equation}
The parameter $\mathcal{B}$ is a measure of the strength of the interaction between the magnetic field and the charged particle.
From \eref{eq:conservedE} and \eref{eq:conservedLz}, we obtain equations of motion for $t$ and $\phi$:
\begin{equation}\label{eq:eomtphi}
\dot{t} = f^{-1} \mathcal{E}; \quad \dot{\phi} = \frac{\mathcal{L}_z}{r^2 \sin^2 \theta} - \mathcal{B}.
\end{equation}
If we restrict the motion of the particle to the equatorial plane, then the two conserved quantities imply that motion along the equatorial plane is completely integrable.
This case has been extensively analyzed in \cite{Frolov2010}.
In the general case of non-equatorial motion however, we would need three conserved quantities for complete integrability. Since the third one is absent, we expect non-equatorial motion to be chaotic in general.

The remaining equations of motion for $r$ and $\theta$ can be obtained from \eref{eq:lorentzforcelaw}:
\begin{equation}\label{eq:eomr}
\eqalign{
\ddot{r}= &\frac{1}{2}\left(2 r-3 r_{g}\right)\left(\dot{\theta}^{2}+\frac{\mathcal{L}_{z}}{r^{4} \sin ^{2} \theta}\right) \\
&+\frac{r_{g}\left(2 \mathcal{L}_{z} \mathcal{B}-1\right)}{2 r^{2}}-\frac{\mathcal{B}^{2}}{2}\left(2 r-r_{g}\right) \sin ^{2} \theta,
}
\end{equation}
\begin{equation}\label{eq:eomtheta}
\ddot{\theta}=-\frac{2}{r} \dot{r} \dot{\theta}+\frac{\mathcal{L}_{z}^{2} \cos \theta}{r^{4} \sin ^{3} \theta}-\mathcal{B}^{2} \sin \theta \cos \theta.
\end{equation}
From the normalization condition of the 4-velocity $u^\mu u_\mu = -1$, we obtain a constraint equation for the specific energy
\begin{equation}\label{eq:energyconstraint}
\mathcal{E}^{2}=\dot{r}^{2}+r^{2} f \dot{\theta}^{2}+U_{\mathrm{eff}}
\end{equation}
where
\begin{equation}\label{eq:effectivepotential}
U_{\mathrm{eff}}(r, \theta)=f\left[1+r^{2} \sin ^{2} \theta\left(\frac{\mathcal{L}_{z}}{r^{2} \sin ^{2} \theta}-\mathcal{B}\right)^{2}\right].
\end{equation}
Trajectories moving under \eref{eq:eomr}\textendash\eref{eq:eomtheta} always satisfy \eref{eq:energyconstraint}\textendash\eref{eq:effectivepotential} at any point in the trajectory.

We note that the equations of motion \eref{eq:eomtphi}\textendash\eref{eq:eomtheta} are invariant under the transformation
\begin{equation}
\phi \rightarrow-\phi, \quad \mathcal{L}_{z} \rightarrow-\mathcal{L}_{z}, \quad \mathcal{B} \rightarrow-\mathcal{B}.
\end{equation}
Since we chose $B>0$, then without loss of generality, we can also choose the particle to have positive charge (and hence $\mathcal{B}>0$).
The motion for the corresponding negatively charged particle can then be obtained via the transformation $\mathcal{L}_z \rightarrow -\mathcal{L}_z$, $\phi \rightarrow -\phi$.
However, having chosen $\mathcal{B}>0$, we need to inspect the cases $\mathcal{L}_z>0$ and $\mathcal{L}_z<0$ separately.
This corresponds to two different directions for the Lorentz force acting on the particle.

\section{Effective potential and exit states}\label{sec:effectivepotential}

\begin{figure}
\begin{center}
\begin{minipage}{0.3\linewidth}
\centering
\includegraphics[width=2in]{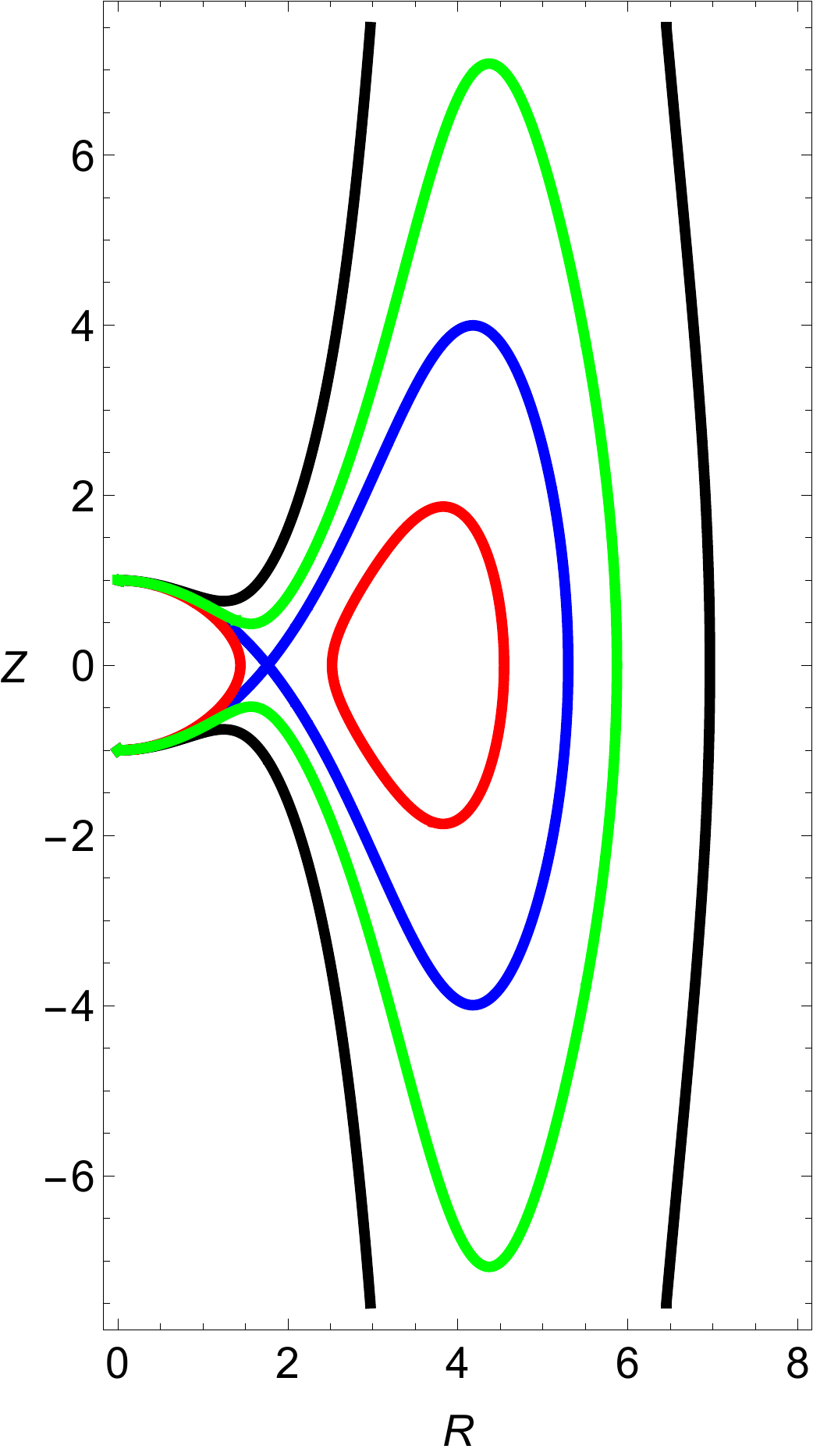}\\
(a) $b=0.1$ and $\ell = 2$
\end{minipage}\hspace{0.05\linewidth}
\begin{minipage}{0.3\linewidth}
\centering
\includegraphics[width=2in]{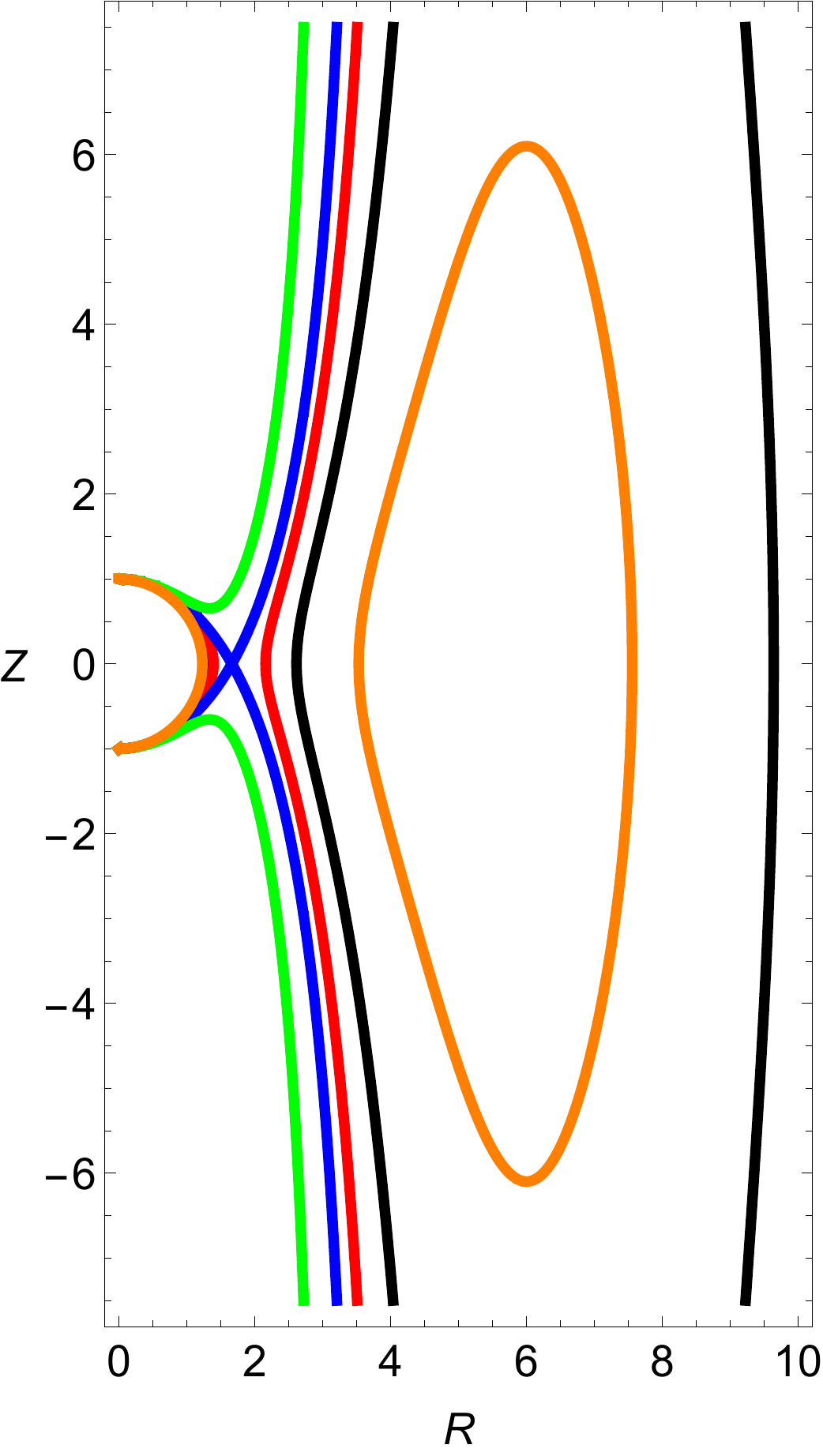}\\
(b) $b=0.1$ and $\ell = -4$
\end{minipage}
\caption{\label{fig:effpotplots} The effective potential. The black and blue curves correspond to $\mathcal{E} = \bar{\mathcal{E}}$ and $\mathcal{E} = \mathcal{E}_\mathrm{cap}$ respectively, while the red and green curves correspond to $\mathcal{E}<\mathcal{E}_\mathrm{cap}$ and $\mathcal{E}>\mathcal{E}_\mathrm{cap}$ respectively. In (a) the red curve corresponds to bounded orbits ($\mathcal{E}<\mathcal{E}_\mathrm{cap}$ and $\mathcal{E}<\bar{\mathcal{E}}$), while in (b) the red curve corresponds to orbits where capture is impossible but escape to $|z|\rightarrow \pm\infty$ is allowed ($\mathcal{E}<\mathcal{E}_\mathrm{cap}$ and $\mathcal{E}>\bar{\mathcal{E}}$). The orange curve in (b) corresponds to bounded orbits in the $\ell < 0$ case.}
\end{center}
\end{figure}

We shall elaborate in this section the possible exit states of our charged particle. 
Introducing the dimensionless quantities
\begin{equation}
\sigma=\frac{\tau}{r_{g}}, \quad \mathcal{T}=\frac{t}{r_g}, \quad \rho=\frac{r}{r_{g}}, \quad l=\frac{\mathcal{L}_{z}}{r_{g}}, \quad b=\mathcal{B} r_{g}
\end{equation}
the equations of motion \eref{eq:eomtphi}\textendash\eref{eq:eomtheta}, as well as the energy constraint \eref{eq:energyconstraint} become
\begin{equation}\label{eq:dlesseomt}
\frac{\mathrm{d}\mathcal{T}}{\mathrm{d}\sigma}=\frac{\mathcal{E}\rho}{\rho-1}
\end{equation}
\begin{equation}\label{eq:dlesseomphi}
\frac{\mathrm{d}\phi}{\mathrm{d}\sigma}=\frac{\ell}{\rho^2\sin^2\theta}-b
\end{equation}
\begin{equation}\label{eq:dlesseomrho}
\eqalign{
\frac{\mathrm{d}^{2} \rho}{\mathrm{d} \sigma^{2}}&=\frac{1}{2}(2 \rho-3)\left(\frac{\mathrm{d} \theta}{\mathrm{d} \sigma}\right)^{2}+\frac{(2 \ell b-1)}{2 \rho^{2}} \\
&+\frac{\ell^{2}(2 \rho-3)}{2 \rho^{4} \sin ^{2} \theta}-\frac{b^{2}}{2}(2 \rho-1) \sin ^{2} \theta
}
\end{equation}
\begin{equation}\label{eq:dlesseomtheta}
\frac{\mathrm{d}^{2} \theta}{\mathrm{d} \sigma^{2}}=-\frac{2}{\rho}\left(\frac{\mathrm{d} \rho}{\mathrm{d} \sigma}\right)\left(\frac{\mathrm{d} \theta}{\mathrm{d} \sigma}\right)+\frac{\ell^{2} \cos \theta}{\rho^{4} \sin ^{3} \theta}-b^{2} \sin \theta \cos \theta
\end{equation}
\begin{equation}\label{eq:dlessenergyconstraint}
\mathcal{E}^{2}=\left(\frac{\mathrm{d} \rho}{\mathrm{d} \sigma}\right)^{2}+\rho(\rho-1)\left(\frac{\mathrm{d} \theta}{\mathrm{d} \sigma}\right)^{2}+U_{\mathrm{eff}}
\end{equation}
\begin{equation}\label{eq:dlesseffectivepotential}
U_{\mathrm{eff}}(\rho,\theta)=\left(1-\frac{1}{\rho}\right)\left[1+\frac{\left(\ell-b \rho^{2} \sin ^{2} \theta\right)^{2}}{\rho^{2} \sin ^{2} \theta}\right].
\end{equation}
We analyze the effective potential  \eref{eq:dlesseffectivepotential} by going to cylindrical coordinates $R=\rho\sin\theta$ and $z=\rho\cos\theta$.
The effective potential in cylindrical coordinates is
\begin{equation}\label{eq:Ueffcyl}
U_{\mathrm{eff}}(R, z)=\left(1-\frac{1}{\sqrt{R^{2}+z^{2}}}\right)\left[1+\frac{\left(\ell-b R^{2}\right)^{2}}{R^{2}}\right].
\end{equation}
From this, we obtain the fixed points of the dynamical system by setting $\nabla U_\mathrm{eff} = 0$.
The fixed points are $(R,z)=(R_c,0)$, where $R_c$ are the roots of the equation
\begin{equation}\label{eq:Rcpoly}
(3-2 R_c) \ell^{2}-2 b R_c^{2} \ell+\left[b^{2} R_c^{2}(2 R_c-1)+1\right] R_c^{2}=0.
\end{equation}
As shown in \cite{Frolov2010}, \eref{eq:Rcpoly} can have at most two real roots.
If we denote these roots $R_{c,1}$ and $R_{c,2}$ with $R_{c,1}<R_{c,2}$, it can be shown that $R_{c,1}$ and $R_{c,2}$ correspond to a local maximum (unstable circular orbit) and a local minimum (stable circular orbit) of the effective potential \eref{eq:Ueffcyl} respectively.
If we denote these local extrema $U_\mathrm{max}$ and $U_\mathrm{min}$, then for energies $\mathcal{E}_\mathrm{min}<\mathcal{E}<\mathcal{E}_\mathrm{max}$ (where $\mathcal{E}=\sqrt{U}$), the contours of the effective potential will be closed along the $R$ direction.
Since $U_\mathrm{eff}\sim R^2$ as $R\rightarrow \infty$, we see that $R$ is bounded from the right regardless of the value of the energy $\mathcal{E}$.
Hence, $\mathcal{E}>\mathcal{E}_\mathrm{max}$ corresponds to the opening of the contours of the effective potential along the direction of black hole capture i.e., black hole capture becomes possible for $\mathcal{E}>\mathcal{E}_\mathrm{max}$.
Because of this, we denote $\mathcal{E}_\mathrm{max}\equiv\mathcal{E}_\mathrm{cap}$.
\Fref{fig:effpotplots} shows different contours of the effective potential for different values of $\mathcal{E}$.
The red, blue and green contours correspond to $\mathcal{E}<\mathcal{E}_\mathrm{cap}$, $\mathcal{E}=\mathcal{E}_\mathrm{cap}$, $\mathcal{E}>\mathcal{E}_\mathrm{cap}$ respectively.

For $b \neq 0$, there exists $\bar{\mathcal{E}} = \sqrt{\bar{U}_\mathrm{eff}}$ such that for $\mathcal{E} < \bar{\mathcal{E}}$, $z \rightarrow \pm \infty$ are forbidden regions (i.e. for $\mathcal{E} < \bar{\mathcal{E}}$, escape to $z\rightarrow \pm \infty$ is impossible).
To see this, note that $U_\mathrm{eff}$ is a monotonically increasing function of $|z|$, and
\begin{equation}\label{eq:Uzinfinity}
\left.U_{\mathrm{eff}} \rightarrow U\right|_{|z|_{\infty}} \equiv\left[1+\frac{\left(\ell-b R^{2}\right)^{2}}{R^{2}}\right]\, \quad (\mathrm{as}\, |z|\rightarrow \infty).
\end{equation}
The value of $\bar{U}_\mathrm{eff}$ is then the minimum of $\left.U\right|_{|z|_\infty}$.
At the minimum,
\begin{equation}
\eqalign{
\frac{\mathrm{d} U_{|z|_{\infty}}}{\mathrm{d} R}=\frac{2 \ell^{2}-2 b^{2} R^{4}}{R^{3}} &=0 \\
\ell^{2}-b^{2} R^{4} &=0 \\
\left(\ell+b R^{2}\right)\left(\ell-b R^{2}\right) &=0
}
\end{equation}
Since we chose (without loss of generality) $b>0$, we have two mutually exclusive cases:
\begin{equation}
\eqalign{
\ell - bR^2 &= 0 \quad (\mathrm{for}\,\ell > 0), \\
\ell + bR^2 &= 0 \quad (\mathrm{for}\,\ell < 0).
}
\end{equation}
Combining this with \eref{eq:Uzinfinity} and with $\bar{\mathcal{E}}=\sqrt{\bar{U}_\mathrm{eff}}$, we get
\begin{equation}\label{eq:Ebar}
\bar{\mathcal{E}} = \cases{
1 & \textrm{if } $\ell > 0$ \\
\sqrt{1 + 4b|\ell|} & \textrm{if } $\ell < 0$
}.
\end{equation}
Note that the $\ell < 0$ case of $\bar{\mathcal{E}}$ in \eref{eq:Ebar} is the minimum energy required for a charged particle to undergo helical motion in a homogeneous magnetic field in flat spacetime (see e.g. \cite{Landau1975}).

The black contours in \fref{fig:effpotplots} correspond to energies $\mathcal{E}=\bar{\mathcal{E}}$.
Also,
\begin{equation}
\lim_{R\rightarrow 0}U_\mathrm{eff} =  \cases{
-\infty & \textrm{if } $|z|<1$ \\
+\infty & \textrm{if } $|z| \geq 1$
}
\end{equation}
Hence, for $|z| \geq 1$, $R$ is bounded from the left regardless of the value of the energy $\mathcal{E}$.
We see this in \fref{fig:effpotplots}.
Thus, our dynamical system has three exits: $R\rightarrow 1$ (black hole capture), and escape to $z\rightarrow \pm \infty$.
Note that in some cases, the contours can be open along $z\rightarrow \pm \infty$ and at the same time be closed along the direction of black hole capture as shown in \fref{fig:effpotplots}b.
This corresponds to energy values $\bar{\mathcal{E}}<\mathcal{E}<\mathcal{E}_\mathrm{cap}$.

\section{Basin plots for a charged particle in a circular orbit}
\label{sec:basinplotscase1}
Exit basins (or basins of attraction in the case of dissipative systems) provide a useful way to analyze chaotic conservative dynamical systems.
This type of analysis has been applied to various dynamical systems such as the paradigmatic H\'{e}non-Heiles system \cite{Blesa2014, Aguirre2001, Seoane2006, Bernal2018}, as well as systems describing black holes \cite{AlZahrani2013,AlZahrani2014,Daza2018, Kopacek2018,FrolovV1999,Liu2017}.
An exit basin $B_i$, associated with a particular exit $E_i$, is the closure of the set of all initial conditions that escape from the bounded region through exit $E_i$.
In our case, since we have three exits (capture, escape to $z\rightarrow +\infty$, and escape to $z\rightarrow -\infty$), we have three basins. Initial conditions leading to bound orbits will also appear in our analysis below, but this set will not be an exit basin.
For chaotic systems, the basin boundaries tend to be fractal, so the end states of initial conditions close to them are highly uncertain.
For an in-depth review of fractal features in basin boundaries, see \cite{Aguirre2009}.

In general, these basins exist as $N$-dimensional volumes in an $N$-dimensional phase space.
In order for these basins to be useful for analysis, we must restrict our attention to a two-dimensional slice of the total phase space.
In this section, we analyze a particular slice of phase space describing particles initially in a circular orbit $\rho_0$ experiencing a purely vertical kick with energy $\mathcal{E}$.
In addition to studying the fractal features of the basin boundaries, this slicing allows us to approximate a ``critical escape energy curve" for charged particles orbiting the black hole.

In \cite{AlZahrani2013}, the authors worked with a particle initially moving in an ISCO (innermost stable circular orbit) of radius $\rho_*$.
This condition restricted the parameters $\ell$ and $b$ to be a function only of $\rho_*$ (see \cite{Frolov2010} for details):
\begin{equation}\label{eq:lbISCO}
\eqalign{
\ell &=\pm \frac{\rho_{*}\left(3 \rho_{*}-1\right)^{1 / 2}}{\sqrt{2}\left(4 \rho_{*}^{2}-9 \rho_{*}+3 \pm \sqrt{\left(3 \rho_{*}-1\right)\left(3-\rho_{*}\right)}\right)^{1 / 2}}, \\
b &=\frac{\sqrt{2}\left(3-\rho_{*}\right)^{1 / 2}}{2 \rho_{*}\left(4 \rho_{*}^{2}-9 \rho_{*}+3 \pm \sqrt{\left(3 \rho_{*}-1\right)\left(3-\rho_{*}\right)}\right)^{1 / 2}}.
}
\end{equation}
In this paper, we relax this ``innermost'' restriction and consider a particle initially moving in a stable (though not necessarily innermost) circular orbit $\rho_0$. We think this is a better starting point for charged particles possibly confined to accretion disks and moving along orbits that are not ISCOs.

Hence, the parameter $b$ is a free parameter. The non-dimensional angular momentum $\ell$ will now be a function of $\rho_0$ and  $b$.
We obtain the function $\ell(b,\rho_0)$ by using the circular orbit constraint \eref{eq:Rcpoly} with $R_c \rightarrow \rho_0$.
Solving for $\ell$, we get
\begin{equation}\label{eq:lpm(rho)}
\eqalign{
\ell_{\pm}(b,\rho_0)&=\frac{\rho_0}{2\rho_0 -3}\bigr[-b\rho_0 \pm \big(b^2\rho_0^2 \\
&-(3-2\rho_0)\{b^2\rho_0^2(2\rho_0-1)+1\}\big)^{1/2}\bigr].
}
\end{equation}
Implicitly, $\rho_{*}^{\pm}(b)$ is given in \eref{eq:lbISCO}, where $\pm$ refers to the different cases $\ell>0$ and $\ell<0$.
It was shown in \cite{Frolov2010} that as $b \rightarrow \infty$, $\rho_{*}^{+}\rightarrow 1$ while $\rho_{*}^{-}\rightarrow (5 + \sqrt{13})/4 \approx 2.15$.
It is straightforward to show that when $\rho_0 \geq \rho_*$, $\ell_\pm$ is real, and that indeed $\ell_+>0$ while $\ell_-<0$.

From a circular orbit $\rho_0$, the particle is initially given a purely upward kick with energy $\mathcal{E}$.
Since the equations of motion is invariant with respect to reflection along the equatorial plane $\theta \rightarrow \pi - \theta$, we can consider only upward kicks without loss of generality. The corresponding downward kick case can be obtained by applying the aforementioned transformation.
Since the kick is purely vertical, it does not induce a change in the azimuthal angular momentum $\ell$.
The initial conditions of the motion are then $(\rho(0),\rho^\prime (0), \theta(0),\theta^\prime (0)) = (\rho_0,0,\pi/2,\omega_0(\mathcal{E}))$ where $\omega_0(\mathcal{E})$ is obtained by solving for $(\mathrm{d}\theta/\mathrm{d}\sigma)$ in \eref{eq:dlessenergyconstraint}.
With these constraints, the only remaining free parameter is $b$, with $\ell$ given by \eref{eq:lpm(rho)}.
We construct the basin plots just as in \cite{AlZahrani2013}, by numerically integrating \eref{eq:dlesseomrho} and \eref{eq:dlesseomtheta} for a $500\times 500$ ($1000 \times 1000$ was used in \cite{AlZahrani2013}) grid of initial conditions $(\rho_0,\mathcal{E})$,
where $\rho \in [\rho_{*},5]$ and $\mathcal{E} \in [0,3]$ for $\ell>0$ while $\mathcal{E} \in [1,6]$ for $\ell<0$, with $\rho_{*}$ given in \eref{eq:lbISCO} for a given $b$.
As was done in \cite{AlZahrani2013}, we used the built-in \textit{Mathematica} 11.2 function \texttt{NDSolve} with a maximum computation time of $\sigma = 10^5$.
Numerically, the final state of the particle at the end of the integration time can be one of four possibilites: 
\begin{itemize}
\item capture by the black hole,
\item escape to $z\rightarrow +\infty$ (upstream with respect to the magnetic field),
\item escape to $z\rightarrow -\infty$ (downstream),
\item neither captured nor escaped.
\end{itemize}
This last possibility is for bound orbits and transiently bound orbits i.e., those whose escapes or captures take place beyond our integration time. We color each pixel of the basin plot depending on its final state: red (capture), yellow (upstream escape), blue (downstream escape), and black (neither capture nor escape). The particle is considered captured if $\rho$ reaches 1, while it is considered to have escaped if $|z| = |\rho\cos\theta|$ reaches $10^3$ at any point in the numerical integration.
If neither state is reached within the maximum computation time, the corresponding pixel is colored black.

We use \eref{eq:dlessenergyconstraint} to calculate the energy $\mathcal{E}(\sigma)$ of the trajectory.
Since the energy $\mathcal{E}$ is a conserved quantity, we can calculate the relative error of $\mathcal{E}$ with respect to the initial energy $\mathcal{E}_0$ of the trajectory:
\begin{equation}
\mathcal{E}_\mathrm{err}(\sigma) = \frac{|\mathcal{E}(\sigma) - \mathcal{E}_0|}{\mathcal{E}_0}.
\end{equation}
Just as in \cite{AlZahrani2013}, we find that the relative error can reach $10^{-2}$ for escape cases.
Although we can increase the working precision of the numerical solver to obtain a more accurate calculation of the trajectories, doing so increases computation time greatly and this became impractical when we needed to calculate $500\times 500$ number of trajectories per basin plot for multiple basin plots.
In the limited cases that we tested, we found that increasing the working precision of the numerical solver did not change the final states of trajectories (i.e. as to which basin the trajectory belonged to) located far from the basin boundaries.
On the other hand, we found that increasing the working precision of the numerical solver changed the final states of trajectories in the neighborhood of the basin boundaries.
This is expected to be the case since the fractal nature of the basin boundaries meant that nearby trajectories are highly uncertain.
In fact, one characteristic feature of fractal regions is that an increase in working precision along these regions does not yield a significant increase in the accuracy of numerical calculations \cite{Grebogi1983}.
However, this is not a problem since we are interested on the structure of the fractal regions themselves, not on whether the trajectories are exact (i.e. errorless).
We found that the overall structure of these fractal regions did not change upon increasing the working precision of the numerical solver.
And despite numerical errors, we can be confident that these fractal structures are real because of the \emph{shadowing lemma} which states that for a numerically calculated trajectory, there exists an exact trajectory with slightly different initial conditions that stays near the numerical trajectory \cite{Ott2002}.

\begin{figure}[t]
\begin{center}
\begin{minipage}{0.32\linewidth}
\centering
\includegraphics[width=\linewidth]{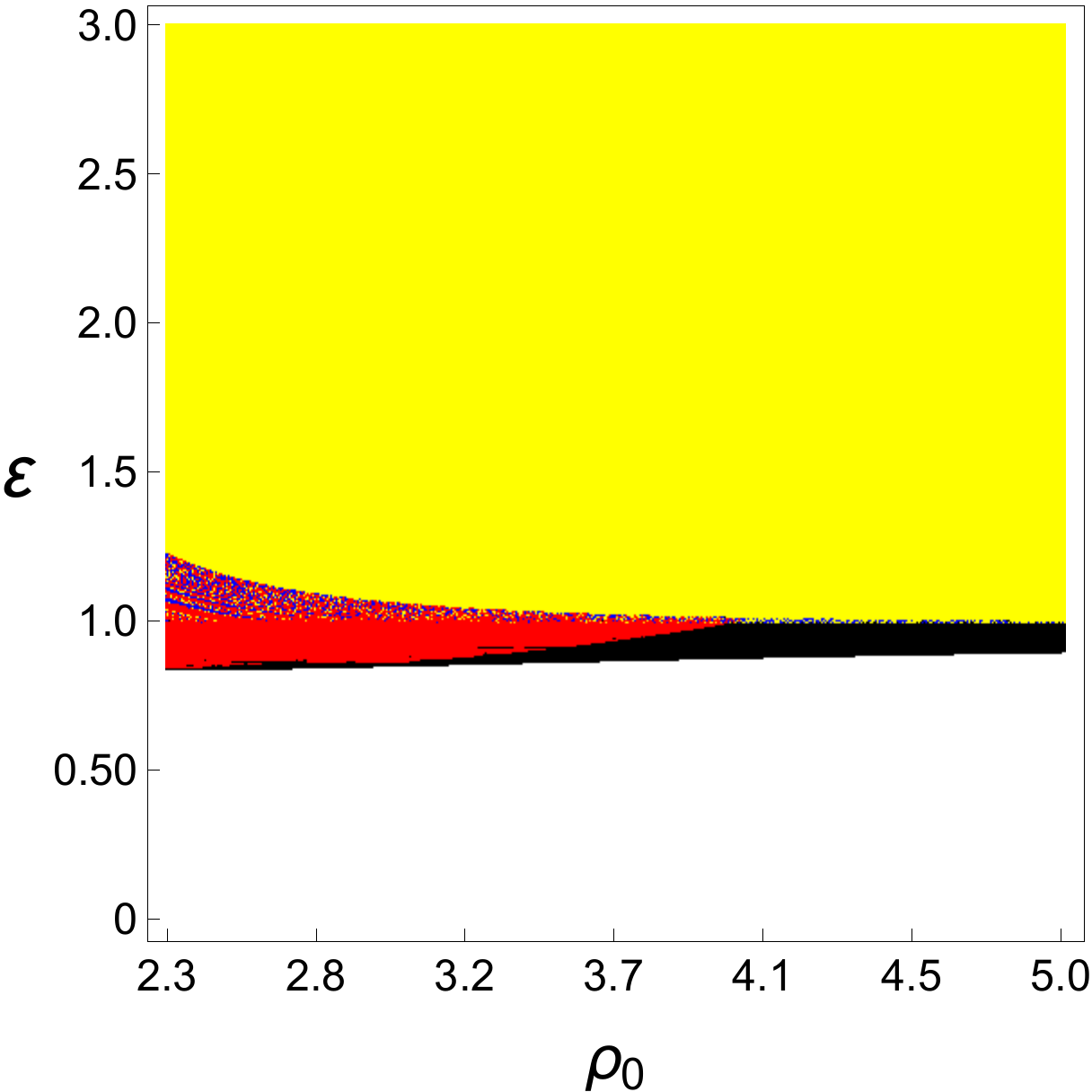}\\
(a) $b=0.1$
\end{minipage}
\begin{minipage}{0.32\linewidth}
\centering
\includegraphics[width=\linewidth]{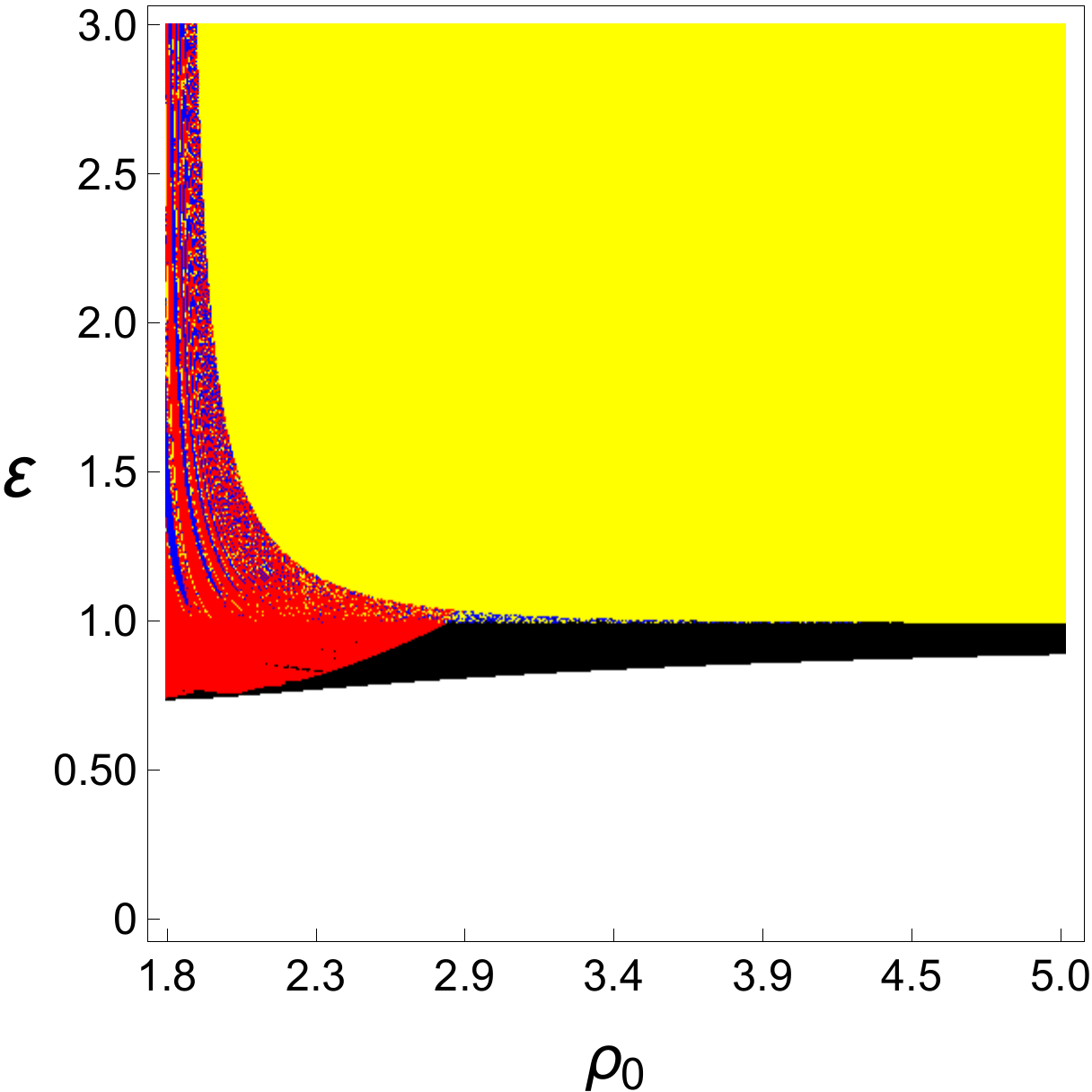}\\
(b) $b=0.3$
\end{minipage}
\begin{minipage}{0.32\linewidth}
\centering
\includegraphics[width=\linewidth]{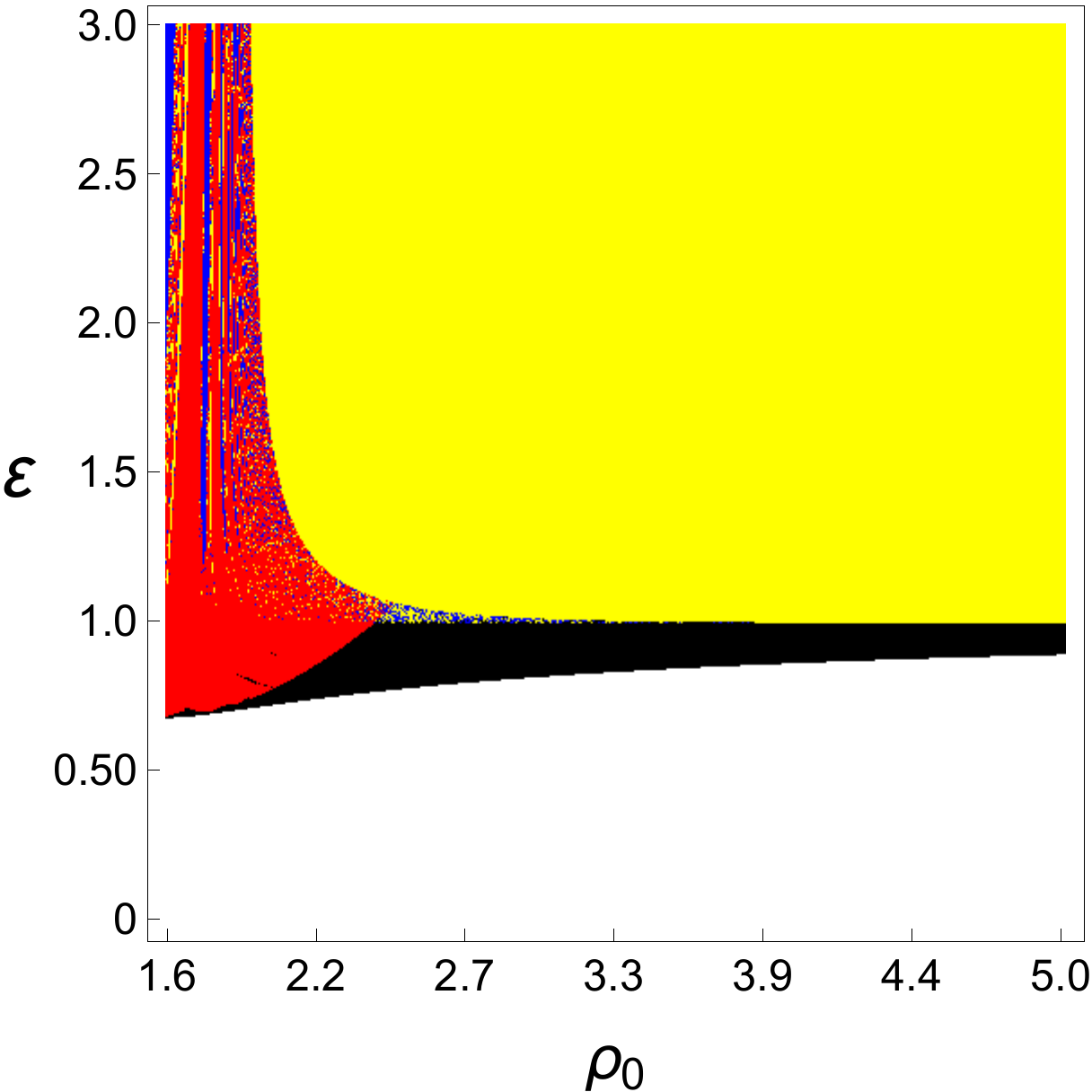}\\
(c) $b=0.5$
\end{minipage}
\caption{\label{fig:initcond1basinplotspl} Basin plots for $\ell>0$ for different values of $b$ with initial conditions described in \sref{sec:basinplotscase1}. Red means capture, yellow means upstream escape, blue means downstream escape, while black means neither capture nor escape.}
\end{center}
\end{figure}
\begin{figure}[t]
\begin{center}
\begin{minipage}{0.32\linewidth}
\centering
\includegraphics[width=0.96\linewidth]{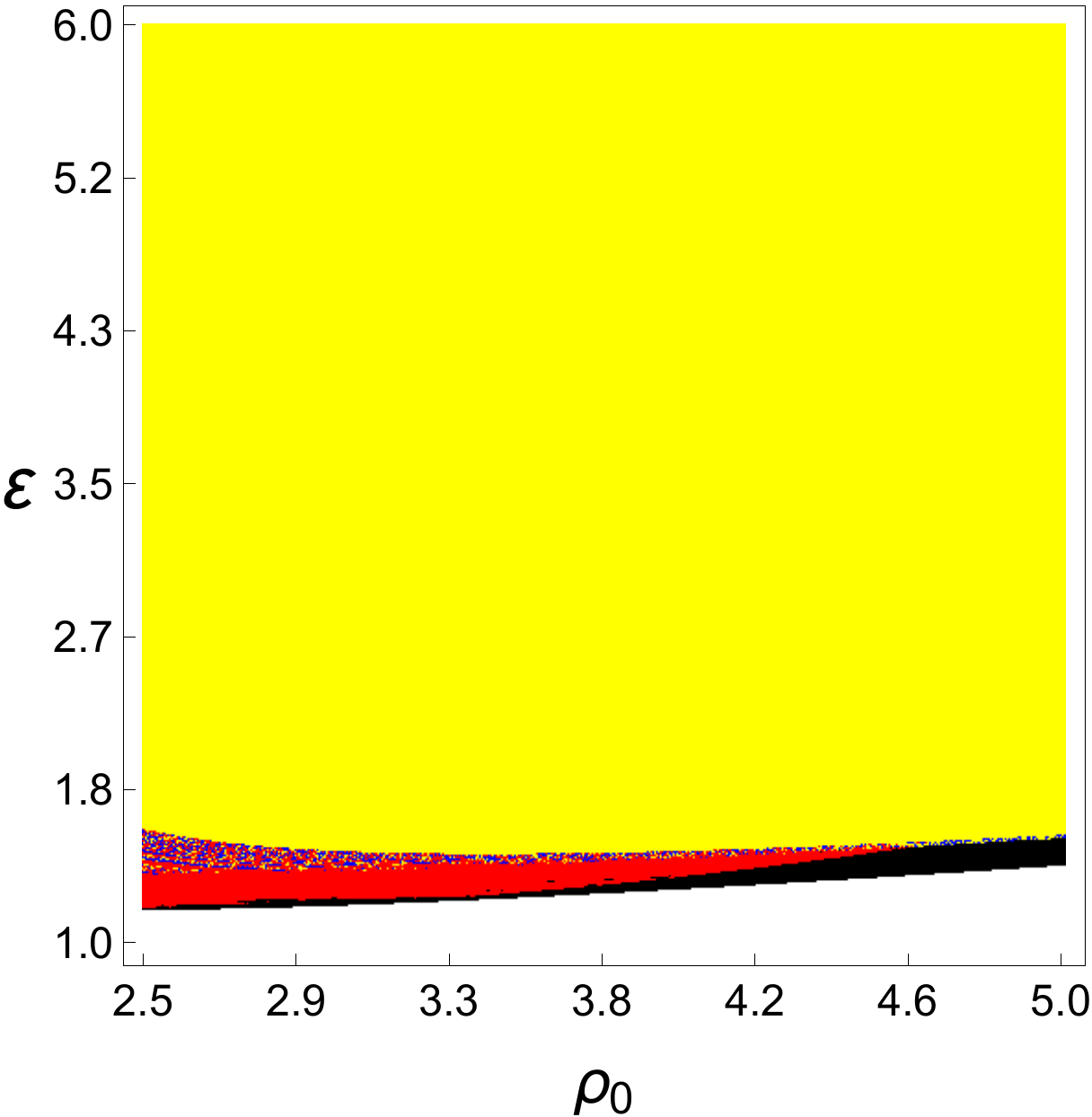}\\
(a) $b=0.1$
\end{minipage}
\begin{minipage}{0.32\linewidth}
\centering
\includegraphics[width=0.96\linewidth]{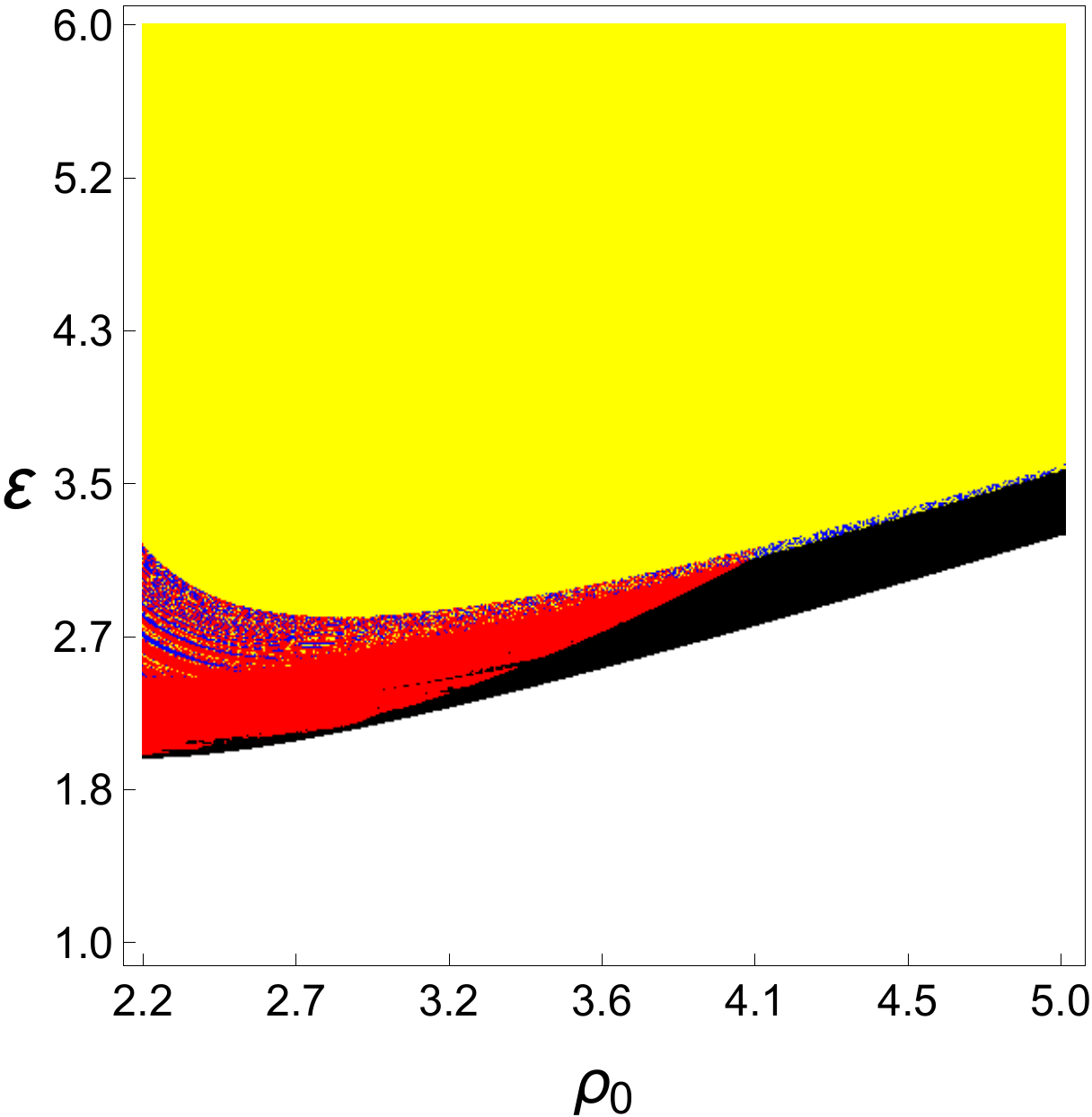}\\
(b) $b=0.3$
\end{minipage}
\begin{minipage}{0.32\linewidth}
\centering
\includegraphics[width=0.96\linewidth]{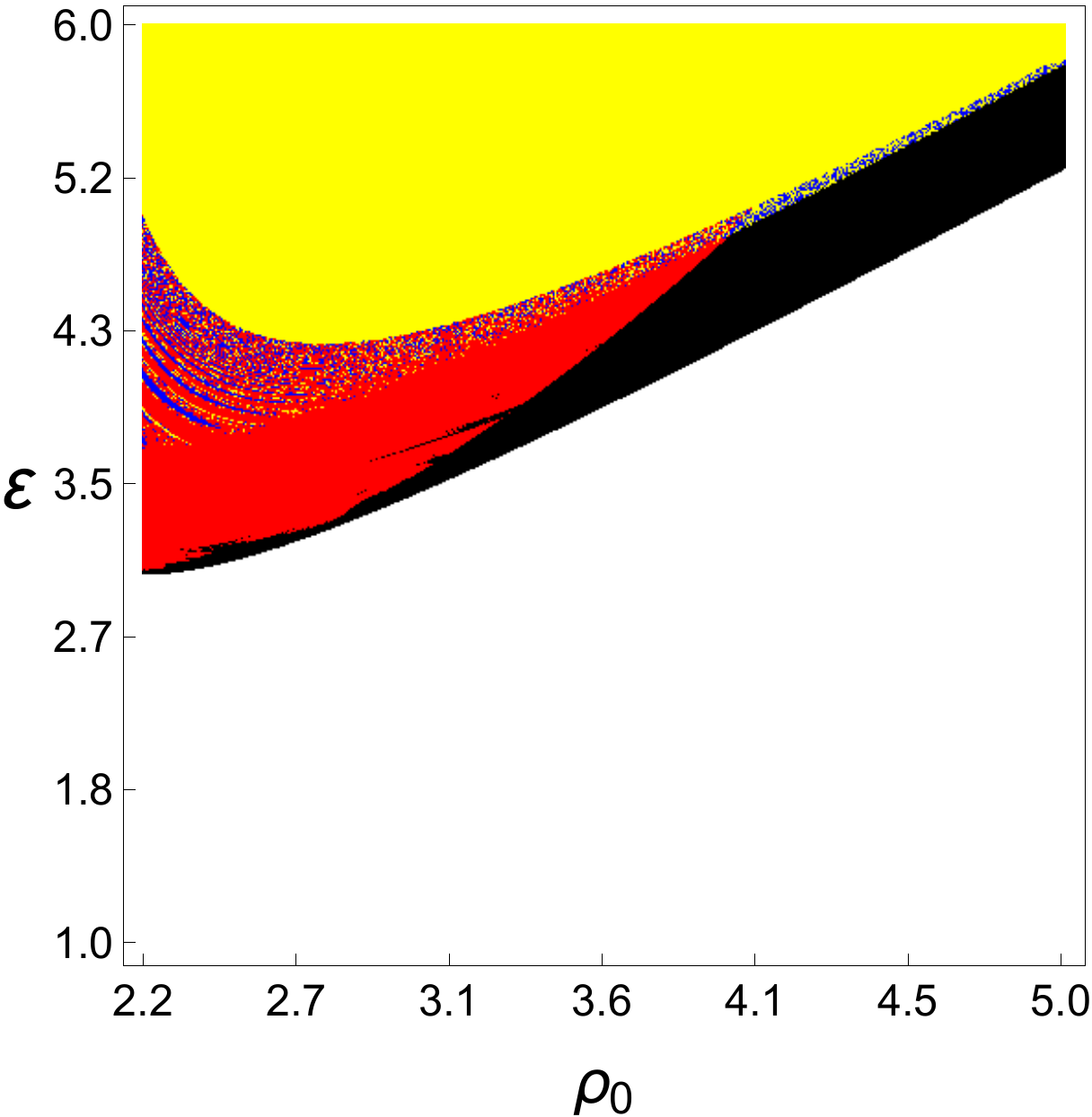}\\
(c) $b=0.5$
\end{minipage}
\caption{\label{fig:initcond1basinplotsnl} Basin plots for $\ell<0$ for different values of $b$ with initial conditions described in \sref{sec:basinplotscase1}. Red means capture, yellow means upstream escape, blue means downstream escape, while black means neither capture nor escape.}
\end{center}
\end{figure}

\begin{figure}[t]
\begin{center}
\begin{minipage}{0.7\linewidth}
\centering
\includegraphics[width=\linewidth]{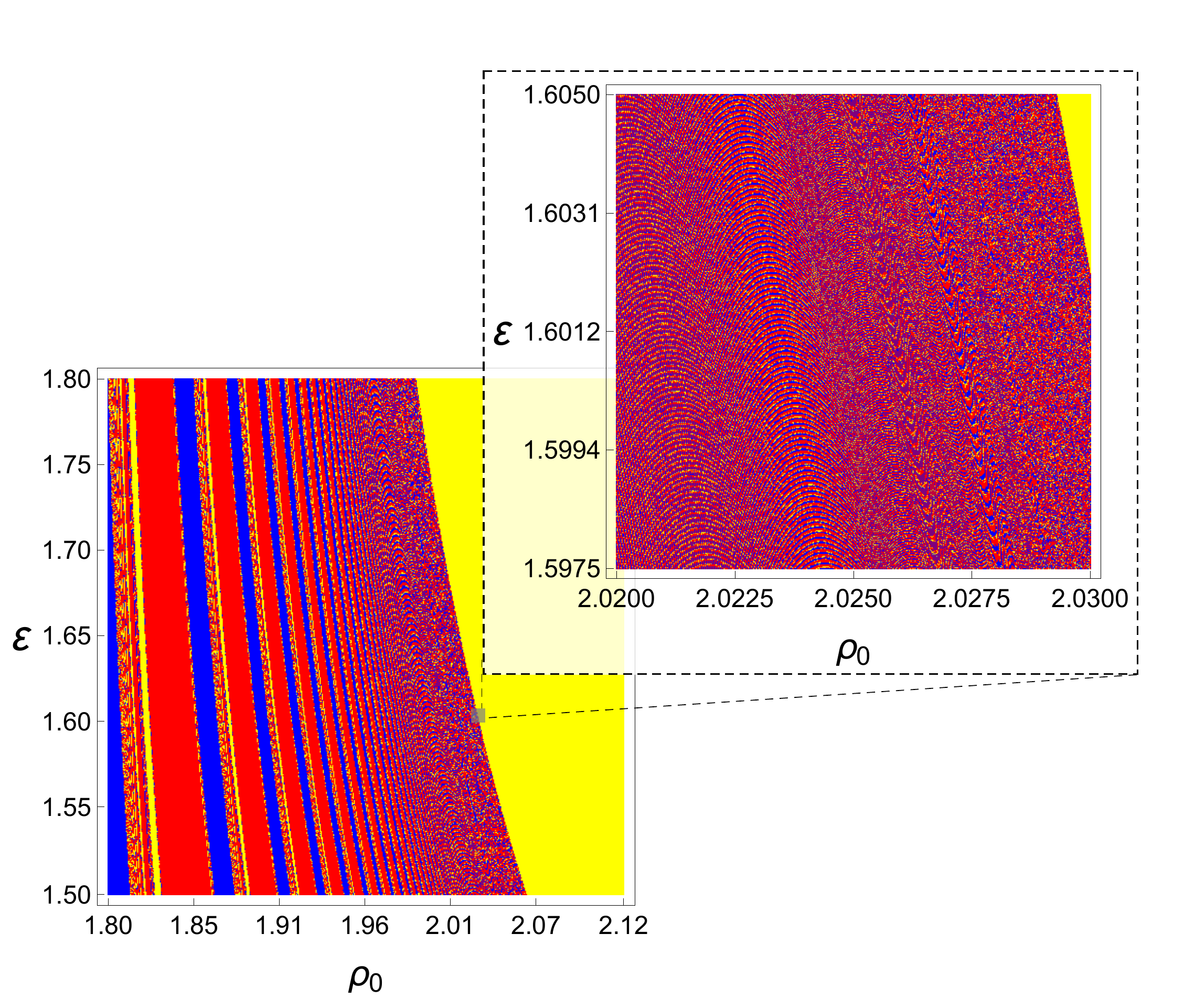}\\
(a) $\ell>0$: $\rho_0 \in [1.80,2.12]$ and $\mathcal{E} \in [1.50,1.80]$
\end{minipage}
\\
\begin{minipage}{0.7\linewidth}
\centering
\includegraphics[width=\linewidth]{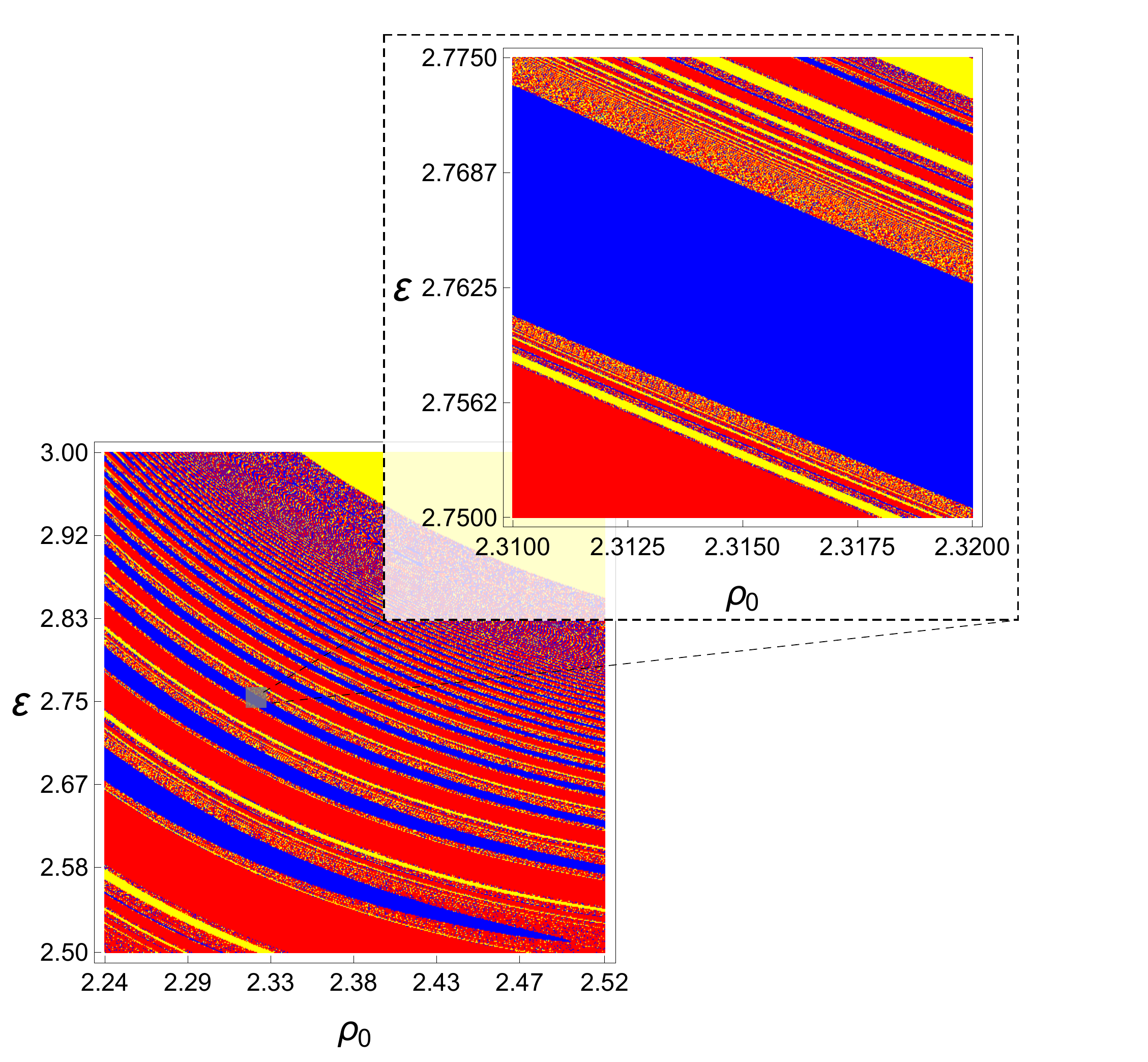}\\
(b) $\ell<0$: $\rho_0 \in [2.24,2.52]$ and $\mathcal{E} \in [2.50,3.00]$
\end{minipage}
\caption{\label{fig:initcond1zooms} Magnified regions of the basin plot for $b=0.3$.}
\end{center}
\end{figure}

We constructed two basin plots for each value of $b$ (one each for $\ell>0$ and $\ell<0$) and we chose the values $b \in \{0.1,0.2,0.3,0.4,0.5\}$.
The plots for $b \in \{0.1,0.3,0.5\}$ are shown in \fref{fig:initcond1basinplotspl} and \fref{fig:initcond1basinplotsnl}.
Immediately, we notice large uniformly yellow regions representing upstream escapes, while below this is a fractal region that mixes black hole captures (red), downstream escapes (blue), and also upstream escapes (yellow) (see \fref{fig:initcond1zooms}). The boundary between these regions was defined in \cite{AlZahrani2013} to be the critical escape energy curve. We emphasize though that this boundary is regular or smooth only far from the black hole, where the gravitational influence of the black hole is negligibly weak and the critical escape energy depends solely on the homogeneous magnetic field. In this region, a vertical kick on the orbiting charge can only result either in an upstream escape (yellow) or another bound orbit (black). The critical energy is $\bar{\mathcal{E}}=1$ for $\ell>0$ and $\bar{\mathcal{E}} = \sqrt{1 + 4b|\ell(b,\rho_0)|} $ for $\ell<0$ (see \eref{eq:Ebar}).

If the starting circular orbit is closer to the black hole, the boundary becomes quite irregular (fractal), as the interplay between strong gravity and electromagnetism allows for more final outcomes that are very sensitive to initial conditions. Thus, the strong-field boundary becomes much less distinct as (yellow) upstream escapes mix with (red) black hole captures and (blue) downstream escapes. Nonetheless, it is still useful to talk about a critical escape energy curve above which there can be only upstream escapes.

Contiguous black regions represent the energies $\mathcal{E}<\mathcal{E}_\mathrm{cap}<\bar{\mathcal{E}}$ and correspond to bound orbits, as discussed in \sref{sec:effectivepotential}.
\Fref{fig:initcond1basinplotspl} and \fref{fig:initcond1basinplotsnl}  also show a few black ``streaks'' that seep through the red capture region. These streaks are transient bound orbits, and are artifacts of our finite integration time. They gradually disappear as the integration time is increased.

\begin{figure}[t]
\begin{center}
\begin{minipage}{0.6\linewidth}
\centering
\includegraphics[width=\linewidth]{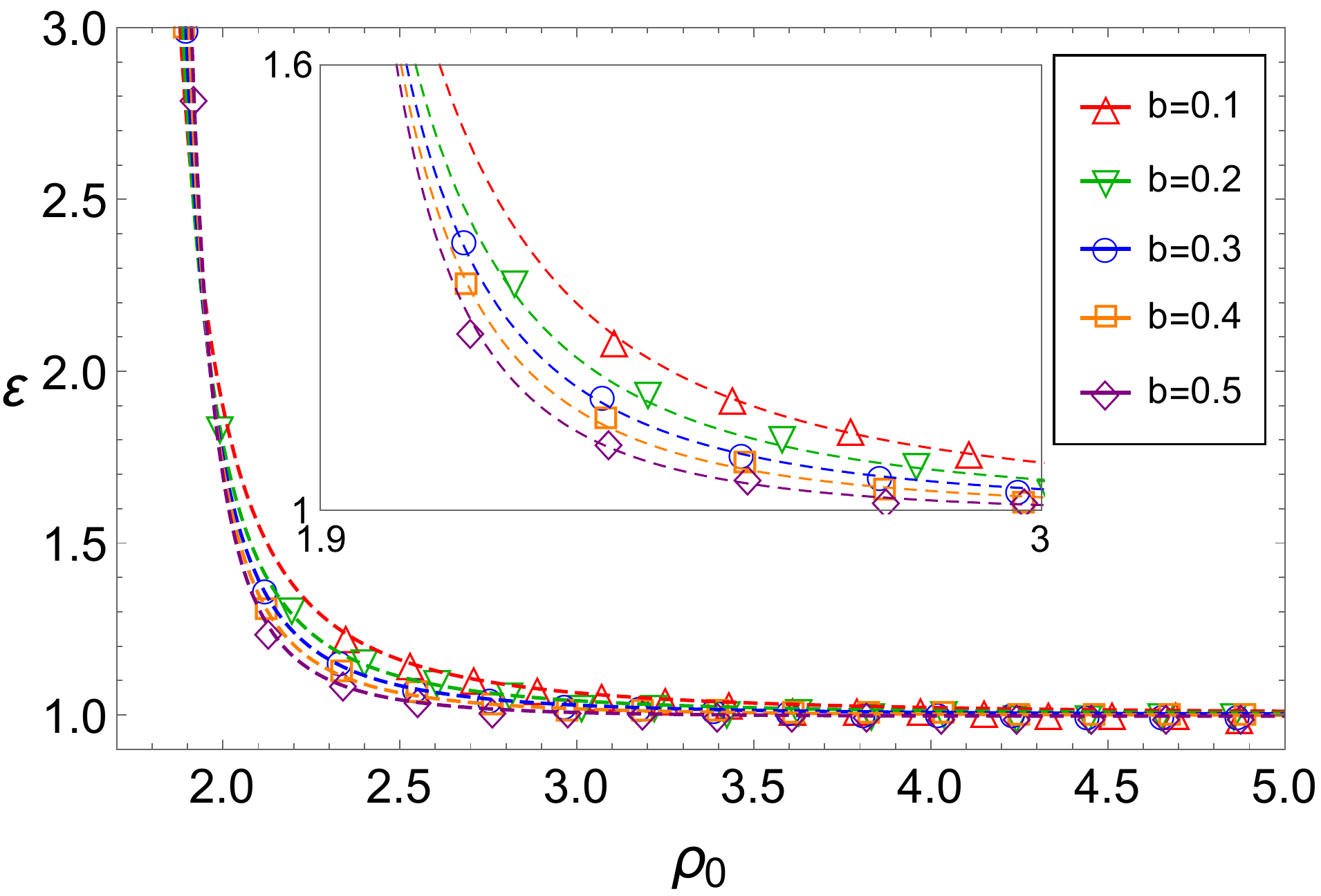}\\
(a) $\ell > 0$
\end{minipage}
\\
\begin{minipage}{0.6\linewidth}
\centering
\includegraphics[width=\linewidth]{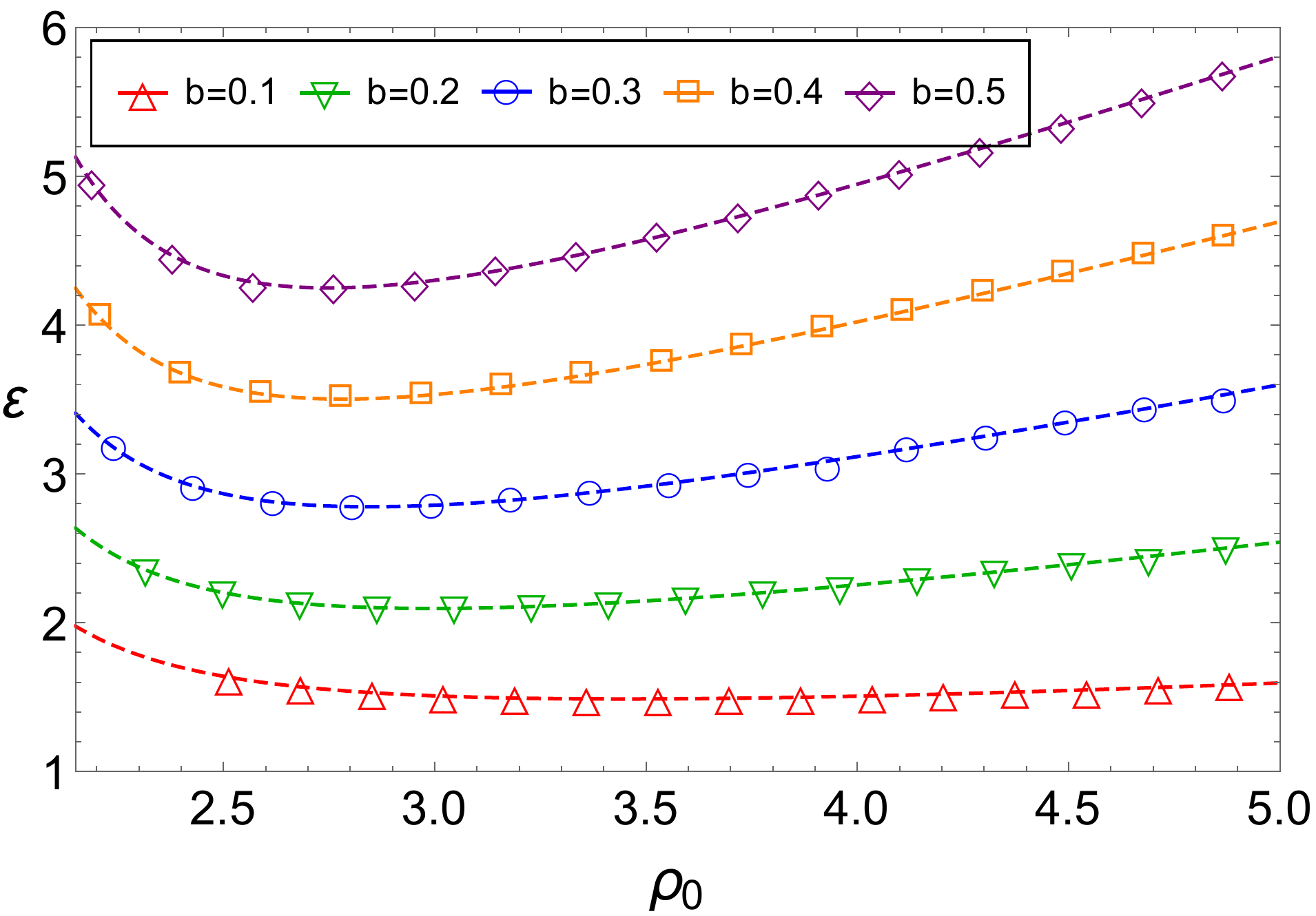}\\
(b) $\ell < 0$
\end{minipage}
\caption{\label{fig:Eesc} The critical escape energy for (a) $\ell>0$ and (b) $\ell<0$. The hollow markers correspond to sampled points of the critical energy escape boundary obtained from \fref{fig:initcond1basinplotspl} and \fref{fig:initcond1basinplotsnl} while dashed lines correspond to the fitted analytic expression in \eref{eq:Eesc}.}
\end{center}
\end{figure}

We can obtain an approximate expression for the critical escape energy. For $\ell>0$ and large enough $b$, this curve seems to diverge at some $\rho_0$. This property and the large-$\rho_0$ asymptotic behavior of the critical escape energy are captured by the form
\begin{equation}
\label{eq:Eesc}
\mathcal{E}_\mathrm{esc}(\rho_0,b) \approx \left(1 + \frac{\alpha_1 b^{\beta_1}}{(\rho_0 - \gamma b^{\delta})}+\frac{\alpha_2 b^{\beta_2}}{(\rho_0 - \gamma b^{\delta})^2}\right)\bar{\mathcal{E}},
\end{equation}
which we can fit reasonably well to our data\footnote{The data can be accessed from \cite{DataGitHub}.}. In the strong-gravity region (small $\rho_0$), we set the escape energy to be the least upper bound energy for captures and downstream escapes\footnote{Operationally, to get $\mathcal{E}_\mathrm{esc}$ for each $\rho_0$ (at fixed $b$), we first look for the non-yellow pixel with the largest $\mathcal{E}$. The slightly higher energy of the pixel immediately next to this is $\mathcal{E}_\mathrm{esc}$ for this $\rho_0$.}. Our best fit is achieved with parameter values $(\alpha_1,\beta_1,\alpha_2,\beta_2,\gamma,\delta) = (-0.09007,1.831,0.02619,-0.6678,1.838,0.05144)$ in the $\ell > 0$ case and $ (-0.02765,-0.3609,0.1471,-0.3290,$ $1.582,0.06902)$ in the $\ell < 0$ case, with $\bar{\mathcal{E}}$ given in \eref{eq:Ebar}.
The fitted formula disagrees with our data at worst by $1.5\%$ (for large $\rho_0$ in the case $\ell> 0$, $b=0.1$), though the agreement is significantly better for all other cases.
\Fref{fig:Eesc} compares the critical escape energy obtained from the basin plots  in \fref{fig:initcond1basinplotspl} and \fref{fig:initcond1basinplotsnl} with our analytic fit in \eref{eq:Eesc}.

We see that for $\ell>0$, the critical escape energy decreases for increasing $b$, while for $\ell<0$, the critical escape energy increases for increasing $b$.
This makes sense intuitively. For the case $\ell>0$ the Lorentz force is repulsive, and increasing $b$, which increases the strength of this Lorentz force, aids in the escape of the charged particle. Conversely, when $\ell<0$ the Lorentz force is attractive, and increasing $b$ makes escaping more difficult.
It is important to note that in the limit $b\rightarrow 0$, the system should reduce to the Schwarzschild case, wherein the escape energy $\mathcal{E}_\mathrm{esc} = 1$.
However, with the aforementioned coefficients for the $\ell>0$ case, \eref{eq:Eesc} yields a function that is singular at $b=0$.
We emphasize that \eref{eq:Eesc} should only be valid for $b\neq 0$.

Beyond mere theoretical interest, our approximate expression was useful for speeding up our code, enabling high-resolution and longer integration runs. Using the approximate expression, we can avoid having to integrate the equations of motion for initial conditions representing energies well above the escape energy, and focus computational resources mainly in the regions close to the escape energy curve.

Zooming in on the diffuse region near the critical escape energy curve (see \fref{fig:initcond1zooms}a), we see Cantor-like fractal features, with repeating patterns of escape and capture regions that increase in frequency as we approach the critical escape energy curve.
The presence of these fractal features indicate that the particles undergo chaotic motion.
The final state of a particle that starts with energies below but near the critical escape energy is thus highly uncertain.
It will be interesting to know whether this ``uncertainty" somehow depends on the parameter $b$.
In \sref{sec:basinentropy}, we will show that indeed the system becomes more uncertain as we increase the parameter $b$, using a recently proposed measure of uncertainty known as the \emph{basin entropy}. 
Another interesting feature of the diffuse region is that the basin boundaries seemingly consist of points belonging to all three basins. 
No matter how close one zooms in on a boundary, one will always see points around it that belong to all of the basins (see \fref{fig:initcond1zooms}b).
We discuss this property in detail in the next section.

\section{Wada property}
\label{sec:wada}
Fractal boundaries in a basin plot can possess an unusual feature called the Wada property. A basin boundary is said to have the Wada property if it is a boundary of more than two basins \cite{Yoneyama1917, Hocking1961}. The basin is then said to be a Wada basin, or is said to possess the Wada property.
Although it is difficult to imagine three or more basins sharing a boundary, this peculiar and uncommon topological property arises in the context of fractal basin boundaries.
Indeed as was shown previously in \fref{fig:initcond1zooms}b, zooming in on boundaries seemingly shared by two basins reveals yet another basin different from the initial two.
Some examples of systems that are known to exhibit the Wada property are the H\'{e}non-Heiles potential \cite{Aguirre2001, Seoane2006}, the Duffing oscillator \cite{Aguirre2002}, the H\'{e}non map, and the forced damped pendulum \cite{Nusse1996a}, as well as binary black hole systems \cite{Daza2018}.
In practice, the Wada property is difficult to verify, because numerical calculations of boundaries can only be up to a finite resolution.
Several authors have proposed methods to check this property, some of which we now discuss.

One of the earliest results proved by Kennedy and Yorke \cite{Kennedy1991} was that the basin boundary possesses the Wada property if the unstable manifold of a periodic orbit on the basin boundary intersects all of the basins, and if its stable manifold is dense in each of the basin boundaries.
This was the basis of the basin cell method proposed by Nusse and Yorke \cite{Nusse1996a}.
However, finding the necessary periodic orbits is very difficult as it required detailed knowledge of the dynamics of the system in question.
This method proved to be difficult to apply except for a select number of systems.
Another method proposed by Daza et al \cite{Daza2015} essentially involves calculating trajectories with finer and finer resolutions about the boundaries of two basins and checking if a trajectory belonging to a third basin appears.
Although this method can be applied to a wider variety of systems, the calculation of additional trajectories makes this method computationally expensive.
We also mention a more recent method developed by Wagemakers et al \cite{Wagemakers2020} that involves the computation of chaotic saddles of the system.
It is able to determine whether a system is Wada without the need to calculate the basins, however it requires detailed knowledge of the dynamics of the system (in particular, a suitable Poincar\'{e} section is required for the method to work and this is not a trivial thing to find in many systems).
The method that will be used in this paper is called the merging method, proposed by Daza et al \cite{Daza2018a}.
It is the least computationally expensive method among the four and it can also be applied to a wide variety of systems since it only requires knowledge of the basin plot itself.

\begin{figure}[t]
\begin{center}
\begin{minipage}{0.3\linewidth}
\centering
\includegraphics[width=\linewidth]{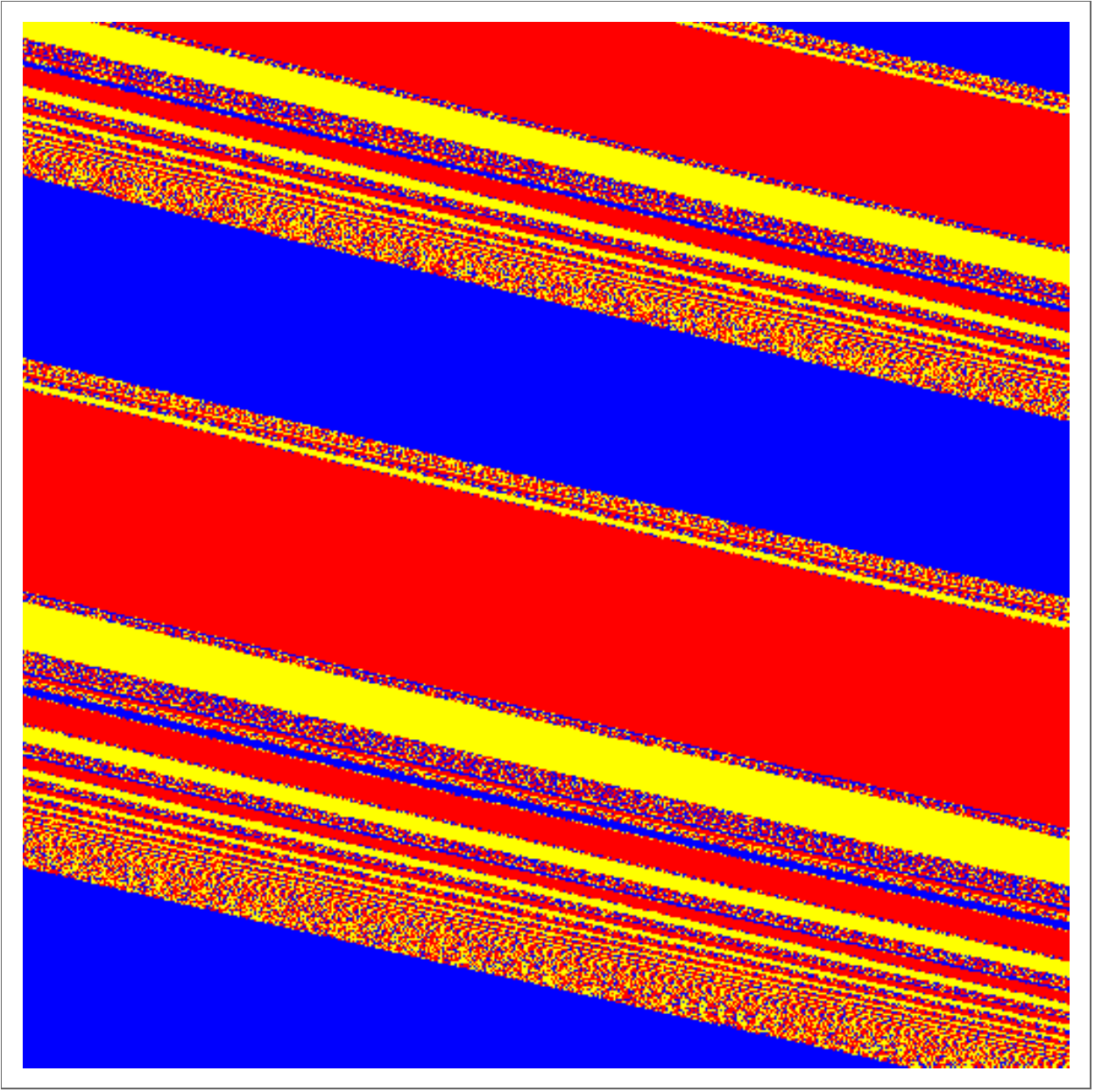}
\end{minipage}
\begin{minipage}{0.3\linewidth}
\centering
\includegraphics[width=\linewidth]{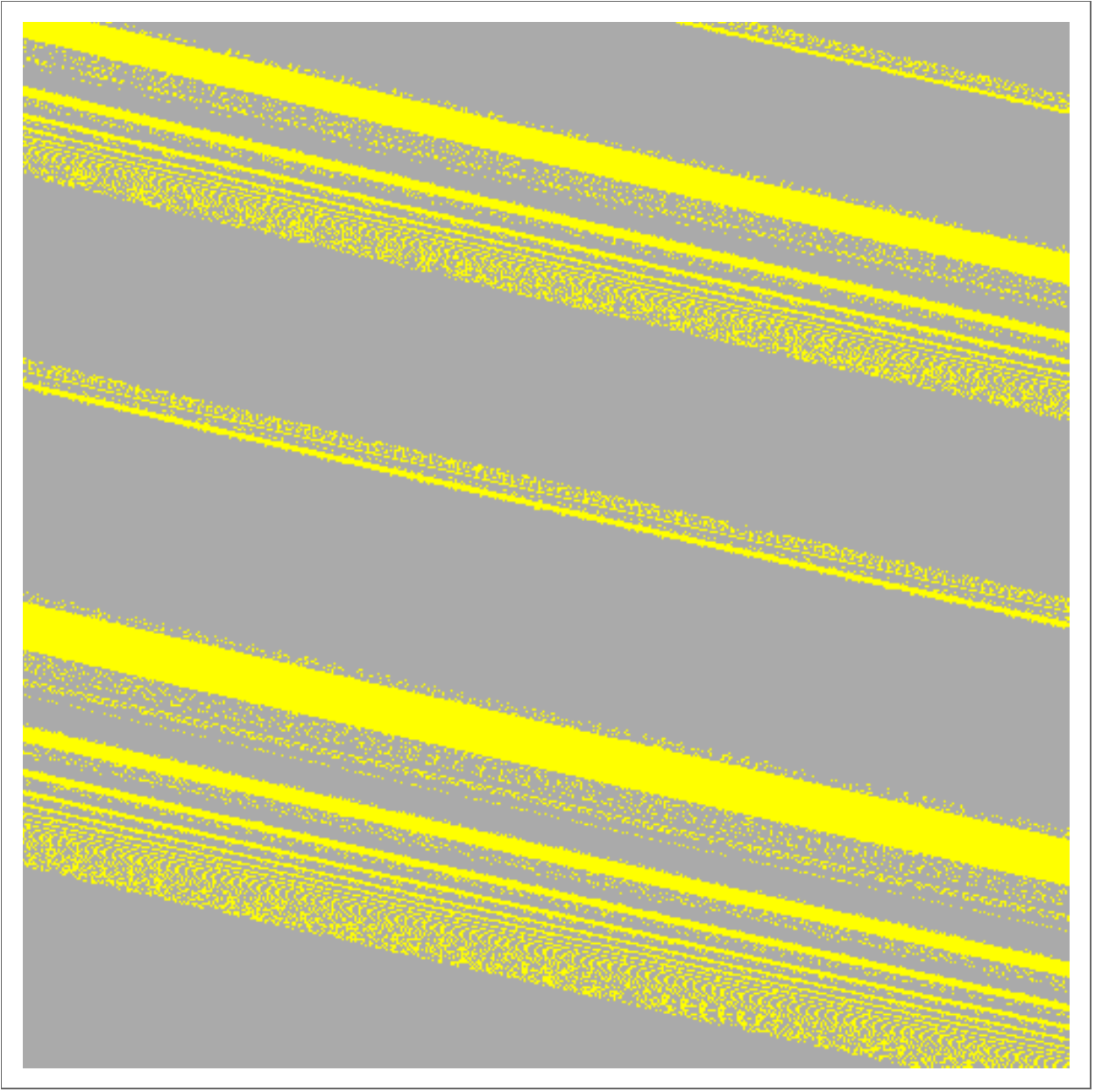}
\end{minipage}
\\
\begin{minipage}{0.3\linewidth}
\centering
\includegraphics[width=\linewidth]{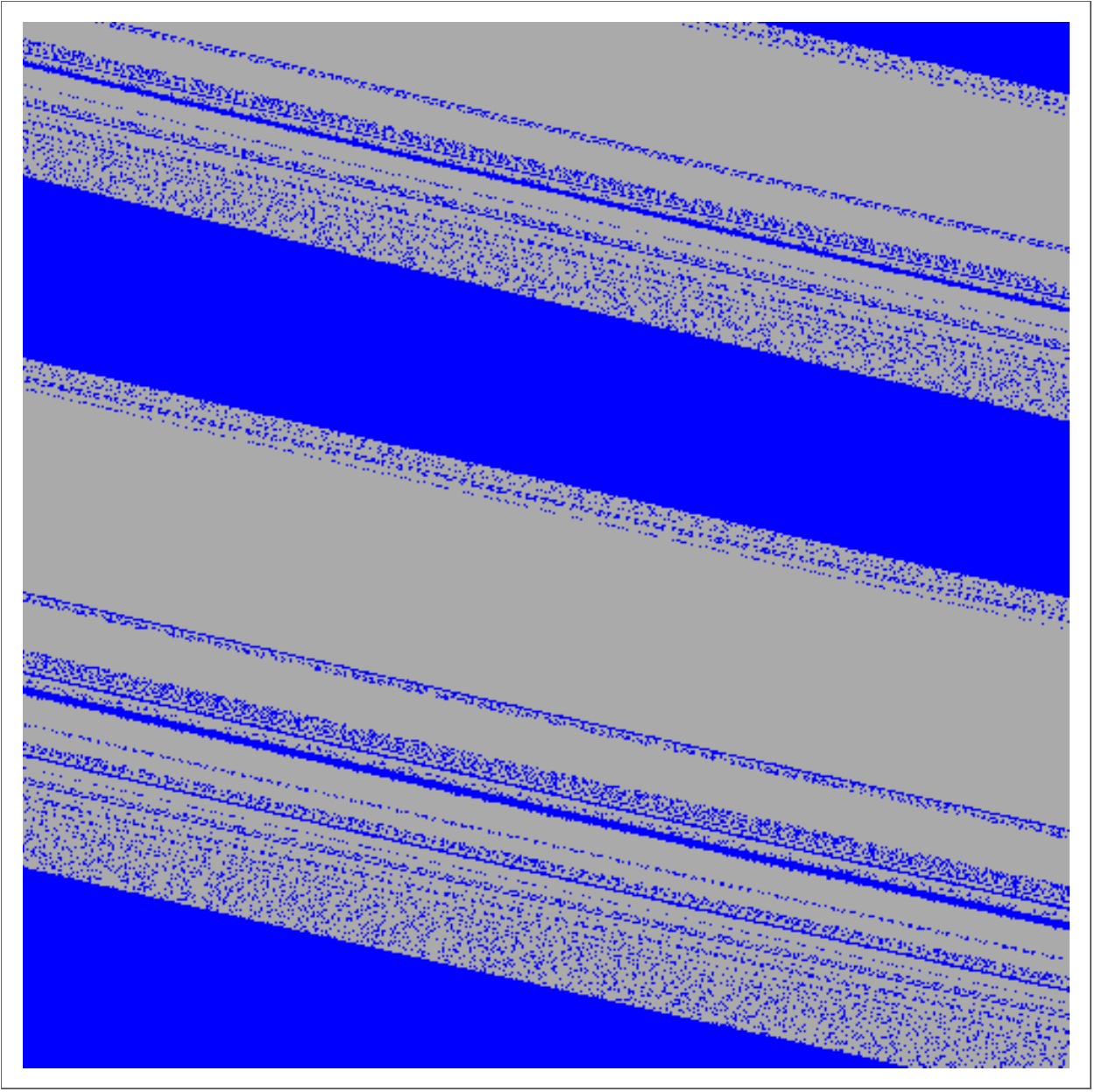}
\end{minipage}
\begin{minipage}{0.3\linewidth}
\centering
\includegraphics[width=\linewidth]{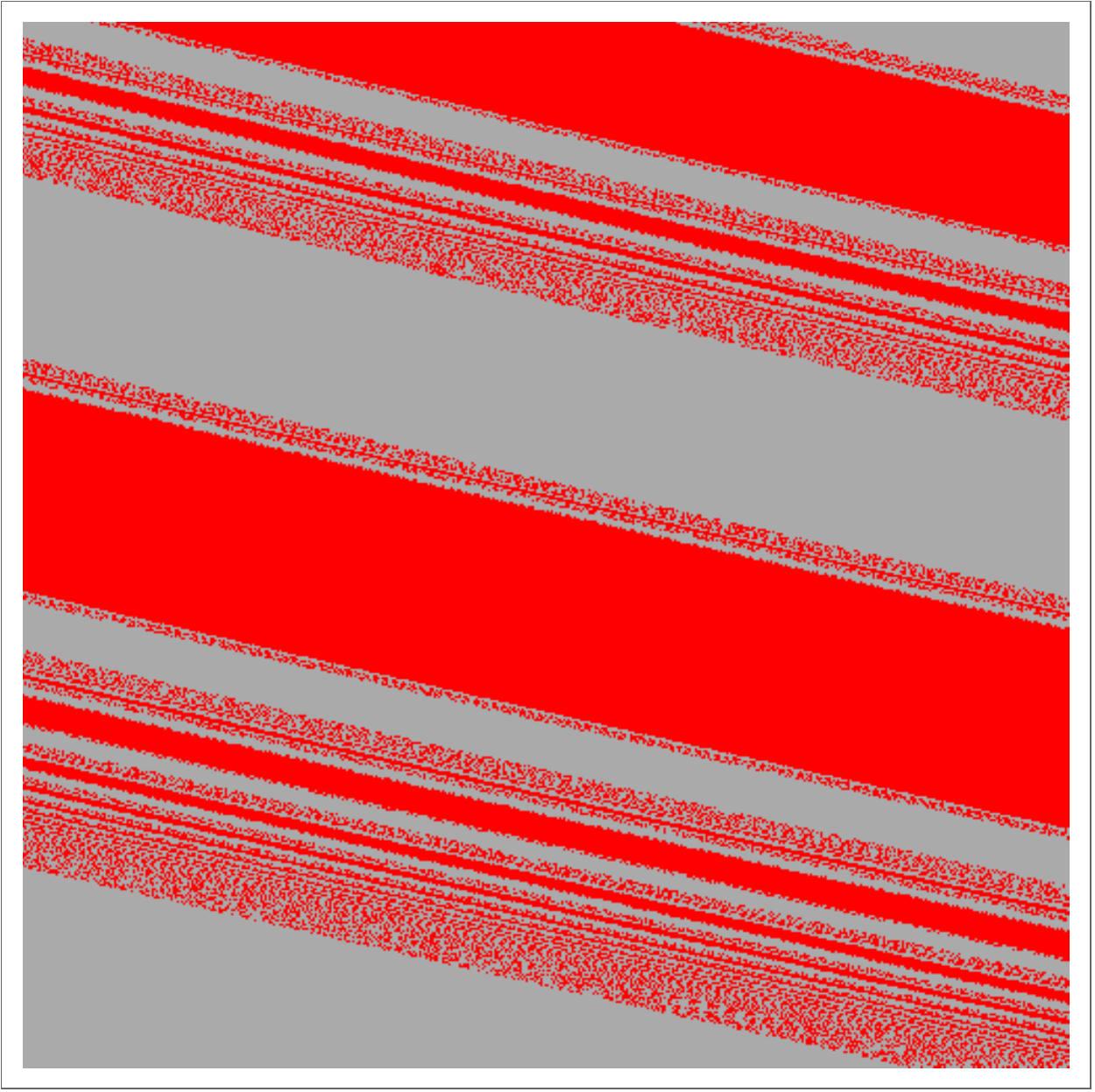}
\end{minipage}
\caption{\label{fig:mergingmethodexample} The main idea behind the merging method. From the original three-colored basin, three two-colored merged basins are generated. Visually it seems that the boundaries of the three merged basins are the same. However, due to the finite resolution of the basin plots, these boundaries are, in fact, slightly different from each other.}
\end{center}
\end{figure}

The main idea of the merging method is the observation that \emph{a Wada boundary is the only one that remains unaltered under the action of merging the basins.}
\Fref{fig:mergingmethodexample} shows a magnified part of the diffuse region in one of the basin plots shown in \sref{sec:basinplotscase1}, as well as merged versions of itself.
Visually we see that the boundaries in the merged versions look similar to each other.
Although in principle the statement is true, actually checking that the basin boundaries remain unaltered would require an arbitrarily high resolution of the basin plots.
As we will see, the merging method was designed to account for this limitation.
The steps for the merging method are summarized as follows:
\begin{enumerate}
\item For each basin $B_i$, $i = 1,2,..., N_a$ where $N_a$ is the total number of basins, we merge the other basins, obtaining a two-colored basin plot made of the original basin $B_i$ and the merged basin $\cup_{j \neq i}B_j$. This yields $N_a$ new basin plots. For example, in \fref{fig:mergingmethodexample}, the original three-colored basin plot generated three two-colored basin plots upon merging.
\item We compute the $N_a$ slim boundaries $\{\partial B_i\}$ of the $N_a$ merged basins. A pixel is considered part of the slim boundary if it has pixels of a different color around itself. In our case, we need to compute three slim boundaries. Although the three slim boundaries appear to be the same visually in \fref{fig:mergingmethodexample}, these slim boundaries will actually be different due to the finite resolution of the basin plots.
\item The obtained $N_a$ slim boundaries $\{\partial B_i\}$ are fattened by turning pixels within a chessboard distance of $r$ around the slim boundaries into boundary pixels to yield $N_a$ fat boundaries $\{\mathbf{\partial B}_i\}$. We initially take the fattening parameter to be $r=1$.
\item We check if the union of all slim boundaries is contained in each of the fat boundaries. In this case, we say that the basins have the Wada property. If not, we record the ratio between the number of pixels that are not contained in the fat boundaries and the total number of pixels to obtain the percentage of non-Wada points. We then repeat steps 3-4, increasing the fattening parameter until we reach a stopping condition $r=r_{\mathrm{max}}$. If $r=r_{\mathrm{max}}$ is reached and the union of slim boundaries is not contained in each fat boundary, then the boundary does not possess the Wada property.
\end{enumerate}
The fattening procedure is meant to account for the finite resolution of the basin plots.
We see then that the merging method can determine whether the system is Wada only up to a certain resolution based on the fattening parameter $r_\mathrm{max}$, and the resolution of the basin plot itself.

\begin{figure}
\begin{center}
\begin{minipage}{0.32\linewidth}
\centering
\includegraphics[width=\linewidth]{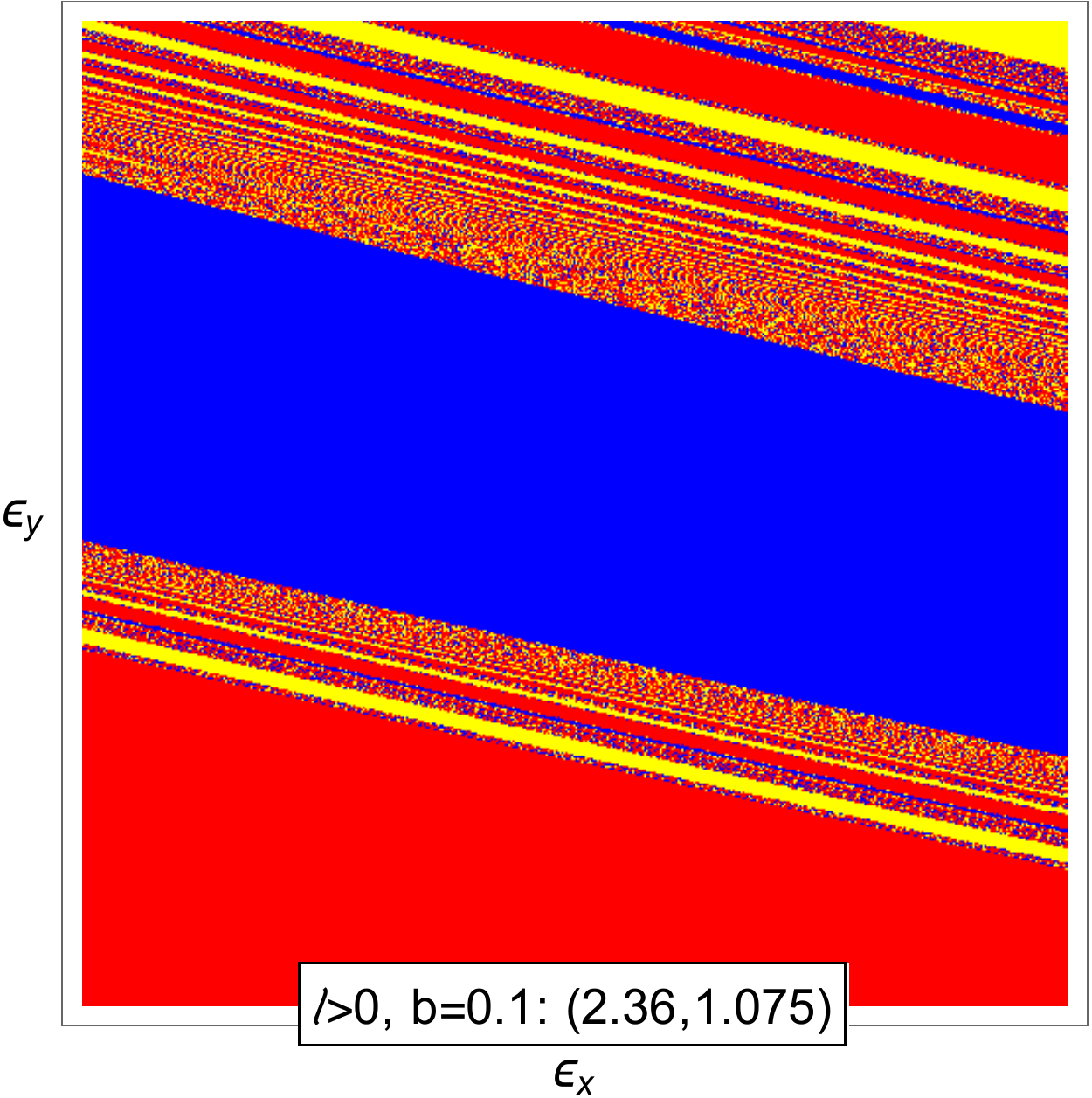}
\end{minipage}
\begin{minipage}{0.32\linewidth}
\centering
\includegraphics[width=\linewidth]{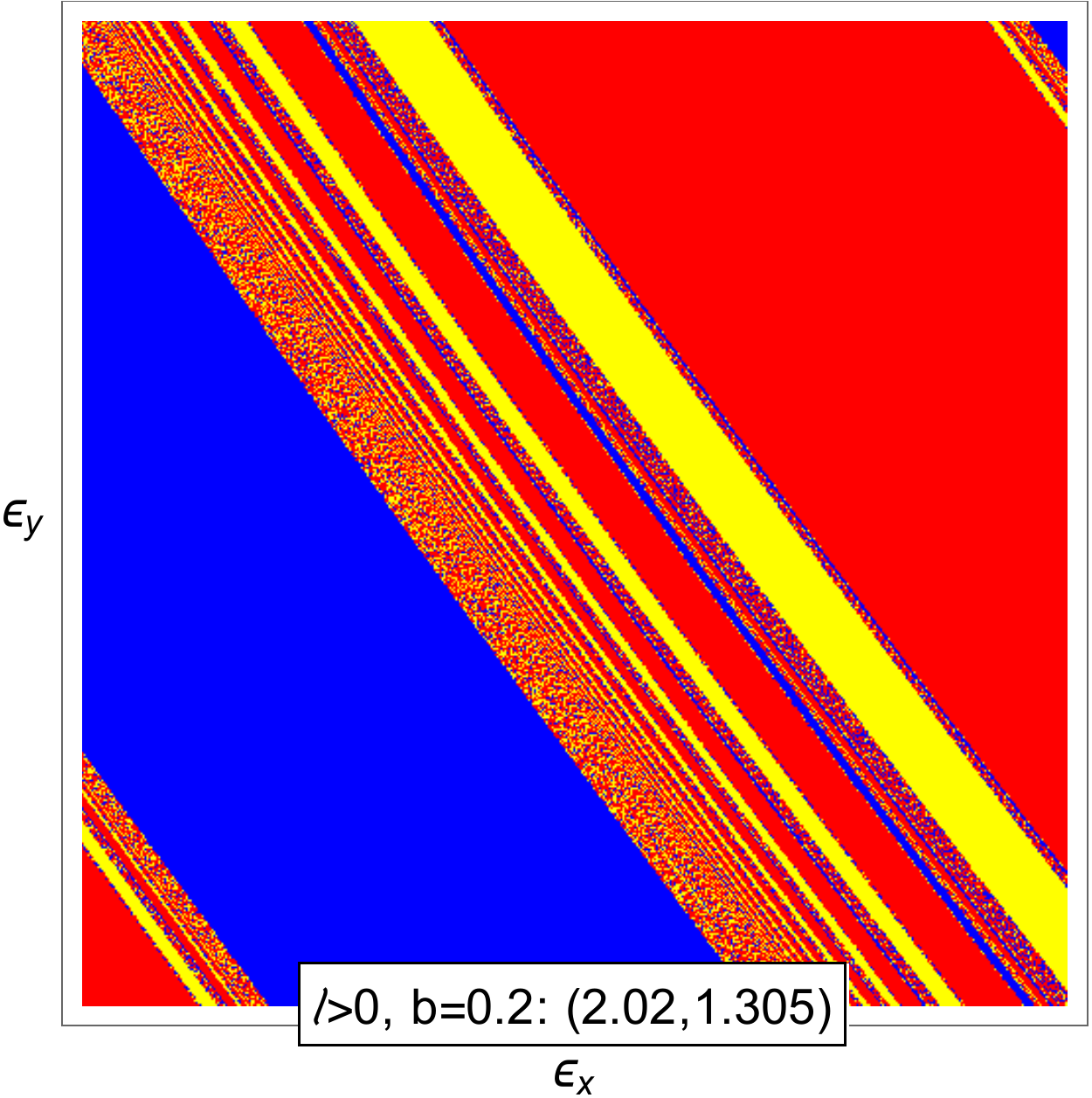}
\end{minipage}
\begin{minipage}{0.32\linewidth}
\centering
\includegraphics[width=\linewidth]{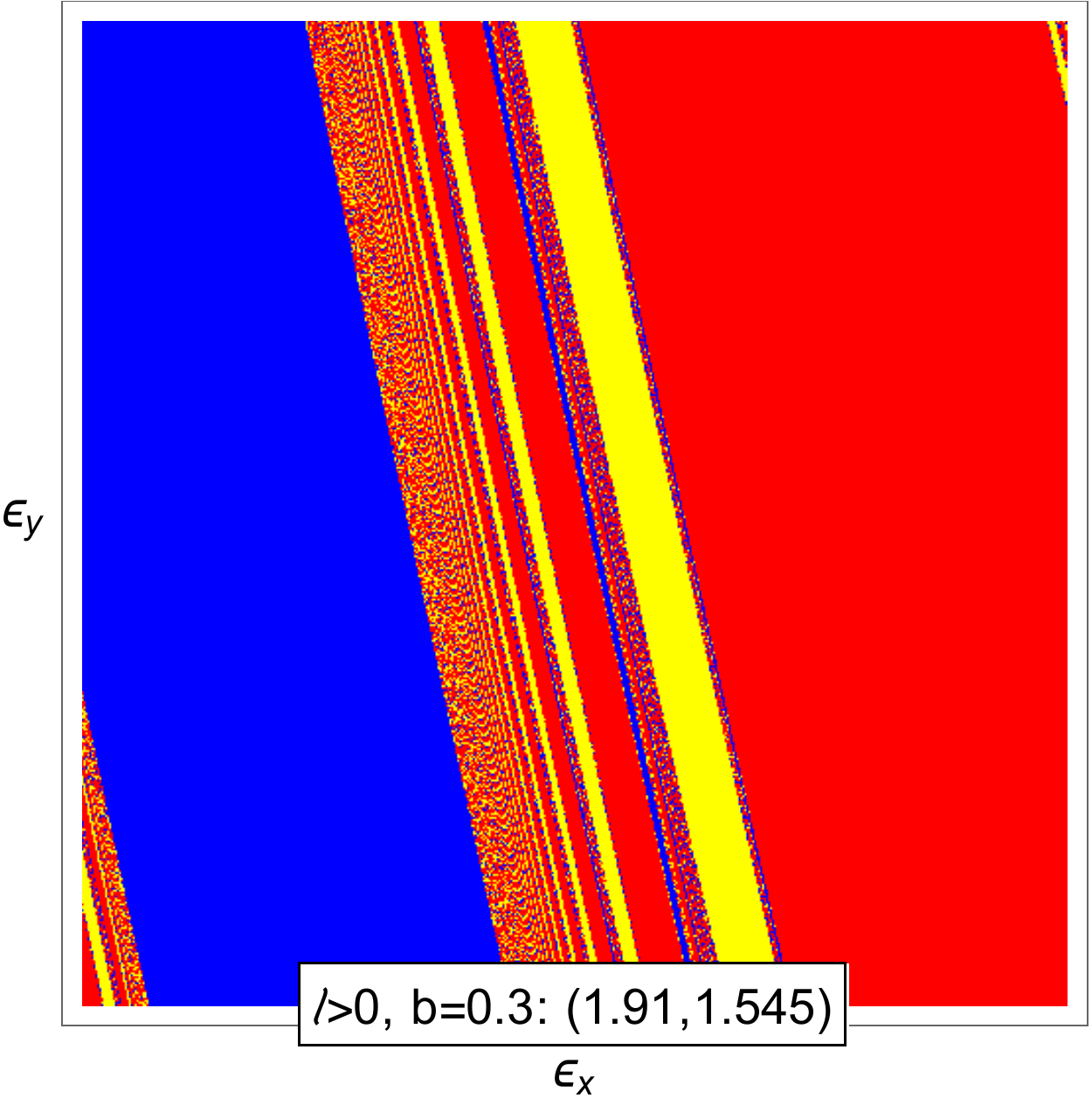}
\end{minipage}
\begin{minipage}{0.32\linewidth}
\centering
\includegraphics[width=\linewidth]{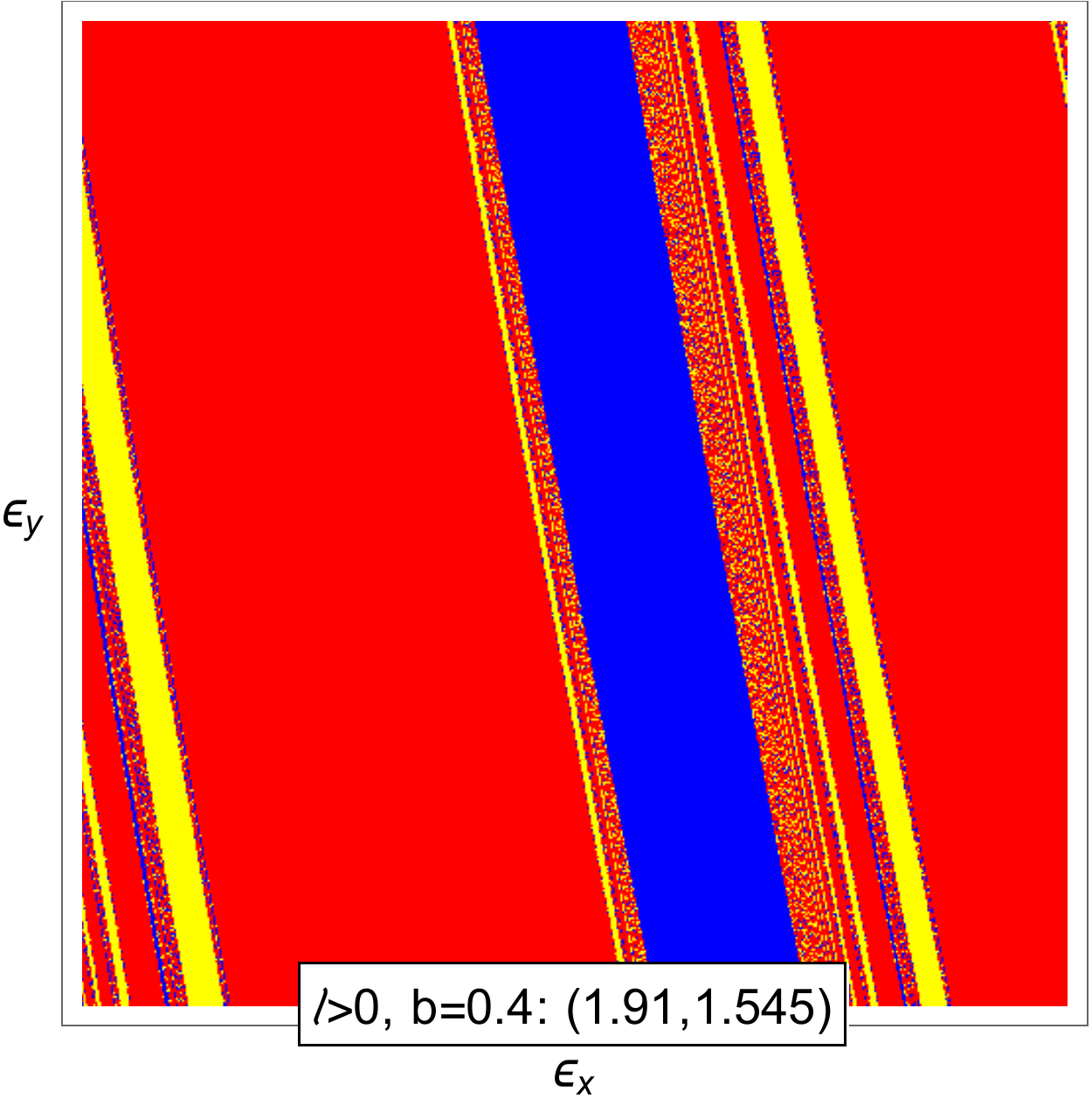}
\end{minipage}
\begin{minipage}{0.32\linewidth}
\centering
\includegraphics[width=\linewidth]{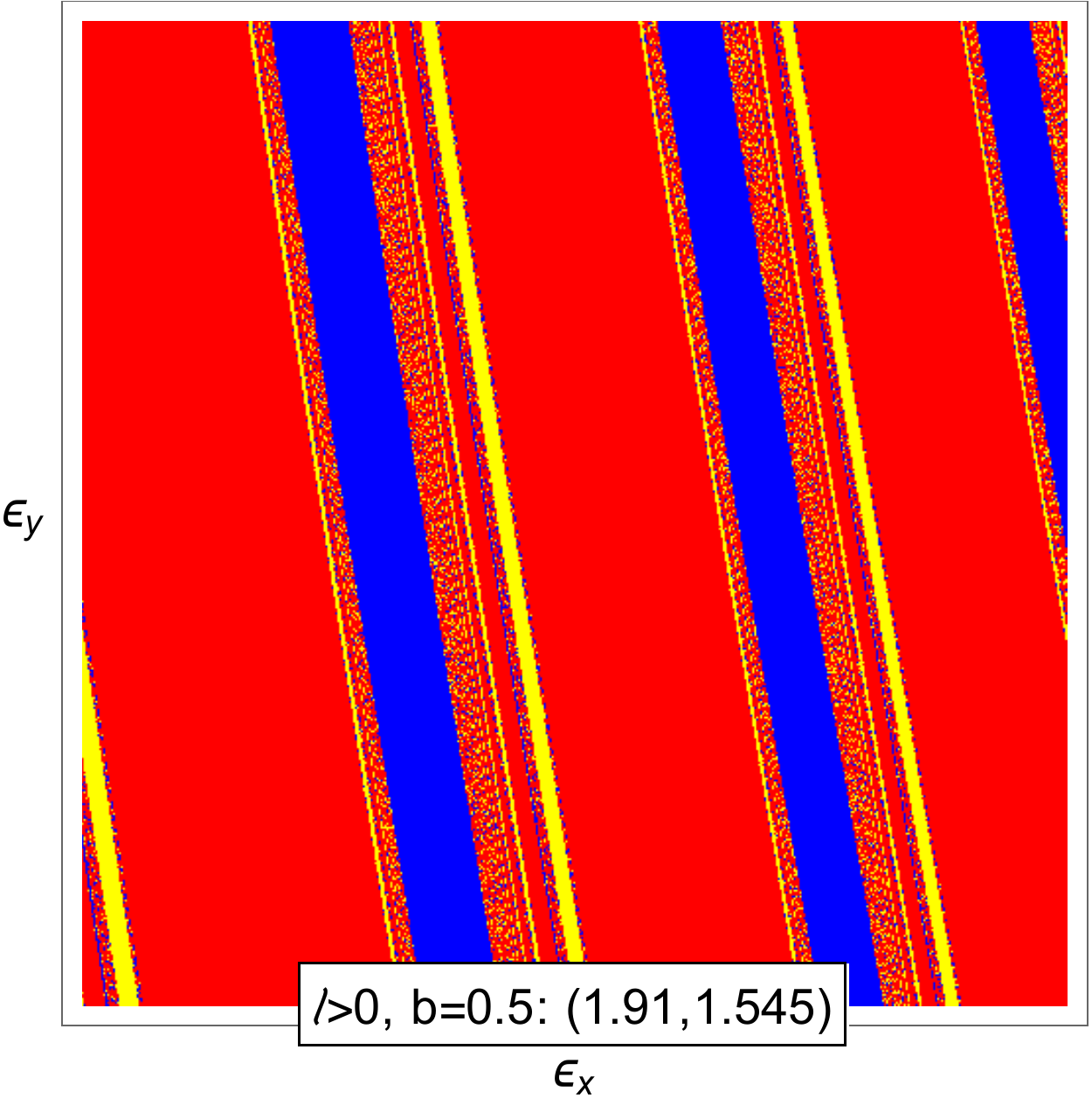}
\end{minipage}
\begin{minipage}{0.32\linewidth}
\centering
\includegraphics[width=\linewidth]{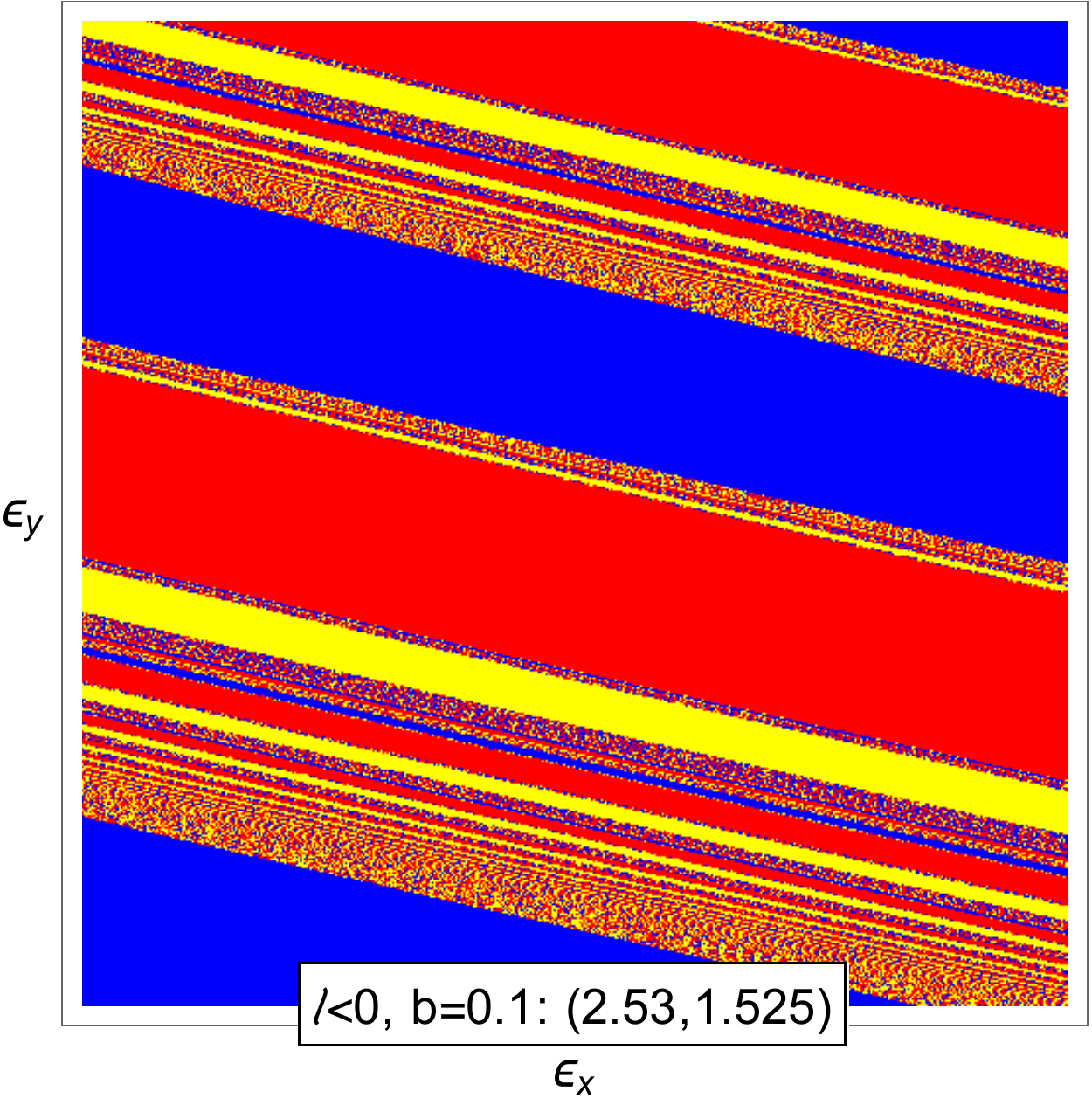}
\end{minipage}
\begin{minipage}{0.32\linewidth}
\centering
\includegraphics[width=\linewidth]{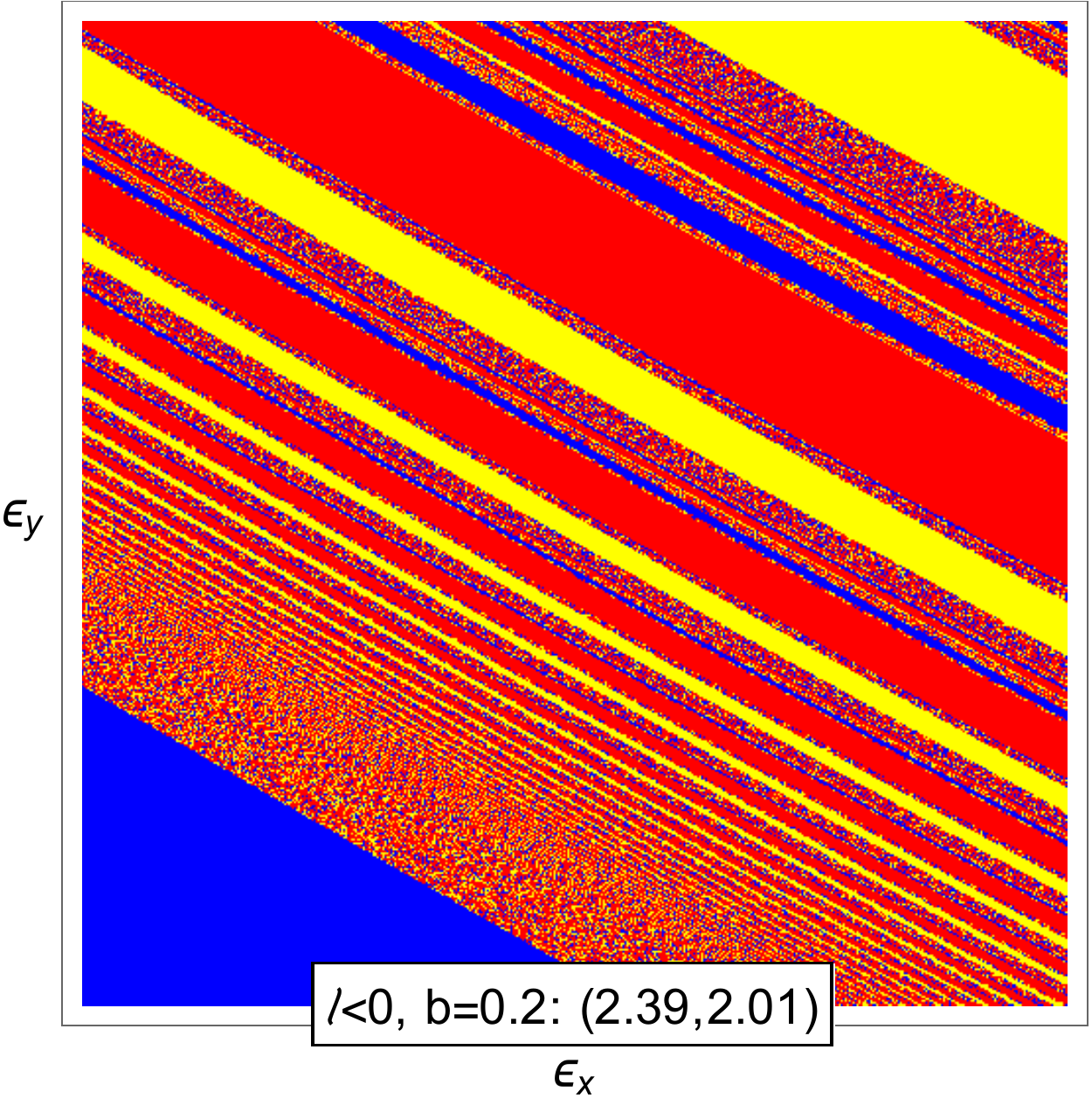}
\end{minipage}
\begin{minipage}{0.32\linewidth}
\centering
\includegraphics[width=\linewidth]{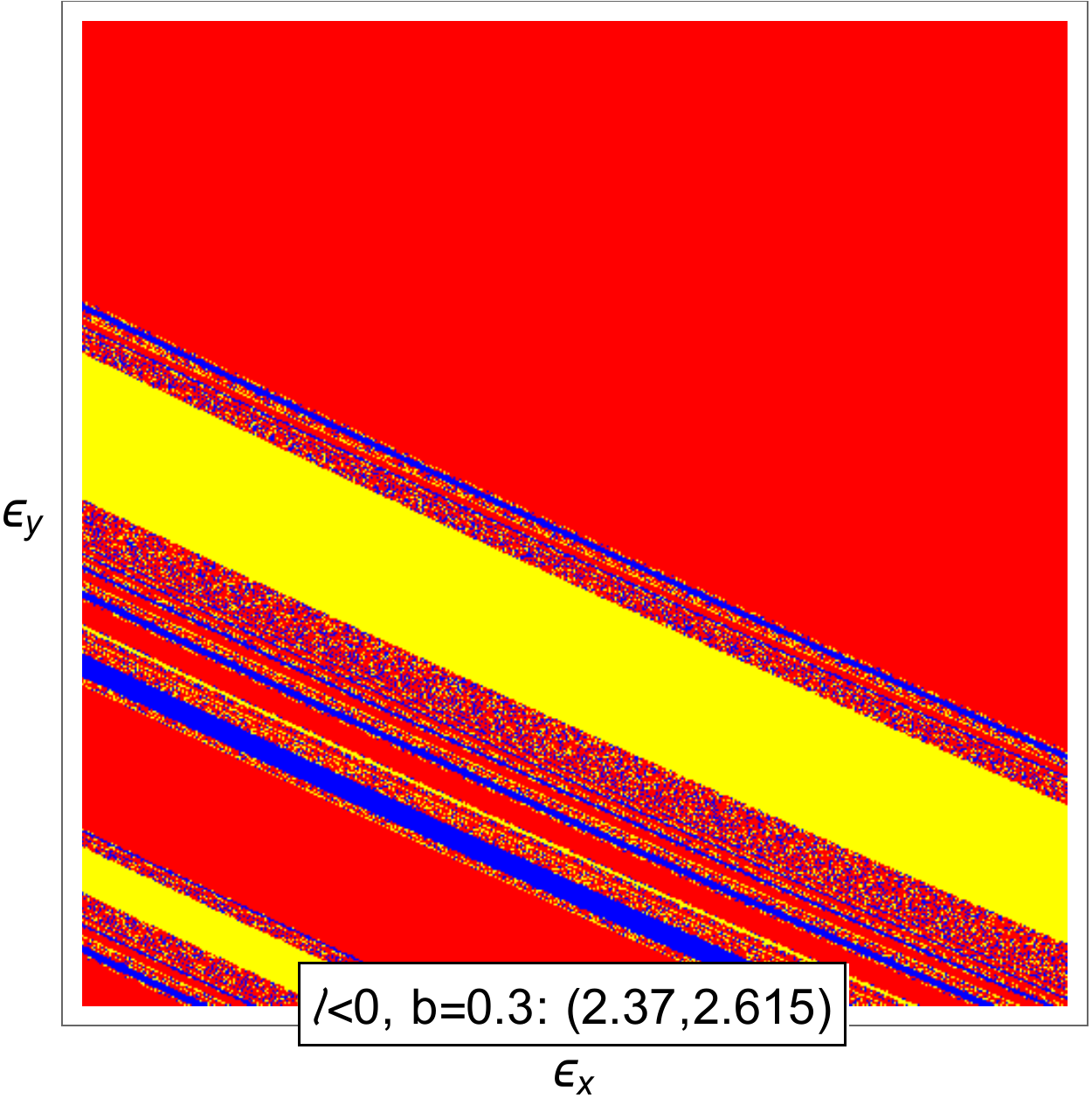}
\end{minipage}
\begin{minipage}{0.32\linewidth}
\centering
\includegraphics[width=\linewidth]{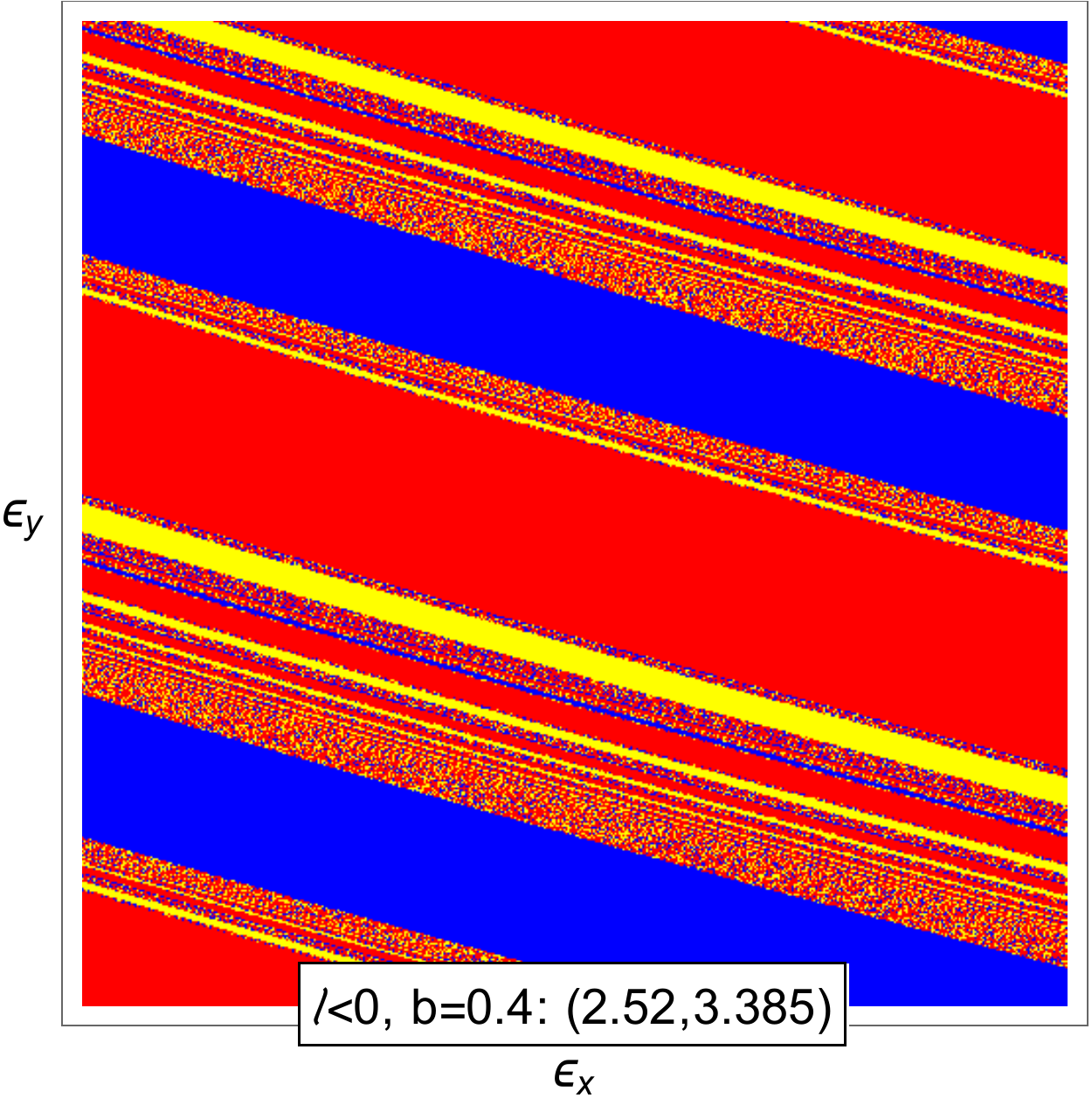}
\end{minipage}
\begin{minipage}{0.32\linewidth}
\centering
\includegraphics[width=\linewidth]{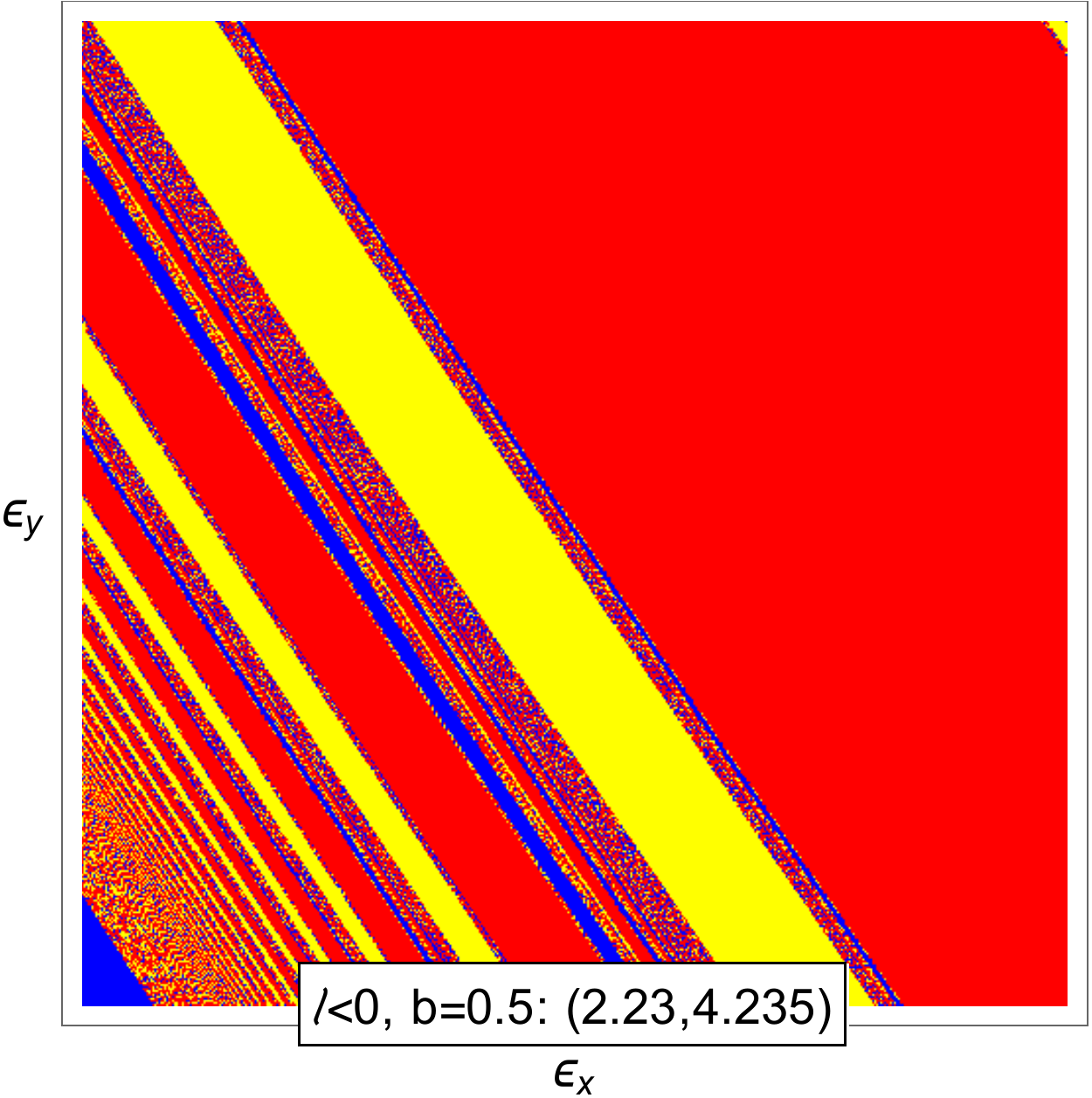}
\end{minipage}
\caption{\label{fig:case1zooms} Magnified parts of the basin plots in \sref{sec:basinplotscase1} around the diffuse region near the critical escape energy. The ordered pair of numbers $(\rho_\mathrm{c},\mathcal{E}_\mathrm{c})$ indicate where the plot is centered.
The axes correspond to the distance from this center point: $\epsilon_x = \rho_0-\rho_\mathrm{c}$, $\epsilon_y = \mathcal{E}-\mathcal{E}_\mathrm{c}$ with $\epsilon_x \in [-0.01,0.01]$ and $\epsilon_y \in [-0.015,0.015]$. All basin plots are $500\times 500$ resolution.}
\end{center}
\end{figure}

\begin{figure}
\begin{center}
\begin{minipage}{0.6\linewidth}
\centering
\includegraphics[width=\linewidth]{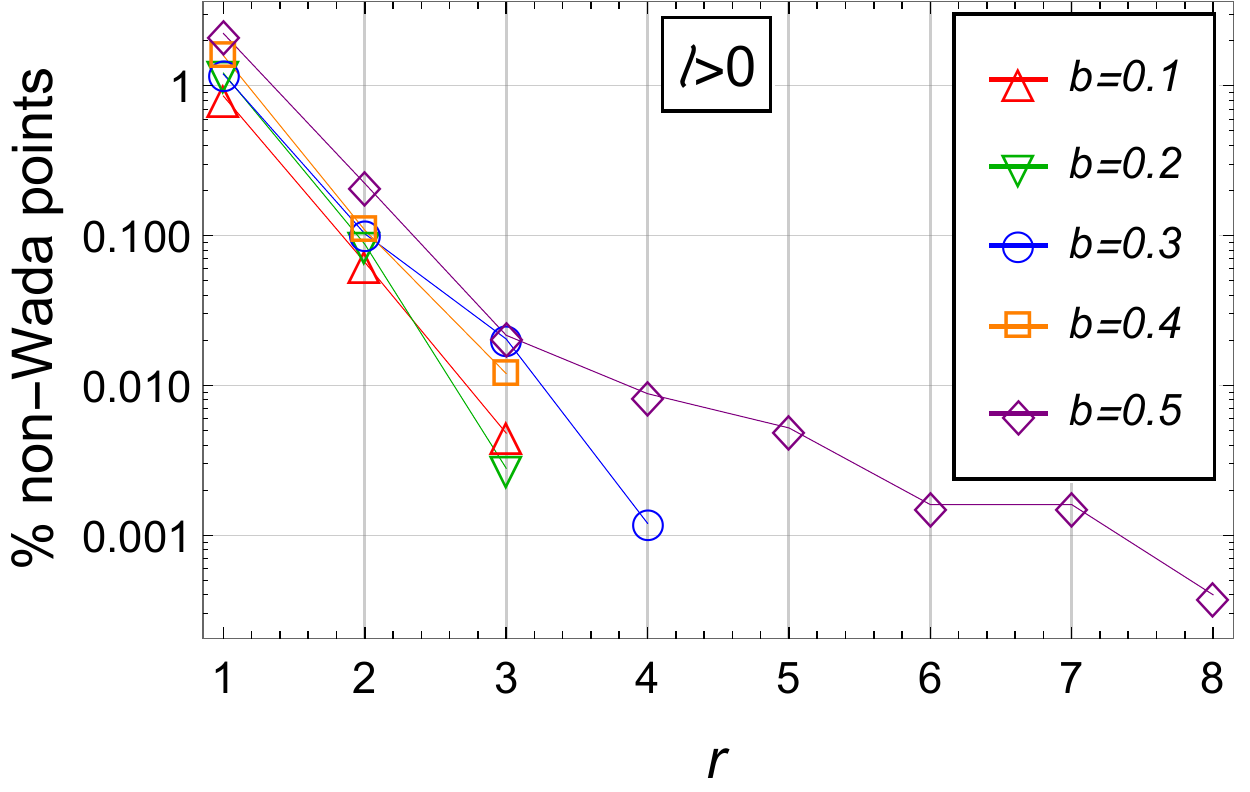}
\end{minipage}
\vskip 0.2cm
\begin{minipage}{0.6\linewidth}
\centering
\includegraphics[width=\linewidth]{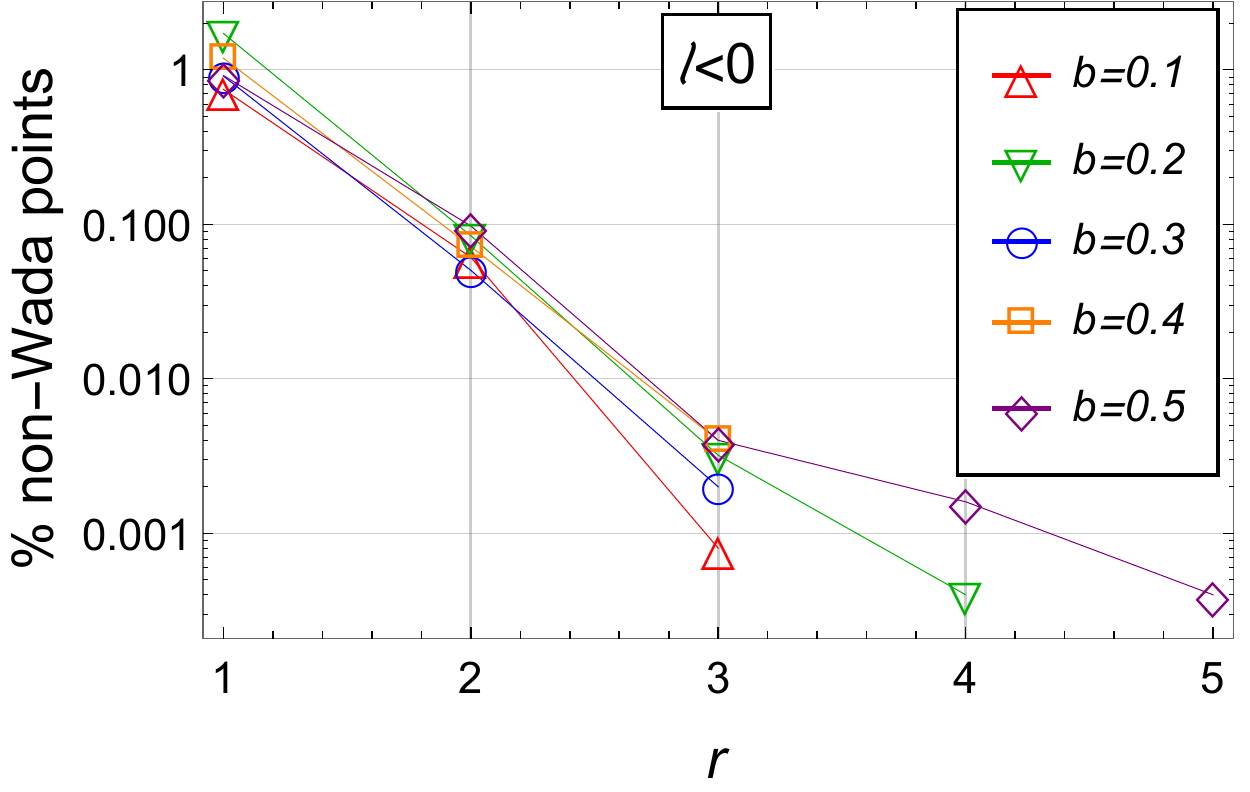}
\end{minipage}
\caption{\label{fig:percentnonwadacase1} Percentage of non-Wada points as a function of the fattening parameter $r$ for selected parts of the diffuse region in the basins obtained in \sref{sec:basinplotscase1}. Note that since the $y-$axis is on a log scale, for each case there is one additional data point $(r_\mathrm{max}+1,0)$.}
\end{center}
\end{figure}

It is important to note that the basins obtained in \fref{fig:initcond1basinplotspl} and \fref{fig:initcond1basinplotsnl} as it stands, cannot exhibit the \emph{full} Wada property.
We see from \fref{fig:initcond1basinplotspl} and \fref{fig:initcond1basinplotsnl} that at large enough $\rho_0$, around the critical escape energy line, the red basin disappears.
Hence at those regions, the basin boundaries are only shared between two basins.
Nevertheless, at the diffuse regions near the critical escape energy line at small $\rho_0$, it is possible for the three basins to share a common boundary and we can then say that the basins exhibit the \emph{partial} Wada property \cite{Zhang2013}.
We applied the merging method to selected magnified parts in the diffuse region of the basin plots obtained in \sref{sec:basinplotscase1}.
We show these magnified basin plots (with $500\times 500$ resolution) in \fref{fig:case1zooms}.
\Fref{fig:percentnonwadacase1} shows the percentage of non-Wada points as a function of the fattening parameter $r$ for the basins in \fref{fig:case1zooms}.
For most of the cases, the merging method was able to classify the basins as Wada for $r\leq 5$.
It is only for the case $b=0.5$ that $r>5$, although we see that even at around $r=5$, the percentage of non-Wada points for $b=0.5$ is already vanishingly small.
In order to have a better idea of the reliability of these results we translate these pixel distances $r$ to actual distances in the phase space we are considering.
We see from \fref{fig:case1zooms} that $(\Delta\epsilon_x,\Delta\epsilon_y) = (0.02,0.03)$.
For a $500\times 500$ resolution basin plot, a pixel of distance in this case would correspond to $\Delta r \sim 10^{-4}$.
Hence, the results in \fref{fig:percentnonwadacase1} indicate that the boundaries in the basins that we considered are the same to within $\sim 10^{-3}$ units of phase space distance.
This is a strong indication that the basins considered in \sref{sec:basinplotscase1} are (partially) Wada in the diffuse region near the critical escape energy.
We show evidence in \ref{sec:wadaresolution} that the results of \fref{fig:percentnonwadacase1} hold for higher resolutions, which strengthens the case for the system being partially Wada.

\section{Basin entropy}\label{sec:basinentropy}
One important question in the study of dynamical systems is how ``uncertain" a particular trajectory or the overall system is. 
That is, given a certain error in choosing initial conditions, how different would the resulting trajectory be? Examples of proposed measures of uncertainty in the context of basin plots are the uncertainty exponent \cite{Grebogi1983} and basin stability \cite{Menck2013}. 
The uncertainty exponent $\alpha$ quantifies the uncertainty by measuring the dimension of the boundaries that separates the basins. The value of $\alpha$ ranges from 0 to 1, with $\alpha = 1$ corresponding to smooth boundaries and $\alpha < 1$ corresponding to fractal boundaries. 
The smaller $\alpha$ is, the more uncertain the system.
One disadvantage of this method is that it does not take into account the size of the boundary, which is related to the relative sizes of the basins.
For example, a system containing two basins with both basins occupying equal volumes in phase space should be more uncertain compared to the case where one basin occupies almost all of phase space.
On the other hand, the basin stability quantifies the uncertainty by measuring the volume in phase space occupied by a particular basin, with larger basins considered more stable (less uncertain).
One main disadvantage of this method however is that it does not take into account how the different basins are mixed.
A highly mixed basin should be more uncertain.
The notion of basin entropy introduced by Daza et al \cite{Daza2016} provides a way to quantify uncertainty that takes into account the fractal dimension of the boundaries, the relative sizes of the boundaries, and the way the basins are mixed.
As we will see, the basin entropy method also takes into account if the basins exhibit the Wada property.

Any numerical procedure that attempts to calculate trajectories in phase space cannot do so with infinite accuracy and precision.
Points in phase space can only be determined up to a finite resolution.
This finite resolution introduces a stochastic element into an otherwise completely deterministic system.
The main idea of the basin entropy method is that by subdividing phase space into a grid with box sizes $\epsilon$ which represents the finite resolution of the experimental or numerical procedure, and considering each box as a random variable (i.e. which basin the box belongs to is random), we can define a Gibbs entropy for a box $i$:
\begin{equation}\label{eq:basinentropy1}
S_i = -\sum_{j=1}^{m_i} p_{i,j} \log{(p_{i,j})},
\end{equation}
where $p_{i,j}$ is the probability that box $i$ is in basin $j$, and $m_i$ is the number of possible basins box $i$ could belong to.
The basin entropy is then defined as
\begin{equation}
S_b = \frac{1}{N}\sum_{i=1}^N S_i
\end{equation}
where $N$ is the total number of boxes.
To a very good approximation, the basin entropy can be expressed as (see \cite{Daza2016} for details)
\begin{equation}\label{eq:basinentropy2}
S_b = \sum_{k=1}^{k_\mathrm{max}} \frac{n_k}{\tilde{n}} \epsilon^{\alpha_k} \log{(m_k)}
\end{equation}
where $\epsilon$ is the box size, $k$ labels different basin boundaries, $m_k$ is the number of possible basins in boundary $k$, $\alpha_k$ is the uncertainty exponent of boundary $k$, and the quantity $(n_k/\tilde{n})$ is a measure of the size occupied by boundary $k$ relative to the total phase space.
\Eref{eq:basinentropy2} makes it clear that the basin entropy increases with the size of the boundaries via the factor $(n_k/\tilde{n})$, the fractality of the boundaries via the uncertainty dimension $\alpha_k$, and the number of basins in a particular boundary via the factor $\log{(m_k)}$.
In the particular case that the basins exhibit the Wada property, then there is just one boundary that separates all the basins, and hence the factor $\log{(m_k)}$ is maximum.

\begin{figure}[t]
\begin{center}
\includegraphics[width=0.6\linewidth]{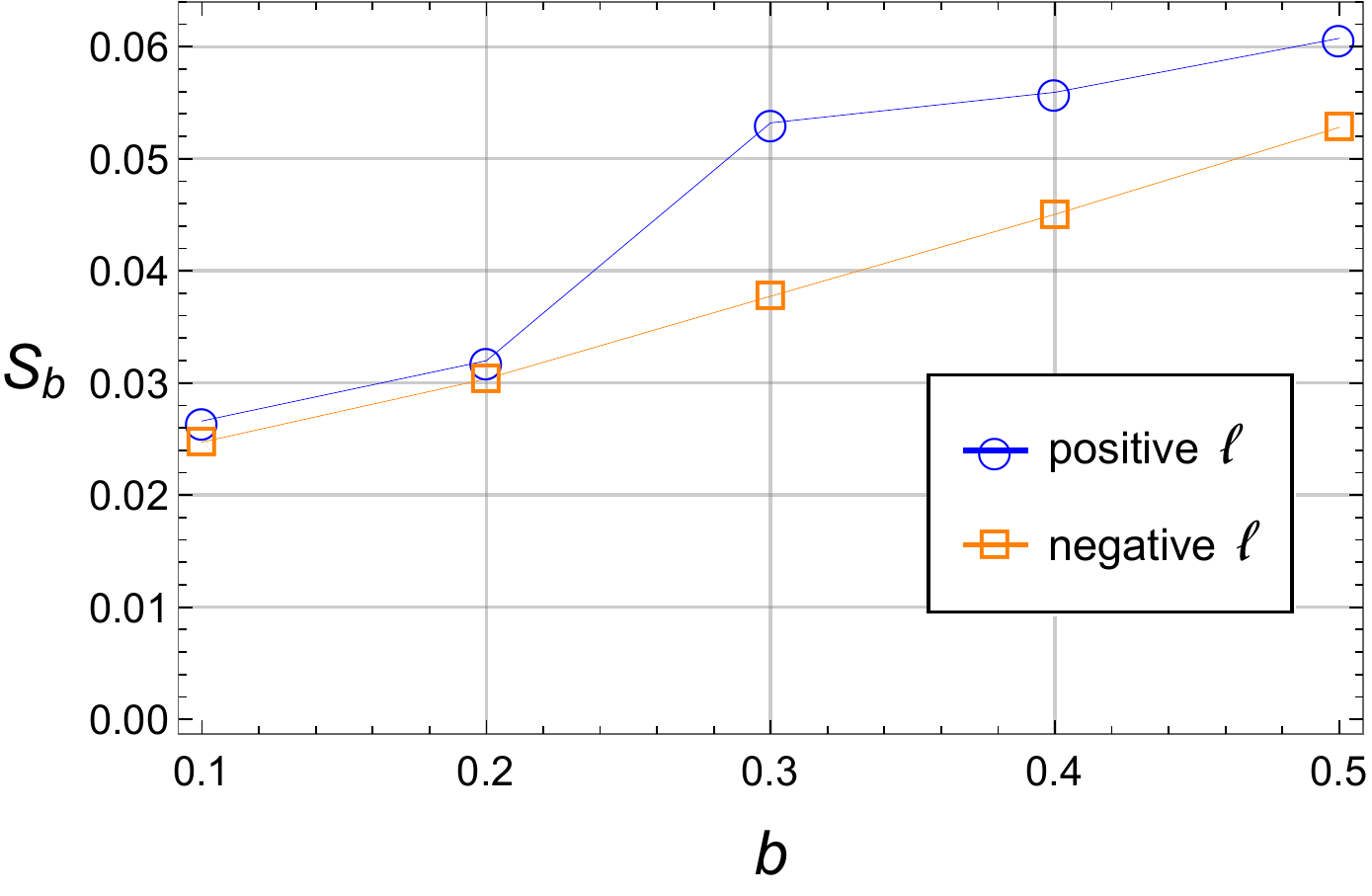}
\caption{\label{fig:basinentropycase1} The basin entropy as a function of the electromagnetic interaction strength $b$ for the basins calculated in \sref{sec:basinplotscase1}.}
\end{center}
\end{figure}

We obtained the basin entropy for the basins calculated in \sref{sec:basinplotscase1} by dividing the $500 \times 500$ pixel basin plots into $5 \times 5$  pixel sized boxes so that we have 25 trajectories per box.
This is the number of trajectories per box recommended by Daza et al in \cite{Daza2016}.
The basin entropy as a function of the electromagnetic interaction strength $b$ is shown in \fref{fig:basinentropycase1}.
We see that in both $\ell>0$ and $\ell<0$ cases, the basin entropy increases as $b$ is increased.
We checked that this overall trend does not change as we lowered the resolution of the basin plots.
We see that for the particular choice of initial conditions in \sref{sec:basinplotscase1}, the trajectories of charged particles are more uncertain as the strength of the electromagnetic interaction increases.

\section{Alternative phase space slicing: zero initial azimuthal velocity}
\label{sec:basinplotscase2}
\begin{figure}
\begin{center}
\includegraphics[width=0.33\linewidth]{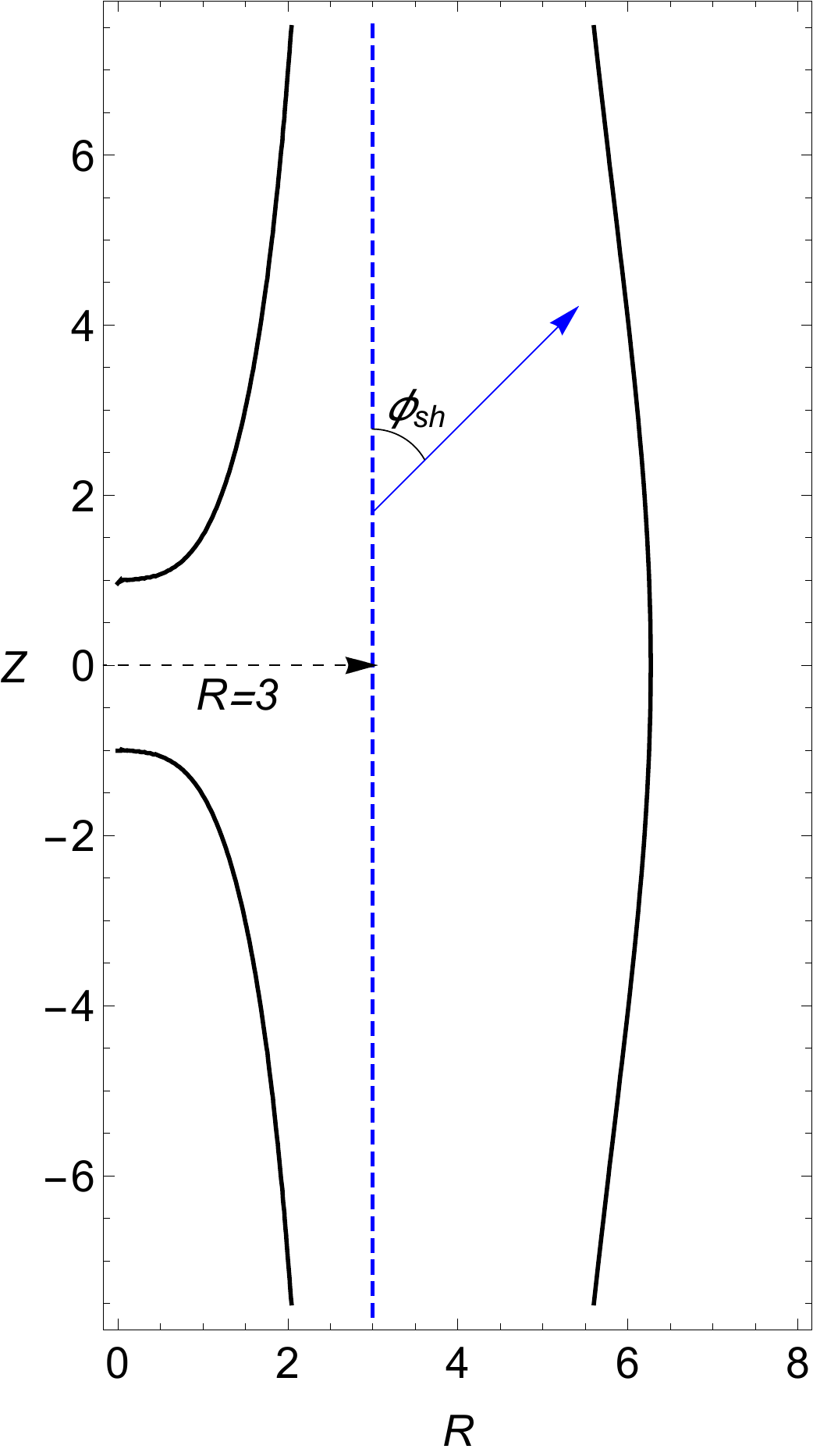}
\caption{\label{fig:initconds2} The particle initially with zero azimuthal velocity at ($R=3$,$z_0$) launched at an angle $\phi_\mathrm{sh}$.}
\end{center}
\end{figure}

\begin{figure}
\begin{center}
\begin{minipage}{0.32\linewidth}
\centering
\includegraphics[width=\linewidth]{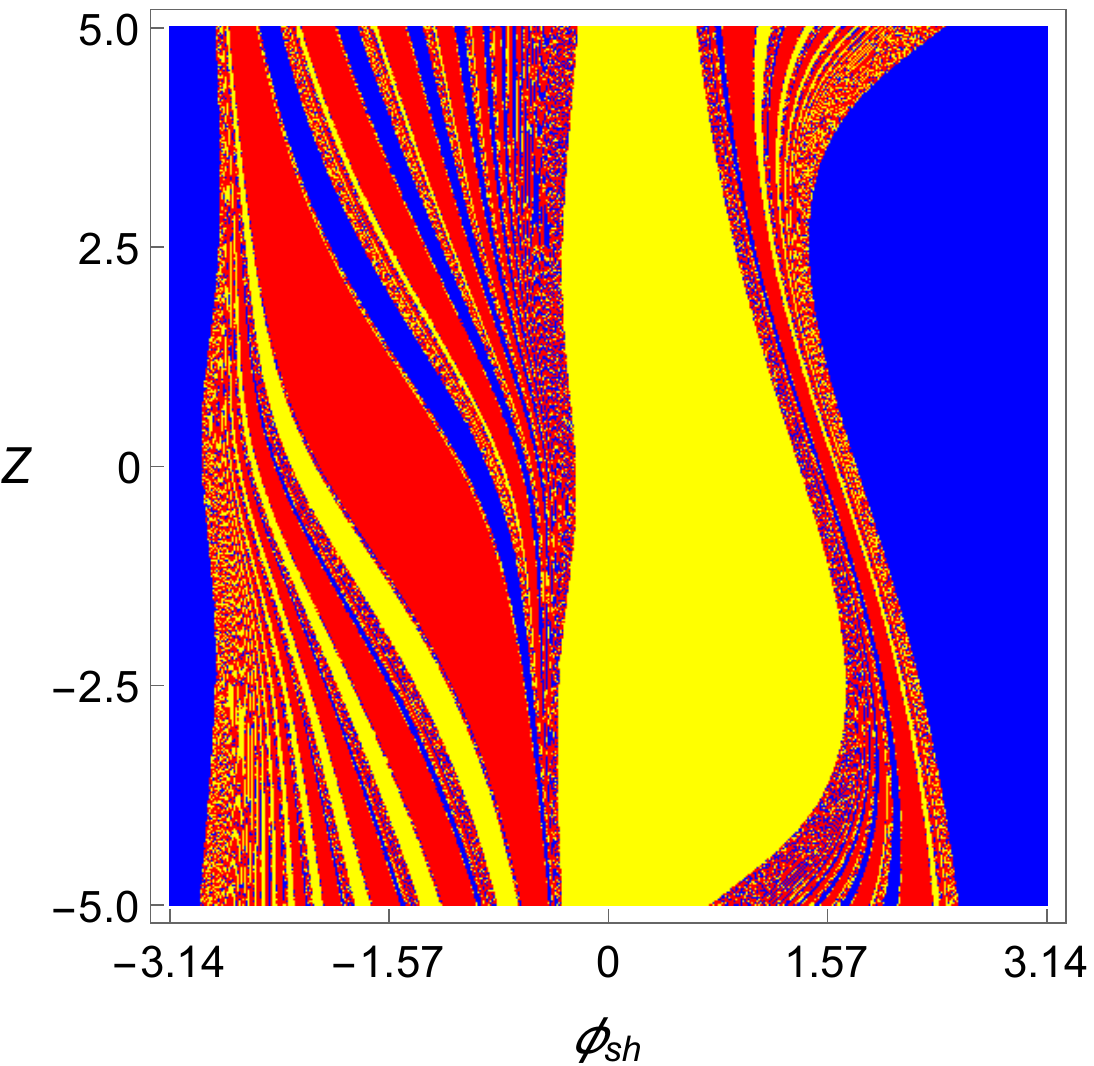}\\
(a) $\mathcal{E}=1.2, b=0.1$
\end{minipage}
\begin{minipage}{0.32\linewidth}
\centering
\includegraphics[width=\linewidth]{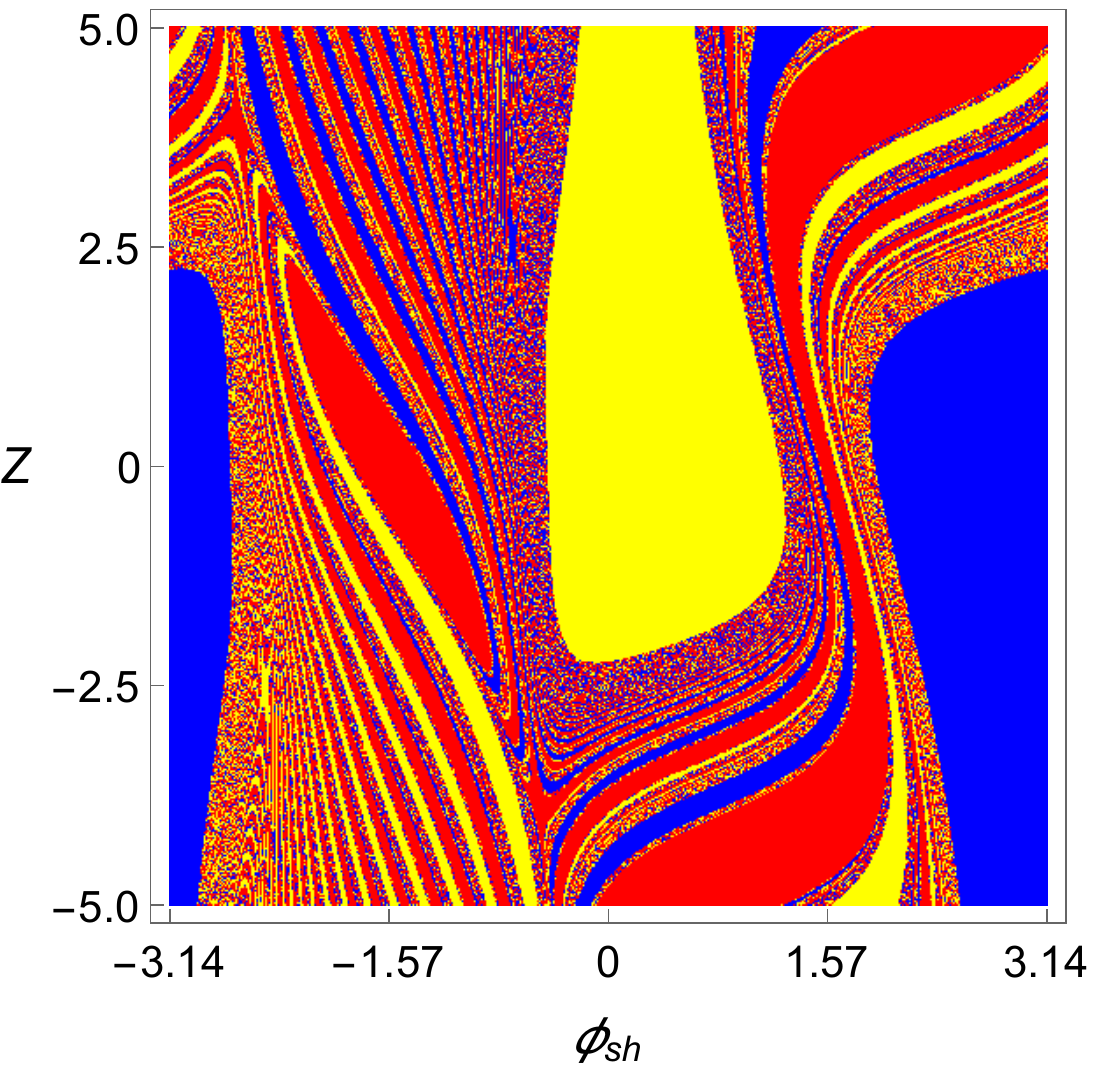}\\
(b) $\mathcal{E}=1.2, b=0.3$
\end{minipage}
\begin{minipage}{0.32\linewidth}
\centering
\includegraphics[width=\linewidth]{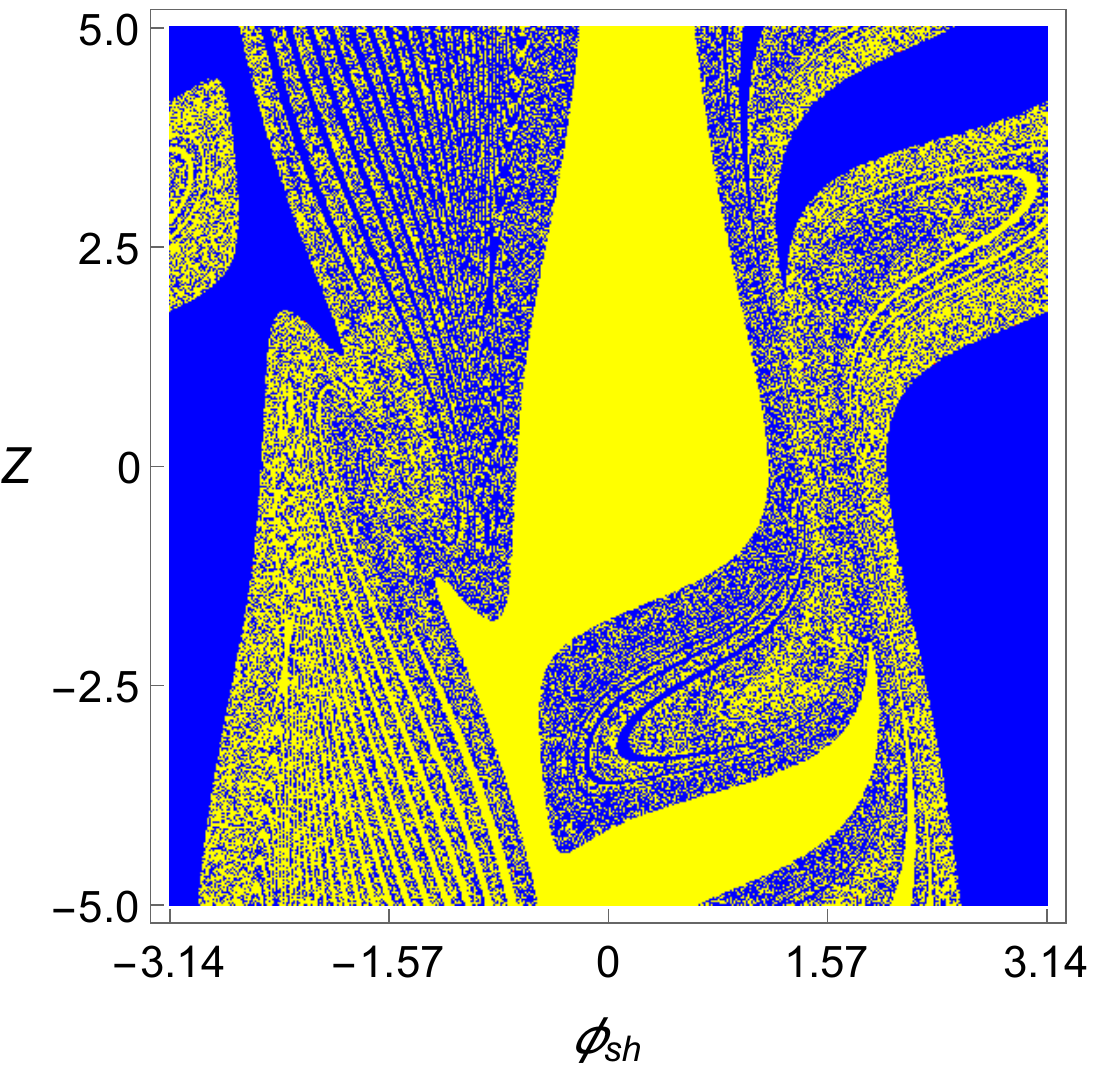}\\
(c) $\mathcal{E}=1.2, b=0.5$
\end{minipage}
\begin{minipage}{0.32\linewidth}
\centering
\includegraphics[width=\linewidth]{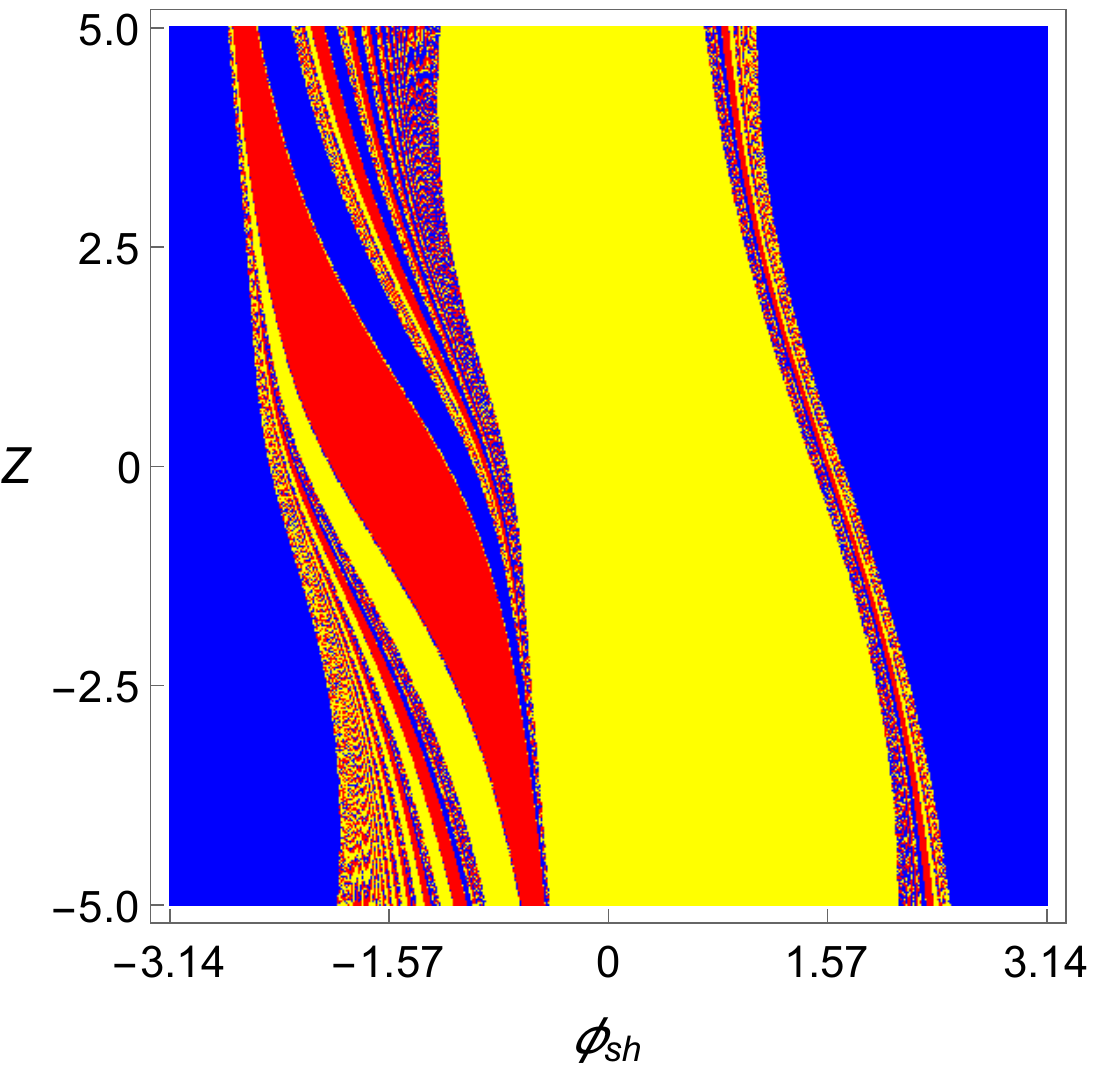}\\
(d) $\mathcal{E}=2.0, b=0.1$
\end{minipage}
\begin{minipage}{0.32\linewidth}
\centering
\includegraphics[width=\linewidth]{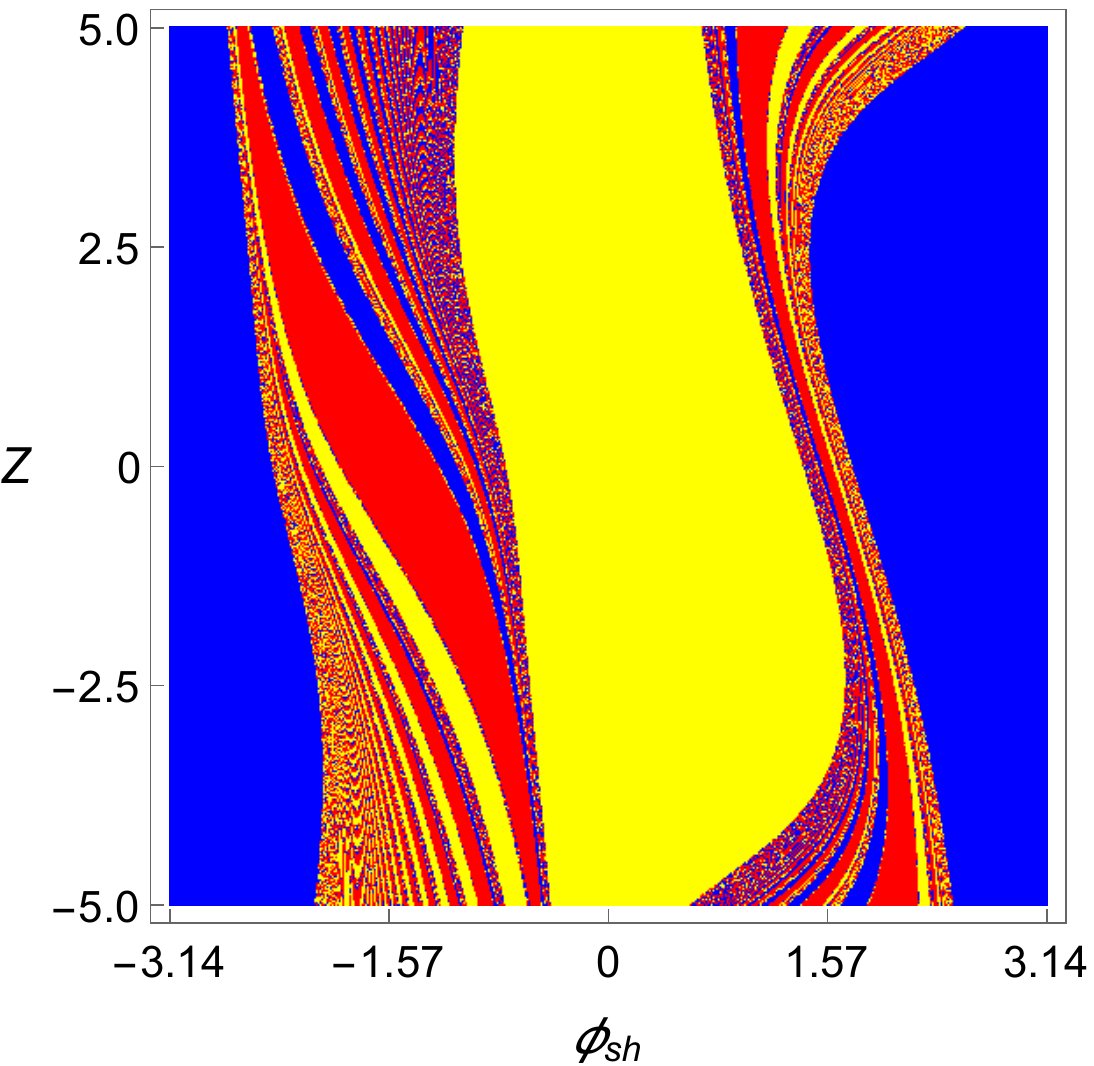}\\
(e) $\mathcal{E}=2.0, b=0.3$
\end{minipage}
\begin{minipage}{0.32\linewidth}
\centering
\includegraphics[width=\linewidth]{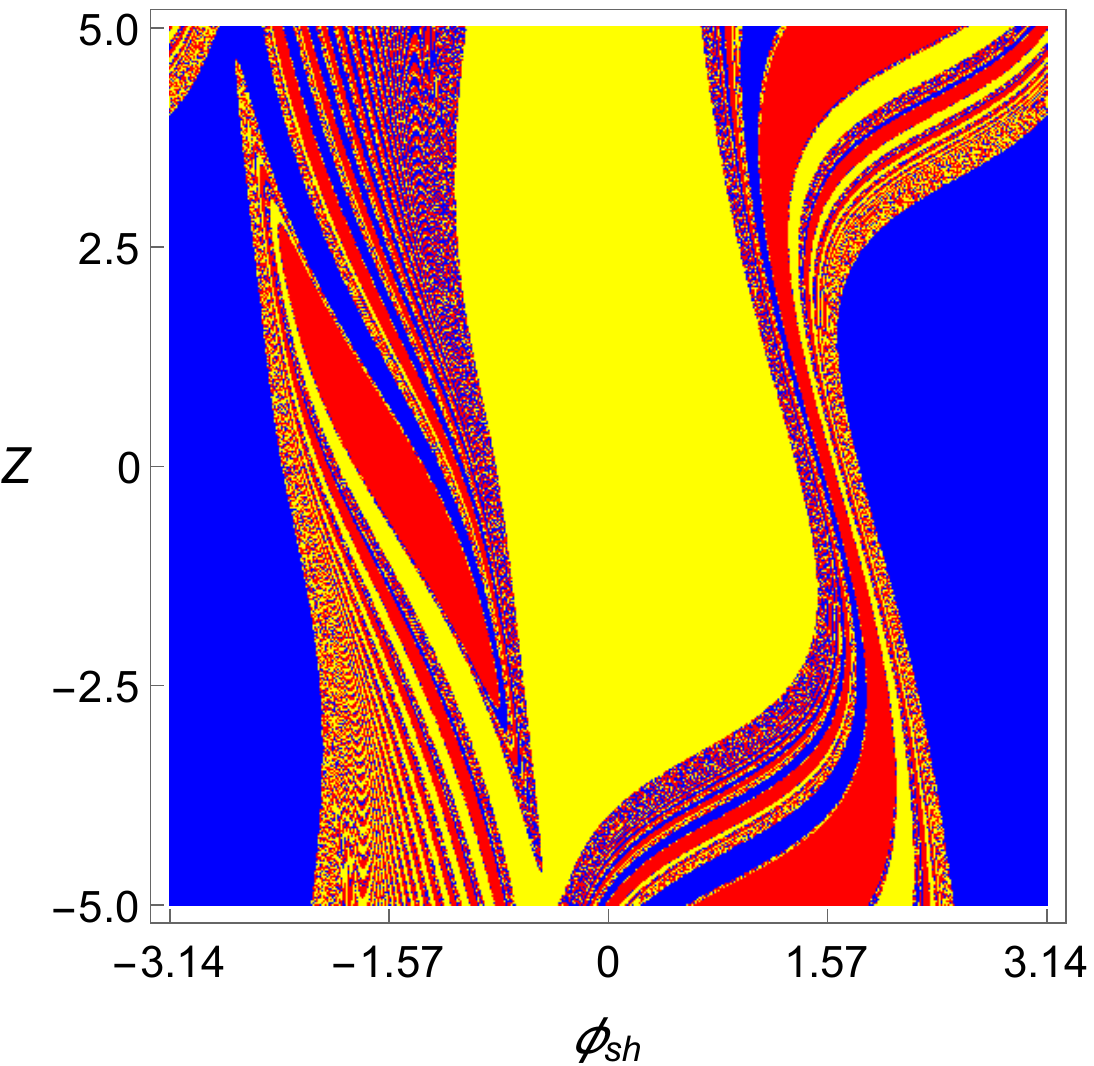}\\
(f) $\mathcal{E}=2.0, b=0.5$
\end{minipage}
\begin{minipage}{0.32\linewidth}
\centering
\includegraphics[width=\linewidth]{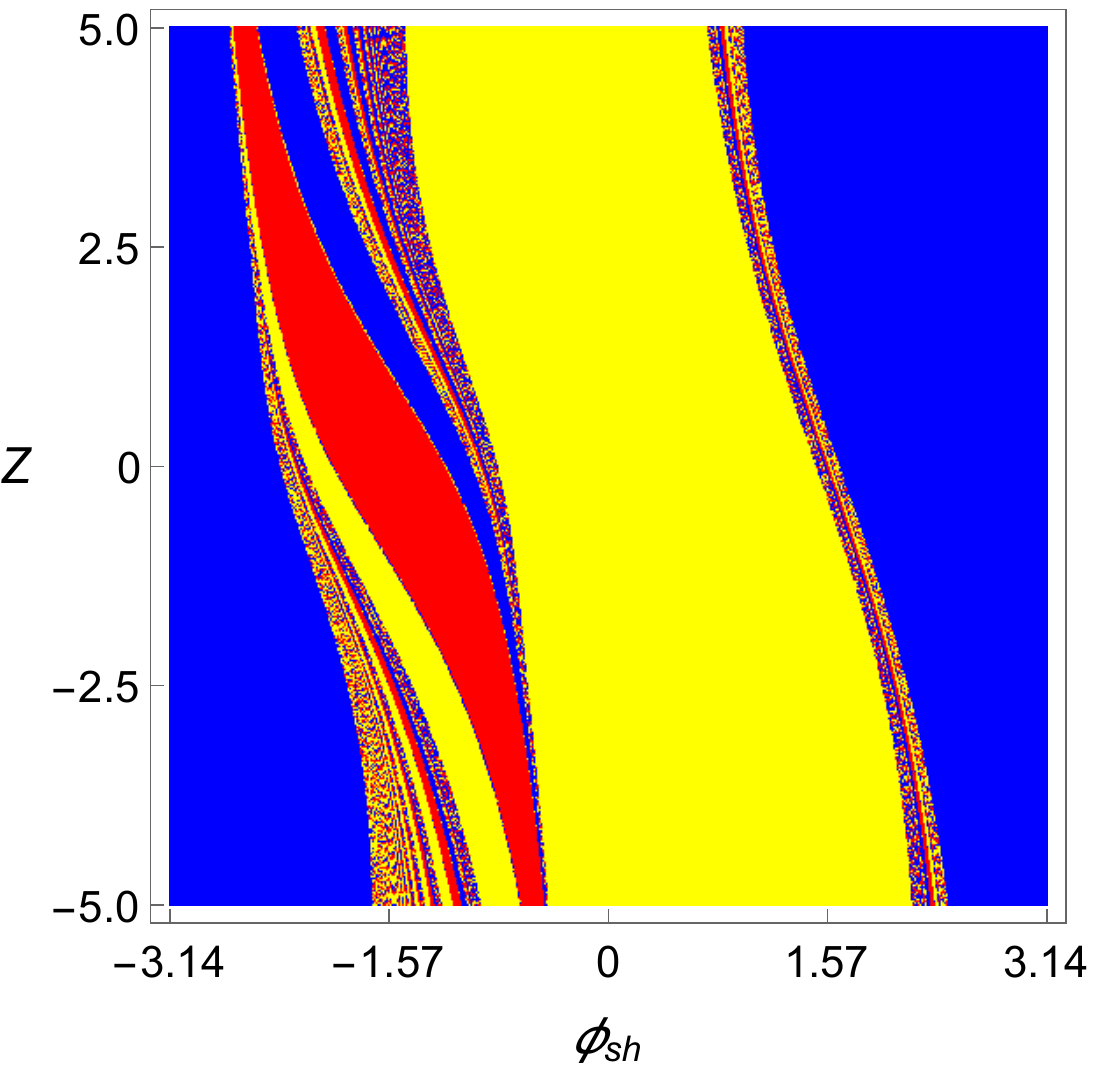}\\
(g) $\mathcal{E}=3.0, b=0.1$
\end{minipage}
\begin{minipage}{0.32\linewidth}
\centering
\includegraphics[width=\linewidth]{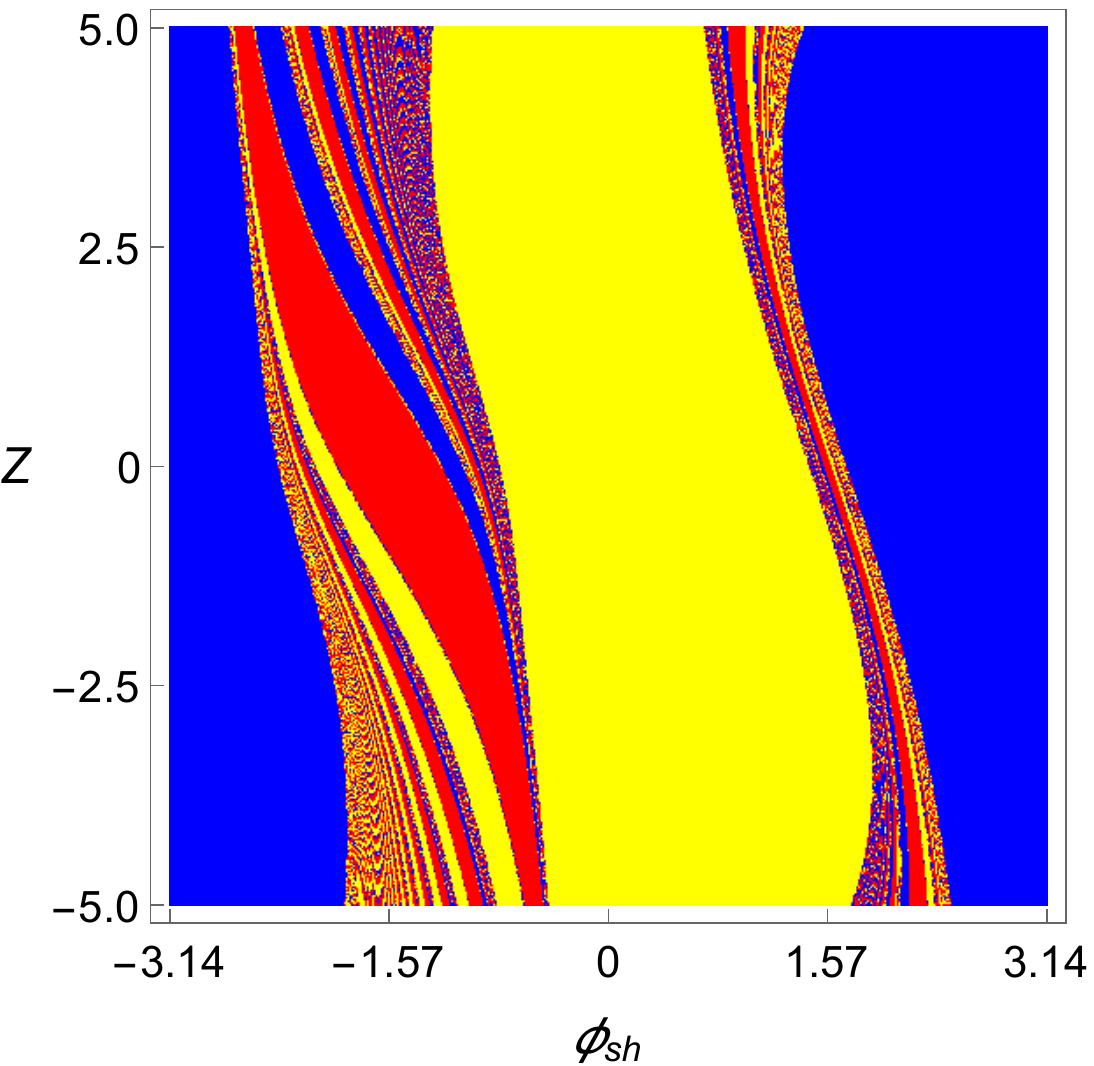}\\
(h) $\mathcal{E}=3.0, b=0.3$
\end{minipage}
\begin{minipage}{0.32\linewidth}
\centering
\includegraphics[width=\linewidth]{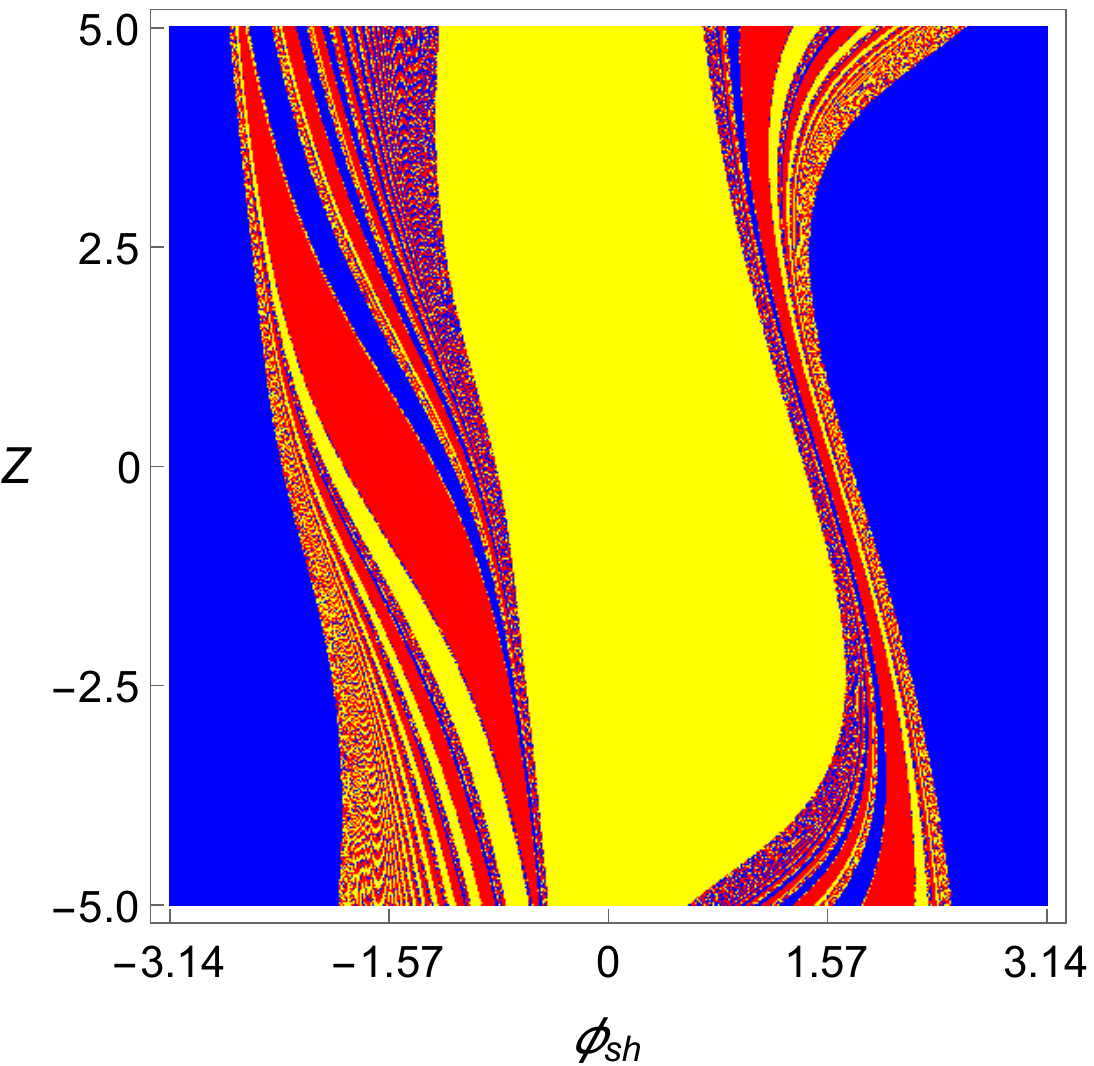}\\
(i) $\mathcal{E}=3.0, b=0.5$
\end{minipage}
\caption{\label{fig:initcond2basinplots} Basin plots for different values of $b$ and $\mathcal{E}$ with initial conditions described in \sref{sec:basinplotscase2}. Red means captures, yellow means upstream escape, blue means downstream escape.}
\end{center}
\end{figure}

In this section we explore a different slice of the phase space to see whether the results obtained in the previous sections hold for more general phase space slices.
We consider a charged particle initially with zero azimuthal velocity along the $R=3$ line launched at an angle $\phi_\mathrm{sh}$ (see \fref{fig:initconds2}).
From \eref{eq:eomtphi}, the zero azimuthal velocity condition, as well as $R=3$ constrains $\ell$ to be
\begin{equation}
\label{eq:l2(b)}
\ell = 3b > 0.
\end{equation}
In this setup, both $b$ and $\mathcal{E}$ become free parameters.
Similar to the previous case of initially circular orbits, we numerically integrate \eref{eq:dlesseomrho} and \eref{eq:dlesseomtheta} for a $500\times 500$ grid of initial conditions $(\phi_\mathrm{sh},z)$ where $\phi_\mathrm{sh} \in [-\pi,\pi]$ and $z \in [-5,5]$.
We also work with values $b \in \{0.1,0.2,0.3,0.4,0.5\}$.
Recall that since $\ell>0$ in this case, from \eref{eq:Ebar}, $\bar{\mathcal{E}}=1$ so that for $\mathcal{E}>1$, upstream and downstream escapes are kinematically possible.
We considered three values for the energy $\mathcal{E}$: $\{1.2, 2.0, 3.0\}$.
The basin plots for $b=\{0.1,0.3,0.5\}$ for these three energies are shown in \fref{fig:initcond2basinplots}.
All the basin plots in \fref{fig:initcond2basinplots} share the same features: a central yellow (upstream escape) region, an outer blue (downstream escape) region, and a fractal region in between where multiple basins coexist.
At lower energies, changes in basin plot features are more sensitive to changes in $b$.
In particular, as $b$ (or $\ell$ via \eref{eq:l2(b)}) is increased, the basins become more mixed.
At a certain critical value of $b \in (0.3,0.4)$, the red basins disappear completely.
This corresponds to the case discussed in \sref{sec:effectivepotential} for which capture becomes impossible but upstream and downstream escapes are possible at certain energies: $\bar{\mathcal{E}} < \mathcal{E} < \mathcal{E}_\mathrm{cap}$ (we recall that $\mathcal{E}_\mathrm{cap}$ increases with $b$).
On the other hand, at higher energies, the basin features become less sensitive to changes in $b$.

\begin{figure}
\begin{center}
\begin{minipage}{0.6\linewidth}
\centering
\includegraphics[width=\linewidth]{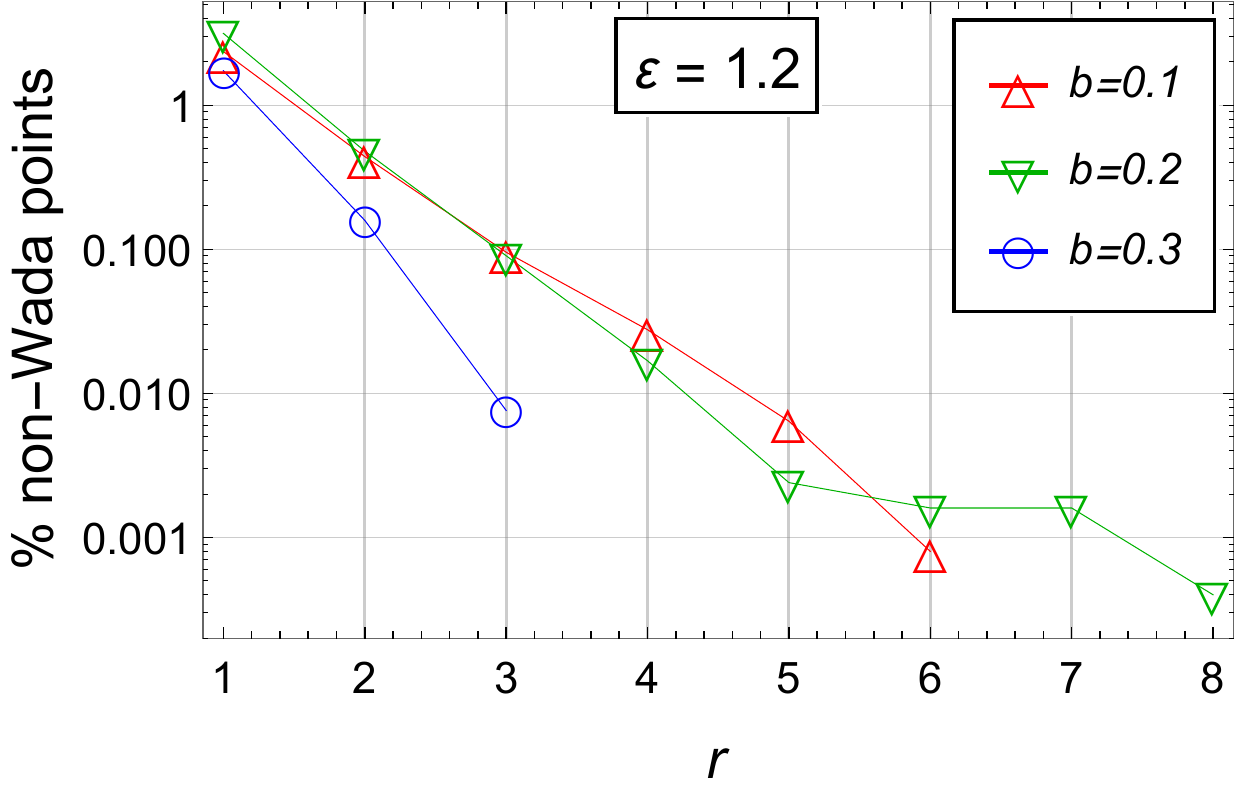}
\end{minipage}
\vskip 0.2cm
\begin{minipage}{0.6\linewidth}
\centering
\includegraphics[width=\linewidth]{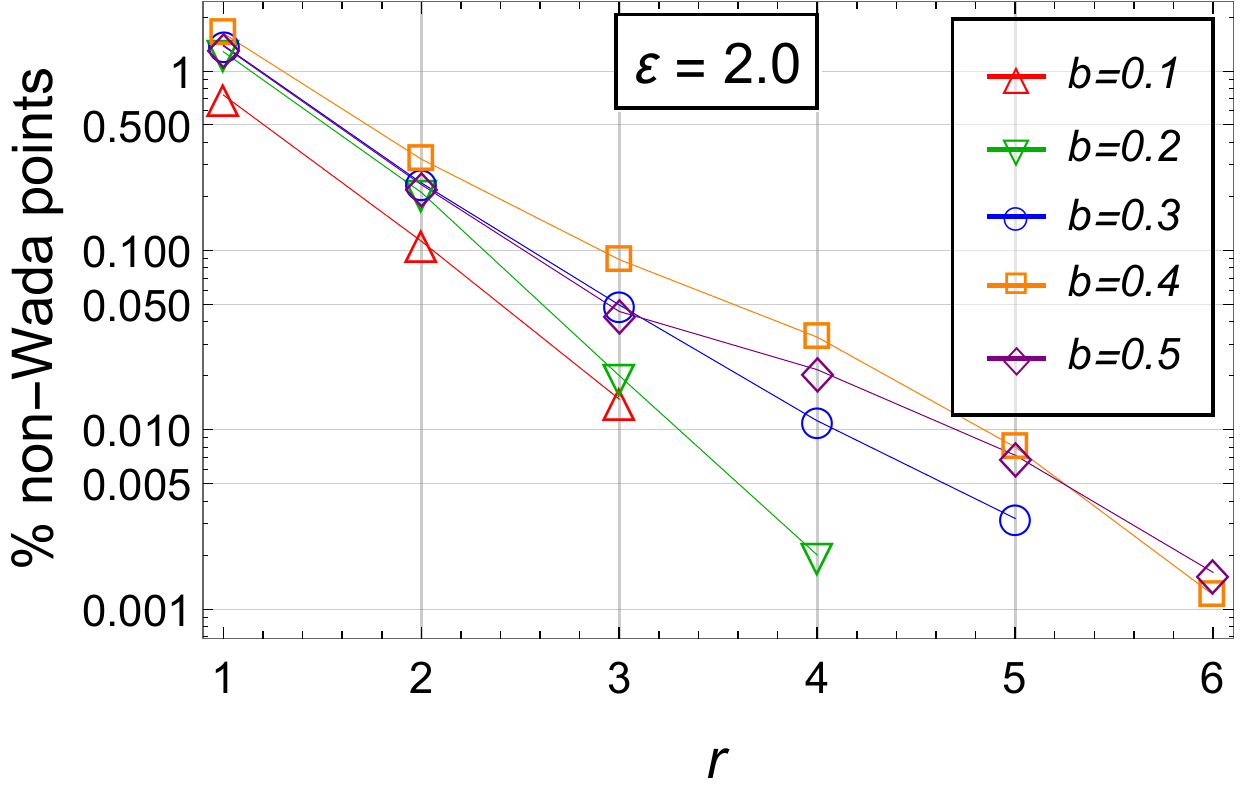}
\end{minipage}
\vskip 0.2cm
\begin{minipage}{0.6\linewidth}
\centering
\includegraphics[width=\linewidth]{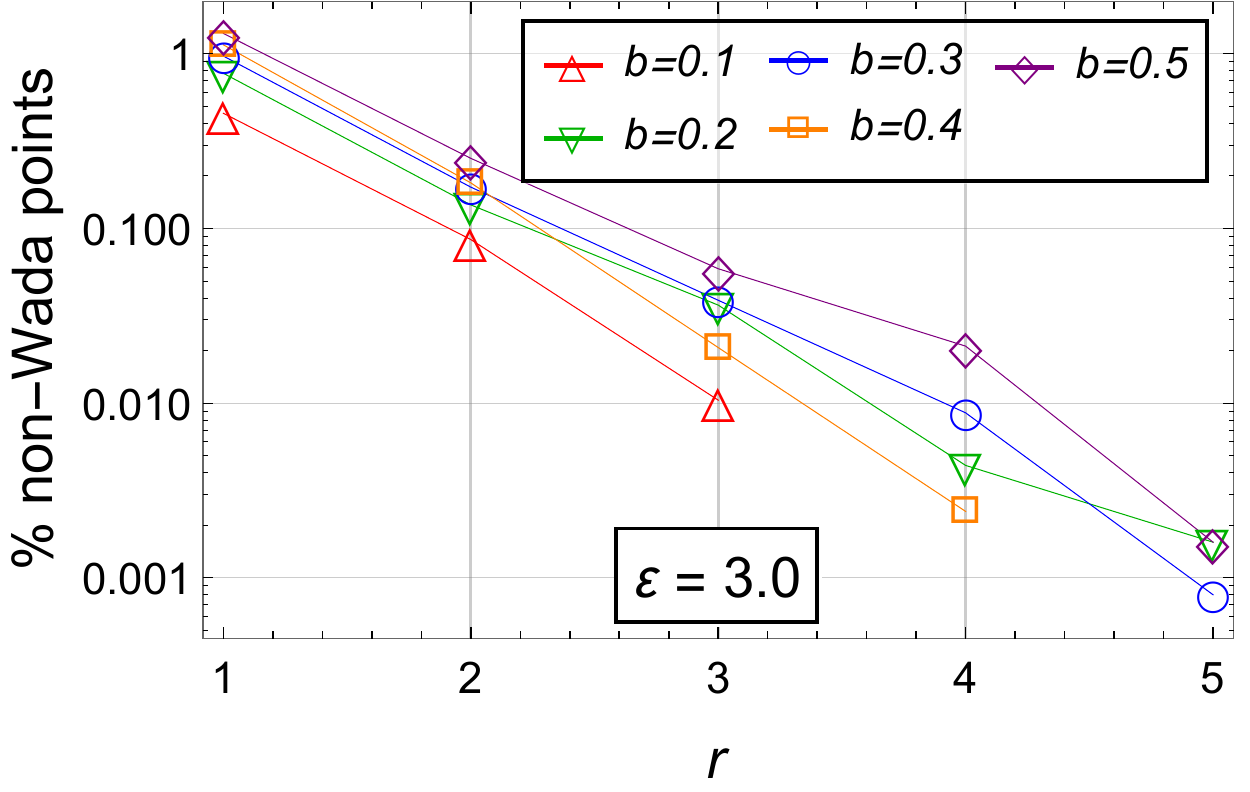}
\end{minipage}
\caption{\label{fig:percentnonwadacase2} Percentage of non-Wada points as a function of the fattening parameter $r$ for the basin plots obtained in \sref{sec:basinplotscase2}. Note that since the $y-$axis is on a log scale, for each case there is one additional data point $(r_\mathrm{max}+1,0)$.}
\end{center}
\end{figure}

In cases where all three of the basins coexist, we used the merging method to test whether the basin boundaries exhibit the Wada property. \Fref{fig:percentnonwadacase2} shows the percentage of non-Wada points as a function of the fattening parameter.
All of the basins were classified as Wada for $r \leq 9$ although for most cases, the percentage of non-Wada points are at most $0.01 \%$ for $r \leq 5$.
As was done in \sref{sec:wada}, we translate pixel distances $r$ to actual distances in the phase space we are considering.
We see from \fref{fig:initcond2basinplots}, we see that $(\Delta\phi_\mathrm{sh},\Delta z)=(2\pi,10)$.
For a $500\times 500$ resolution basin plot, a pixel of distance in this case would correspond to $\Delta r \sim 0.0236$.
Our results would indicate then that the boundaries in the basins we considered are the same to within $\sim 0.21$ units of phase space distance.
We show evidence in \ref{sec:wadaresolution} that the results of \fref{fig:percentnonwadacase2} hold for higher resolutions.

\begin{figure}
\begin{center}
\includegraphics[width=0.6\linewidth]{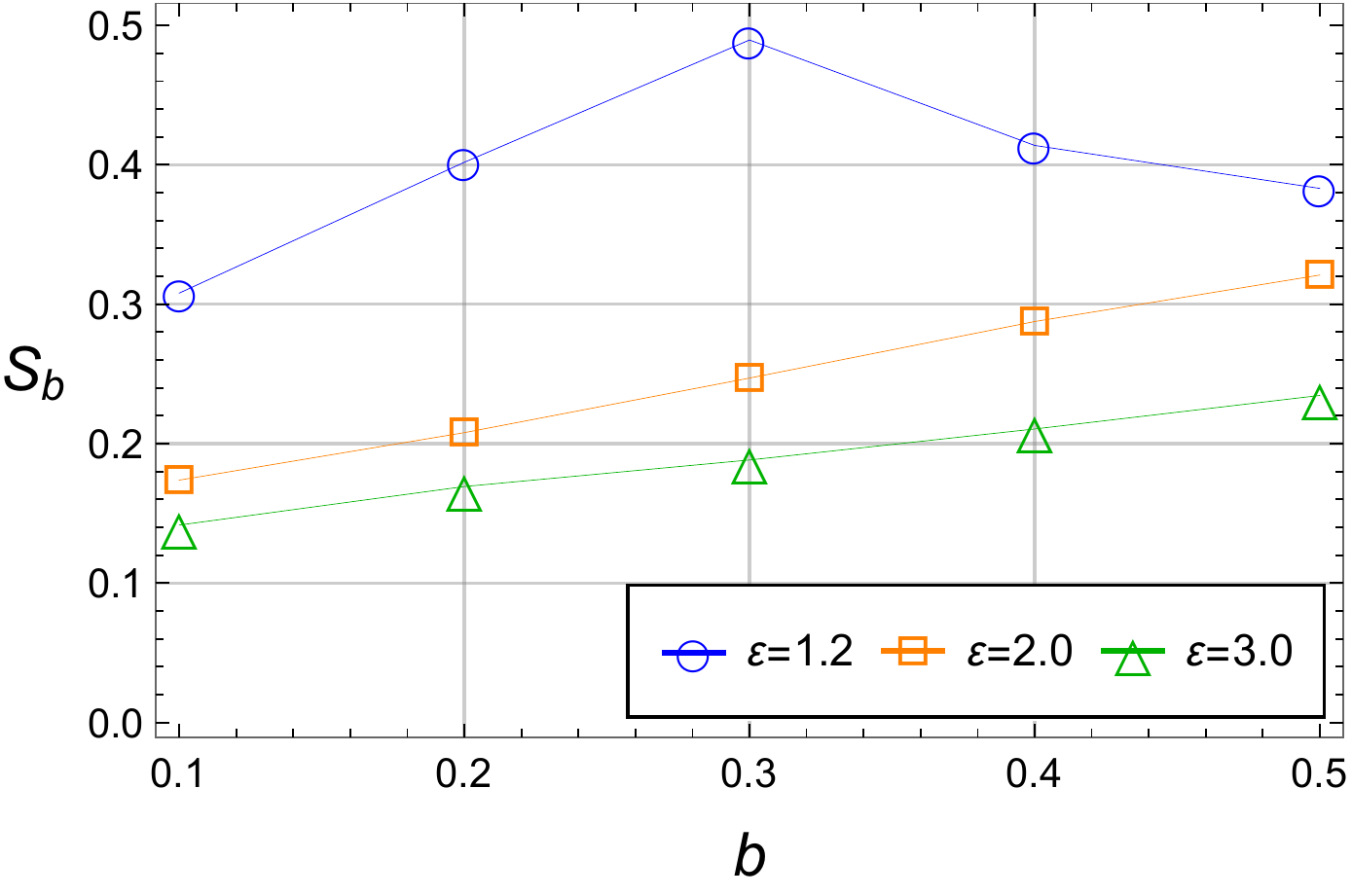}
\caption{\label{fig:basinentropycase2} The basin entropy as a function of the electromagnetic interaction strength $b$ for different values of $\mathcal{E}$ for the basins calculated in \sref{sec:basinplotscase2}.}
\end{center}
\end{figure}

Next, we obtained the basin entropy for the basins calculated in this section.
The basin entropy as a function of the electromagnetic interaction strength $b$ is shown in \fref{fig:basinentropycase2}.
We see that in cases where all three of the basins are present, the basin entropy increases with the strength of the electromagnetic interaction $b$.
This is in line with what we see from \fref{fig:initcond2basinplots} wherein the basins become more mixed as $b$ increases.
In the particular case of $\mathcal{E} = 1.2$, from $b=0.3$ to $b=0.4$, the basin entropy decreased.
This is because the basin entropy measure is sensitive to the number of basins present: the higher the number of basins, the higher the basin entropy.
In the same case, we notice a further decrease in basin entropy from $b=0.4$ to $b=0.5$.
The trend of increasing basin entropy as $b$ is increased might not hold in the vicinity of these ``phase transitions".

\section{Summary}
\label{sec:summary}
This paper was concerned with the motion of a charged particle in the vicinity of a weakly magnetized Schwarzschild black hole from the viewpoint of chaotic scattering and nonlinear dynamics. In the vicinity of the black hole, the charged particle undergoes transient chaotic motion and it can scatter off through either of three exits: $z\rightarrow \pm\infty$ or black hole capture. The system phase space can be partitioned into exit basins, according to the final state of each phase space point, and particular two-dimensional slices of these partitions can be visualized and analyzed with basin plots. We have confirmed the known result that the basin boundaries are fractal. In the case of a charged particle vertically kicked from a circular equatorial orbit, we have computed the critical escape energy, and determined a more general approximate expression for its dependence on orbital radius and magnetic field strength. 

More importantly, we have demonstrated that charged particle motion in this weakly magnetized environment exhibits the Wada property: its basin boundaries are shared by more than two basins. This implies extreme sensitivity to uncertainties in initial conditions. And so, despite the motion being fully deterministic, it is very challenging to predict the final state of initial conditions that belong to these Wada basins. Morever, we have shown that this Wada (or partial Wada) property persists in at least two different ways of slicing the phase space. This suggests that the Wada property might be a generic property of this phase space, independent of the way one chooses to slice it. This is another example of a relativistic system possessing the Wada property, similar to the case of a photon escaping from a Majumdar-Papapetrou black hole binary \cite{Daza2018}. It would be interesting to see the Wada property verified in other relativistic dynamical systems. 

Finally, we have quantified the uncertainty in the Wada basins using the notion of basin entropy. Our numerical experiments show that the basin entropy generally increases as we increase the magnetic interaction strength via the parameter $b$. Exceptions arise only when an increase $b$ results in the disappearance of a basin of attraction. i.e. when this renders an exit state kinematically forbidden.

\ack
The authors are grateful to Sam Dolan for helpful comments on an earlier draft, and to Sean Fortuna and Marc Perez for pointers during the early stages of this work. We also thank Martin Kolo\v{s} and Vojt\v{e}ch Witzany for drawing our attention to references we inadvertently missed in an earlier draft. J.B. thanks the Department of Science and Technology Advanced Science and Technology Human Resources Development Program - National Science Consortium (DOST ASTHRDP-NSC) for financial support. This research is supported by the University of the Philippines Diliman Office of the Vice Chancellor for Research and Development through Project No. 191937 ORG.

\appendix
\section{Dependence of the merging method on basin resolution}
\label{sec:wadaresolution}
\begin{figure}[b]
\begin{center}
\includegraphics[width=0.6\linewidth]{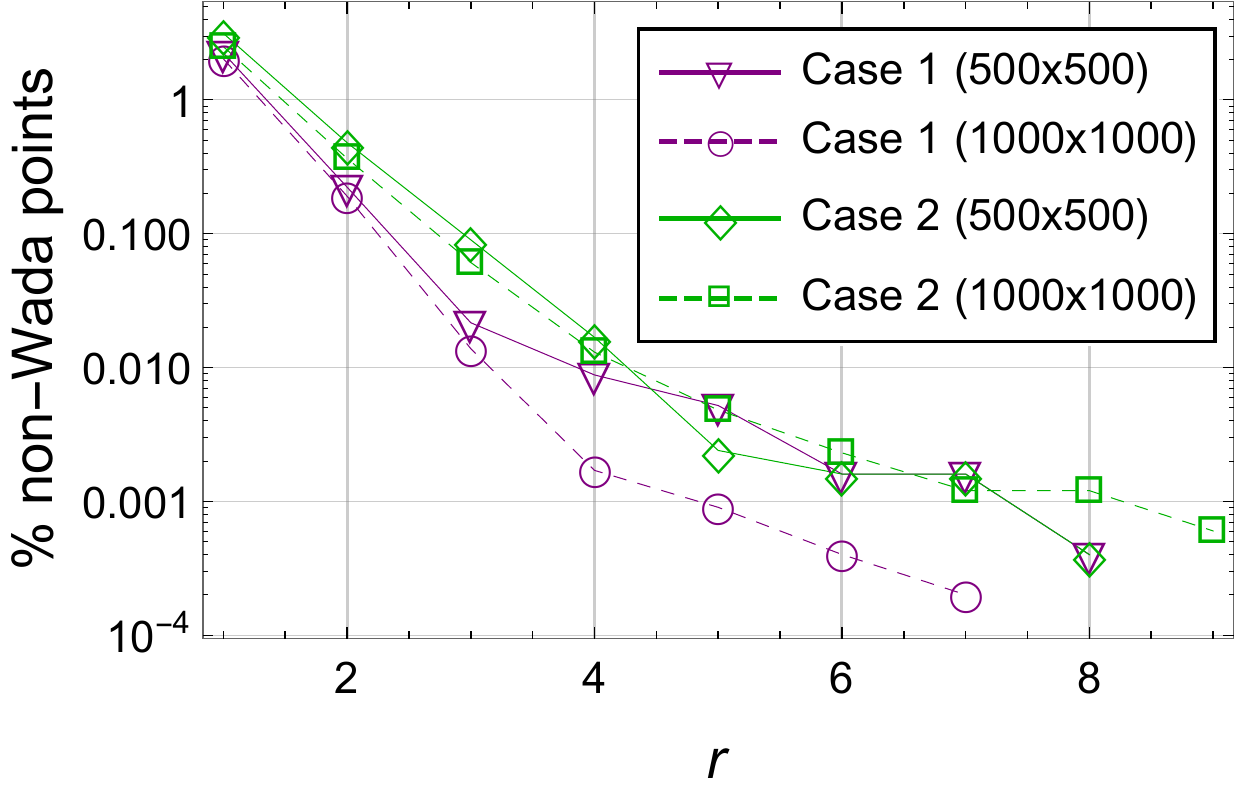}
\caption{\label{fig:wadaresocomparison} Percentage of non-Wada points as a function of the fattening parameter $r$ for at different resolutions. Case 1 corresponds to the the initial conditions in \sref{sec:basinplotscase1} with parameters $(\ell>0, b=0.5)$. Case 2 corresponds to the initial conditions in \sref{sec:basinplotscase2} with parameters $(\mathcal{E}=1.2, b=0.2)$. Note that since the $y-$axis is on a log scale, for each case there is one additional data point $(r_\mathrm{max}+1,0)$.}
\end{center}
\end{figure}

The authors of \cite{Daza2018a} report that results obtained using the merging method (i.e. the percentage of non-Wada points as a function of the fattening parameter $r$) does not change regardless of the resolution of the basin plot.
Nevertheless, results obtained using the merging method are expected to be more reliable for higher basin resolutions.
Here we check how conclusions based on the merging method depend on resolution. In two of the basin plots previously discussed, one from \sref{sec:basinplotscase1} and one from \sref{sec:basinplotscase2}, we increase the resolution from $500\times 500$ to $1000\times 1000$. \Fref{fig:wadaresocomparison} shows the lower resolution results being very similar to the higher resolution results, verifying the claims of \cite{Daza2018a}. This shows that the Wada property persists at higher resolution. 

\clearpage

\section*{References}
\bibliography{References}

\begin{thebibliography}{10}

\bibitem{Zeraoulia2012}
Elhadj Zeraoulia.
\newblock {\em Models and applications of chaos theory in modern sciences}.
\newblock CRC Press, 2012.

\bibitem{Seoane2013}
Jes{\'{u}}s~M. Seoane and Miguel A~F Sanju{\'{a}}n.
\newblock New developments in classical chaotic scattering.
\newblock {\em Reports Prog. Phys.}, 76(1):016001, 2013.

\bibitem{Barrio2010}
R.~Barrio, F.~Blesa, and S.~Serrano.
\newblock Bifurcations and chaos in {H}amiltonian systems.
\newblock {\em Int. J. Bifurc. Chaos}, 20(05):1293--1319, 2010.

\bibitem{Bleher1990}
Siegfried Bleher, Celso Grebogi, and Edward Ott.
\newblock Bifurcation to chaotic scattering.
\newblock {\em Phys. D Nonlinear Phenom.}, 46(1):87--121, 1990.

\bibitem{Gaspard1998}
Pierre Gaspard.
\newblock {\em Chaos, Scattering and Statistical Mechanics}.
\newblock Nonlinear Science Series. Cambridge University Press, 1998.

\bibitem{AlZahrani2013}
A.~M. {Al Zahrani}, Valeri~P. Frolov, and Andrey~A. Shoom.
\newblock Critical escape velocity for a charged particle moving around a
  weakly magnetized {S}chwarzschild black hole.
\newblock {\em Phys. Rev. D - Part. Fields, Gravit. Cosmol.}, 87(8):1--12,
  2013.

\bibitem{AlZahrani2014}
A.~M. {Al Zahrani}.
\newblock Escape of charged particles moving around a weakly magnetized {K}err
  black hole.
\newblock {\em Phys. Rev. D - Part. Fields, Gravit. Cosmol.}, 90(4):1--13,
  2014.

\bibitem{Daza2016}
Alvar Daza, Alexandre Wagemakers, Bertrand Georgeot, David Gu{\'{e}}ry-Odelin,
  and Miguel~A.F. Sanju{\'{a}}n.
\newblock Basin entropy: A new tool to analyze uncertainty in dynamical
  systems.
\newblock {\em Sci. Rep.}, 6:1--10, 2016.

\bibitem{Daza2018}
{\'{A}}lvar Daza, Jake~O. Shipley, Sam~R. Dolan, and Miguel~A.F. Sanju{\'{a}}n.
\newblock {W}ada structures in a binary black hole system.
\newblock {\em Phys. Rev. D}, 98(8):1--13, 2018.

\bibitem{Stuchlik2016}
Zden\v{e}k Stuchlík and Martin Kolo\v{s}.
\newblock Acceleration of the charged particles due to chaotic scattering in
  the combined black hole gravitational field and asymptotically uniform
  magnetic field.
\newblock {\em The European Physical Journal C}, 76(1), Jan 2016.

\bibitem{Kopacek2018}
Ond\v{r}ej Kop\'{a}\v{c}ek and Vladimír Karas.
\newblock Near-horizon structure of escape zones of electrically charged
  particles around weakly magnetized rotating black hole.
\newblock {\em The Astrophysical Journal}, 853(1):53, Jan 2018.

\bibitem{Tursunov2018}
Arman Tursunov, Martin Kolo\v{s}, Zden\'{e}k Stuchlík, and Dmitri~V.
  Gal’tsov.
\newblock Radiation reaction of charged particles orbiting a magnetized
  {S}chwarzschild black hole.
\newblock {\em The Astrophysical Journal}, 861(1):2, Jun 2018.

\bibitem{Horowitz1999}
Gary~T. Horowitz and Saul~A. Teukolsky.
\newblock Black holes.
\newblock {\em Rev. Mod. Phys.}, 71(2):S180--S186, 1999.

\bibitem{Punsly2009}
Brian Punsly.
\newblock {\em Black hole gravitohydromagnetics}.
\newblock Astrophysics and Space Science Library 355. Springer-Verlag Berlin
  Heidelberg, 2 edition, 2009.

\bibitem{Misner1973}
Charles~W. Misner, Kip~S. Thorne, and John~Archibald Wheeler.
\newblock {\em Gravitation}.
\newblock Physics Series. W. H. Freeman, first edition edition, 1973.

\bibitem{Aliev1989}
A~N Aliev and D~V Gal'tsov.
\newblock ``{M}agnetized" black holes.
\newblock {\em Sov. Phys. Uspekhi}, 32(1):75--92, 1989.

\bibitem{Wald1974}
Robert~M. Wald.
\newblock Black hole in a uniform magnetic field.
\newblock {\em Phys. Rev. D}, 10(6):1680--1685, 1974.

\bibitem{Frolov2010}
Valeri~P. Frolov and Andrey~A. Shoom.
\newblock Motion of charged particles near weakly magnetized {S}chwarzschild
  black hole.
\newblock {\em Phys. Rev. D - Part. Fields, Gravit. Cosmol.}, 82(8), 2010.

\bibitem{Landau1975}
L.D. Landau and E.M. Lifshitz.
\newblock {\em The Classical Theory of Fields (Fourth Edition)}, volume~2 of
  {\em Course of Theoretical Physics}.
\newblock Pergamon, 4 edition, 1975.

\bibitem{Blesa2014}
Fernando Blesa, Jes\'{u}s~M. Seoane, Roberto Barrio, and Miguel~A.F.
  Sanju\'{a}n.
\newblock Effects of periodic forcing in chaotic scattering.
\newblock {\em Phys. Rev. E - Stat. Nonlinear, Soft Matter Phys.}, 89(4):1--9,
  2014.

\bibitem{Aguirre2001}
Jacobo Aguirre, Juan~C. Vallejo, and Miguel~A.F. Sanju{\'{a}}n.
\newblock {W}ada basins and chaotic invariant sets in the {H{\'{e}}non-Heiles}
  system.
\newblock {\em Phys. Rev. E - Stat. Physics, Plasmas, Fluids, Relat.
  Interdiscip. Top.}, 64(6):11, 2001.

\bibitem{Seoane2006}
Jes{\'{u}}s~M. Seoane, Jacobo Aguirre, Miguel~A.F. Sanju{\'{a}}n, and
  Ying~Cheng Lai.
\newblock Basin topology in dissipative chaotic scattering.
\newblock {\em Chaos}, 16(2), 2006.

\bibitem{Bernal2018}
Juan~D. Bernal, Jes\'{u}s~M. Seoane, and Miguel~A.F. Sanju\'{a}n.
\newblock Uncertainty dimension and basin entropy in relativistic chaotic
  scattering.
\newblock {\em Phys. Rev. E}, 97(4):1--9, 2018.

\bibitem{FrolovV1999}
Andrei {Frolov V} and Arne~L. Larsen.
\newblock Chaotic scattering and capture of strings by a black hole.
\newblock {\em Class. Quantum Gravity}, 16(11):3717--3724, 1999.

\bibitem{Liu2017}
Chen-Yu Liu, Da-Shin Lee, and Chi-Yong Lin.
\newblock Geodesic motion of neutral particles around a {Kerr–Newman} black
  hole.
\newblock {\em Class. Quantum Gravity}, 34(23):235008, dec 2017.

\bibitem{Aguirre2009}
Jacobo Aguirre, Ricardo~L. Viana, and Miguel~A.F. Sanju{\'{a}}n.
\newblock Fractal structures in nonlinear dynamics.
\newblock {\em Rev. Mod. Phys.}, 81(1):333--386, 2009.

\bibitem{Grebogi1983}
Celso Grebogi, Steven~W. McDonald, Edward Ott, and James~A. Yorke.
\newblock Final state sensitivity: An obstruction to predictability.
\newblock {\em Phys. Lett. A}, 99(9):415--418, 1983.

\bibitem{Ott2002}
Edward Ott.
\newblock {\em Chaos in Dynamical Systems}.
\newblock Cambridge University Press, 2002.

\bibitem{DataGitHub}
Joshua Bautista.
\newblock Critical escape energy data for a charged particle in a weakly
  magnetized {S}chwarzschild black hole.
\newblock \url{https://github.com/josh0429/charge-in-schd-esc-energy}.

\bibitem{Yoneyama1917}
Kunizo Yoneyama.
\newblock Theory of continuous set of points (not finished).
\newblock {\em Tohoku Math. J., First Series}, 12:43--158, 1917.

\bibitem{Hocking1961}
John~G. Hocking and Gail~S. Young.
\newblock {\em Topology}.
\newblock 1961.

\bibitem{Aguirre2002}
Jacobo Aguirre and Miguel~A.F. Sanju{\'{a}}n.
\newblock Unpredictable behavior in the duffing oscillator: {W}ada basins.
\newblock {\em Phys. D Nonlinear Phenom.}, 171(1-2):41--51, 2002.

\bibitem{Nusse1996a}
Helena~E. Nusse and James~A. Yorke.
\newblock {W}ada basin boundaries and basin cells.
\newblock {\em Phys. D Nonlinear Phenom.}, 90(3):242--261, 1996.

\bibitem{Kennedy1991}
Judy Kennedy and James~A. Yorke.
\newblock Basins of {W}ada.
\newblock {\em Physica D}, 51(1-3):213--225, 1991.

\bibitem{Daza2015}
Alvar Daza, Alexandre Wagemakers, Miguel~A.F. Sanju{\'{a}}n, and James~A.
  Yorke.
\newblock Testing for basins of {W}ada.
\newblock {\em Sci. Rep.}, 5:1--7, 2015.

\bibitem{Wagemakers2020}
Alexandre Wagemakers, Alvar Daza, and Miguel~A.F. Sanju{\'{a}}n.
\newblock The saddle-straddle method to test for {W}ada basins.
\newblock {\em Commun. Nonlinear Sci. Numer. Simul.}, 84:1--9, 2020.

\bibitem{Daza2018a}
Alvar Daza, Alexandre Wagemakers, and Miguel~A.F. Sanju{\'{a}}n.
\newblock Ascertaining when a basin is {W}ada: the merging method.
\newblock {\em Sci. Rep.}, 8(1):1--8, 2018.

\bibitem{Zhang2013}
Yongxiang Zhang and Guanwei Luo.
\newblock {W}ada bifurcations and partially {W}ada basin boundaries in a
  two-dimensional cubic map.
\newblock {\em Phys. Lett. Sect. A Gen. At. Solid State Phys.},
  377(18):1274--1281, 2013.

\bibitem{Menck2013}
Peter~J. Menck, Jobst Heitzig, Norbert Marwan, and J{\"{u}}rgen Kurths.
\newblock How basin stability complements the linear-stability paradigm.
\newblock {\em Nat. Phys.}, 9(2):89--92, 2013.

\end{thebibliography}
\bibliographystyle{unsrt}

\end{document}